\documentclass[8pt,a4paper]{article}

\usepackage[english]{babel}

\usepackage{natbib}

\usepackage{amsfonts,amsbsy,amssymb,amsmath,graphicx,float,authblk} 

\usepackage[paperwidth=18cm,paperheight=29.7cm,hmargin=2cm,vmargin=3cm]{geometry}

\usepackage[linktoc=all,colorlinks=true,citecolor=blue]{hyperref}

\begin{document}
	
\title{Exploring Isomerization Dynamics on a \\ Potential Energy Surface with an Index-2 Saddle \\ using Lagrangian Descriptors}
	
\author{V. J. Garc\'ia-Garrido$^{1}$, M. Agaoglou$^{2}$, S. Wiggins$^{1}$ \\
$^1$ Departamento de F\'{i}sica y Matem\'{a}ticas, \\ Universidad de Alcal\'{a}, 28871, Alcal\'{a} de Henares, Spain. \\ vjose.garcia@uah.es \\
\vspace{.3cm} $^2$School of Mathematics, University of Bristol, \\ Fry Building, Woodland Road, Bristol, BS8 1UG, United Kingdom. \\ makrina.agaoglou@bristol.ac.uk , s.wiggins@bristol.ac.uk}

\maketitle

\begin{abstract}

In this paper we explore the phase space structures governing isomerization dynamics on a potential energy surface with four wells and an index-2 saddle. For this model, we analyze the influence that coupling both degrees of freedom of the system and breaking the symmetry of the problem have on the geometrical template of phase space structures that characterizes reaction. To achieve this goal we apply the method of Lagrangian descriptors, a technique with the capability of unveiling the key invariant manifolds that determine transport processes in nonlinear dynamical systems. This approach reveals with extraordinary detail the intricate geometry of the isomerization routes interconnecting  the different potential wells, and provides us with valuable information to distinguish between initial conditions that undergo sequential and concerted isomerization.   

\end{abstract}


% \tableofcontents

\section{Introduction}
\label{sec:intro}

The simplest description of an isomerization reaction is as a transition between two minima (wells) on the potential energy surface (PES) through an index-1 saddle point. More precisely, the transition occurs via a trajectory moving along a path on the PES  from the index-1 saddle to the minimum that minimizes the potential energy (``minimum energy path’’, MEP). However, if the PES contains multiple minima and, in addition to index-1 saddles, index-2 and higher order saddles, isomerization can become a very complex dynamical phenomenon. For example, it may be possible for trajectories to transition between wells that are not separated by a single index-1 saddle. Instead, the wells may be connected by a sequence of MEPs constructed from more than one index-1 saddle (sequential isomerization) or, if the PES has an index two saddle, it may be possible for a trajectory to transition from one well to another by bypassing all of the traditional MEPs associated with index-1 saddles and transition via a non-traditional path created by the higher index saddle (concerted isomerization).

In the context of chemical reaction dynamics, phase space plays a central role as it allows to rigorously establish some of the fundamental concepts introduced by Transition State Theory to measure the reactivity flux of a chemical reaction, and derive from them reaction rates. The long-sought dividing surface (DS) with the no-recrossing property  separating reactants from products that any reactive trajectory in phase space has to cross, is a geometrical structure in phase space that can be constructed from normally hyperbolic invariant manifolds (NHIMs) associated to index-1 saddles of the PES. But the energy landscape of a PES can be rather complicated, and the presence of higher order saddles is typical, so that the study of the phase space structures induced by them and how they impact the dynamical behavior of the system is fundamental to understand the underlying mechanisms of more complex isomerization reactions. Recently, \cite{Ezra2009} has described the phase space structures related to index-2 saddles of a PES, and their influence on reaction dynamics for Hamiltonian systems with $N$ degrees-of-freedom (DoF) was addressed in \cite{collins2011}. While higher index saddles have not received a great deal of attention in the study of reaction mechanisms,  \cite{collins2011} pointed out that they are central for the understanding of a number of reaction mechanisms in diverse settings, such as organic reactions, clusters, and the classical dynamics of the ionization of helium in a constant electric field.  Since this work, new applications of index-2 saddles to reaction dynamics have appeared. For example, \cite{Nagata2013a} studied the possibility of extracting the reactivity boundaries, which are vital for revealing the mechanism of reactions associated to index-2 saddles. In \cite{Quapp2015}, a generalized gentlest ascent algorithm was applied to determine an index-2 saddle and study the PES of the ring opening of Cyclobutene. Recently, \cite{pradhan2019can} discusses a computational study of the mechanisms of the denitrogenation of 1-pyrazoline and its substituted analogue and show that an index-2 saddle plays a central role in ``organizing’’ the reactions paths and explaining the nature of non-traditional reaction paths. Studies in \cite{shepler2011roaming} and \cite{harding2012separability} of the recently discovered ‘’roaming mechanism’’ for chemical reactions indicate that an index-2 saddle might play a role in explaining this mechanism for certain chemical reactions. Another example where the importance of index-2 and higher order saddles for chemical reactions is analyzed is \cite{mauguiere2013bond}, where a Morse chain under tension is studied to demonstrate that higher order saddles are relevant for the simultaneous breaking of bonds. A detailed background on the phase space approach to reaction dynamics from a dynamical systems point of view can be found in \cite{Agaoglou2019}.

%and later in the paper \cite{Nagata2013b} the reactivity boundaries associated with higher-index and multiple saddles. Their studies reveals which initial conditions are effective in making the reaction happen and explain the reaction mechanism. 

In this paper we continue the study of \cite{collins2011} by analyzing the phase space structures that govern different reaction pathways in a model PES with four wells separated by index-1 saddles, and an index-2 saddle at the origin. We deal both with the symmetric case introduced in \cite{collins2011} and also explore the influence of including a symmetry-breaking perturbation on the system that gives rise to interesting saddle-node bifurcations. Our goal is to analyze, in terms of the model parameters, the geometrical template of phase space structures, i.e. the underlying isomerization pathways, that characterizes the dynamical behavior of the system. We do so by applying the method of Lagrangian descriptors (LDs) which is a scalar diagnostic developed in the context of nonlinear dynamics to reveal the invariant manifolds in phase space. We will show how this technique allows us to extract a complete \textit{phase space tomography} of the dynamical skeleton of isomerization chemical reactions.

The contents of this paper are outlined as follows: In Section \ref{sec:sec1} we introduce the Hamiltonian system for which we study the phase space structures that govern isomerization dynamics on a potential energy surface with four wells and an index-2 saddle. Section \ref{sec:sec2} is devoted to describing our results in the context of the uncoupled system, and Section \ref{sec:sec3} discusses the effects that  coupling among the degrees-of-freedom has on the system dynamics. Finally, in Section \ref{sec:conc} we present a summary of the conclusions of this work. The reader can find additional material on the method of Lagrangian descriptors in Appendix \ref{sec:appA}, where a detailed explanation is given on how this technique can be applied to reveal the geometrical template of invariant manifolds in the high-dimensional phase space of Hamiltonian systems. To conclude, we provide in Appendix \ref{sec:appB} an analysis of the equilibrium points of the Hamiltonian system in terms of the model parameters.

\section{Hamiltonian Model for Isomerization Dynamics}
\label{sec:sec1}

In this section we describe the Hamiltonian model with two degrees-of-freedom (DoF) that is the main focus of this work. Our starting point is the Hamiltonian system introduced in Collins et al. \cite{collins2011} to analyze isomerization dynamics on a PES with an index-2 saddle at the origin, and four symmetric potential wells separated by index-1 saddles. We recall that, for a two-dimensional PES, an index-1 saddle is a critical point of saddle nature, a potential well is a local minimum, and an index-2 saddle corresponds to a local maximun, i.e. a ``hilltop''. We will give more details about the phase space dynamical significance of these critical points of the PES in Section \ref{sec:sec2}. The potential energy of the model system described in Collins et al. \cite{collins2011} is given by:
\begin{equation}
\mathcal{V}(x,y) = x^4 - \alpha x^2 + y^4 - y^2 + \beta  x^2 y^2 \;,
\label{pes_modelCollins}
\end{equation}
where the potentials in the $x$ and $y$ DoF have a double well structure, and both DoF are coupled. In this setup, the model parameter $\alpha$ measures the barrier height corresponding to the potential of the $x$ DoF, and $\beta$ represents the coupling strength between both DoF in the system. In this work, we study a slightly modified version of the model PES in Eq.  \eqref{pes_modelCollins} where we introduce the effect of asymmetry in the double well potential of the $x$ DoF following the approach described in \cite{brizzard2017}. Consider the classical Hamiltonian obtained as the sum of kinetic plus potential energy:
\begin{equation}
H(x,y,p_x,p_y) = \dfrac{1}{2} \, p_x^2  + \dfrac{1}{2} \, p_y^2  + V(x,y) \;,
\label{hamiltonian}
\end{equation}
where we have supposed for simplicity that the mass in each DoF is $m_x = m_y = 1$. If we define $\delta$ as the model parameter representing the asymmetry in the double well potential of the $x$ DoF, the PES can be described as follows:
\begin{equation}
V(x,y) = x^4 - \alpha x^2 - \delta x + y^4 - y^2 + \beta x^2 y^2 \;.
\label{pes_model}
\end{equation}
For all the results discussed in this paper we have fixed $\alpha = 1$, so that the double well potentials in the $x$ and $y$ DoF have the same quadratic term. The dynamical evolution of the Hamiltonian system in Eq. \eqref{hamiltonian} takes place in a four-dimensional phase space, and is determined by Hamilton's equations of motion:
\begin{equation}
\begin{cases}
\dot{x} = \dfrac{\partial H}{\partial p_x} = p_x \\[.4cm]
\dot{y} = \dfrac{\partial H}{\partial p_y} = p_y \\[.4cm]
\dot{p}_x = -\dfrac{\partial H}{\partial x} = -\dfrac{\partial V}{\partial x} = -4 x^3 + 2 \alpha x + \delta - 2 \beta x y^2 \\[.4cm]
\dot{p}_y = -\dfrac{\partial H}{\partial y} = -\dfrac{\partial V}{\partial y} = -4 y^3 + 2 y - 2 \beta x^2 y
\end{cases}
\;.
\label{ham_eqs}
\end{equation}
Moreover, due to energy conservation, the system dynamics is restricted to a three-dimensional energy hypersurface embedded in the full phase space. 

Our goal in this paper is to study isomerization dynamics on the PES described in Eq. \eqref{pes_model} in terms of the asymmetry and coupling parameters, looking at several energy values for the Hamiltonian system. The question that we address with this work is the following: Given an initial condition located on the lower-left well of the PES, can this trajectory reach at some point along its evolution the upper-right well region? Moreover, if this is the case, can we identify the initial conditions that follow this path and what are the underlying phase space structures that govern this dynamical mechanism? We provide a positive answer to these questions. 

\begin{figure}[htbp]
	\begin{center}
	\includegraphics[scale=0.25]{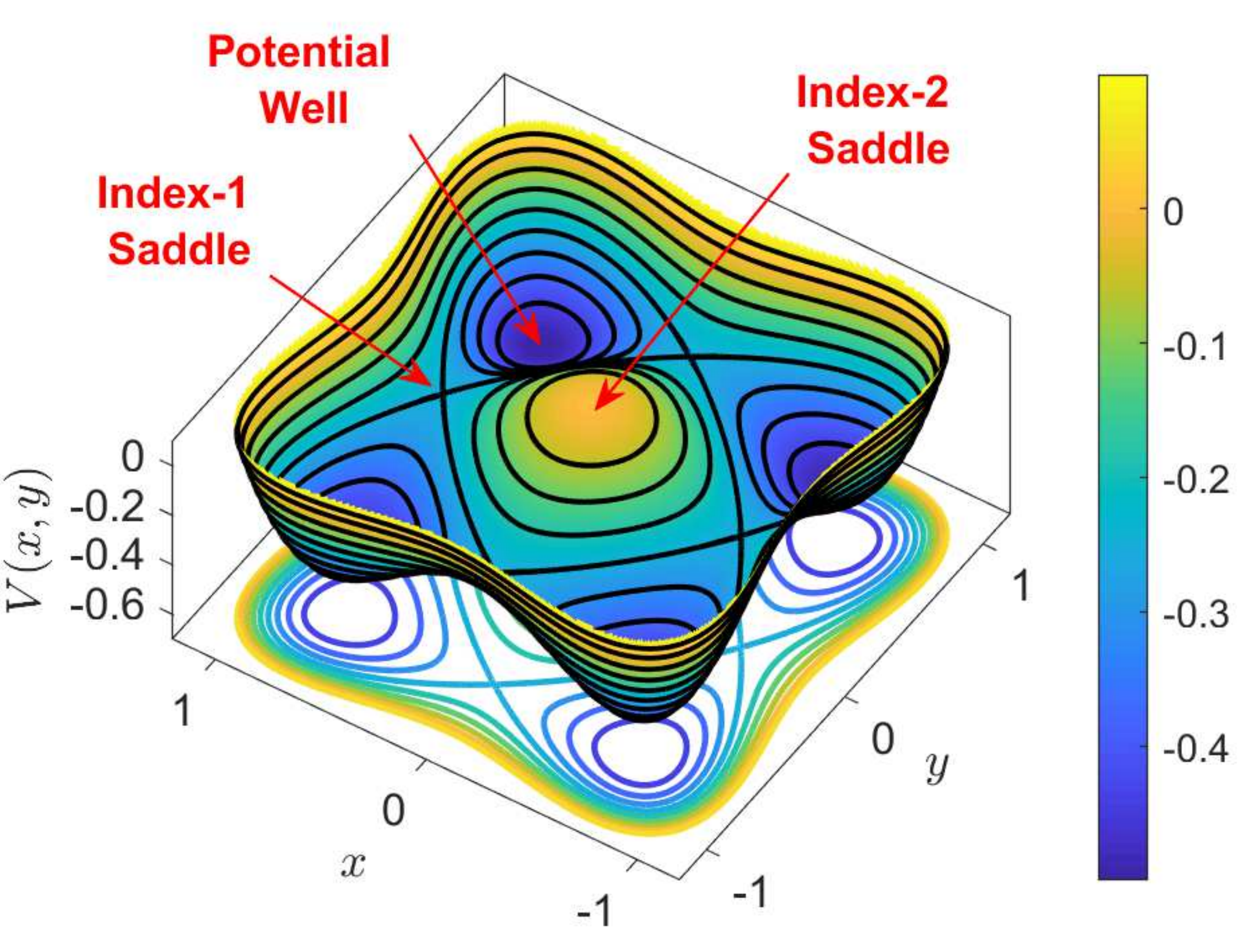}
	\includegraphics[scale=0.29]{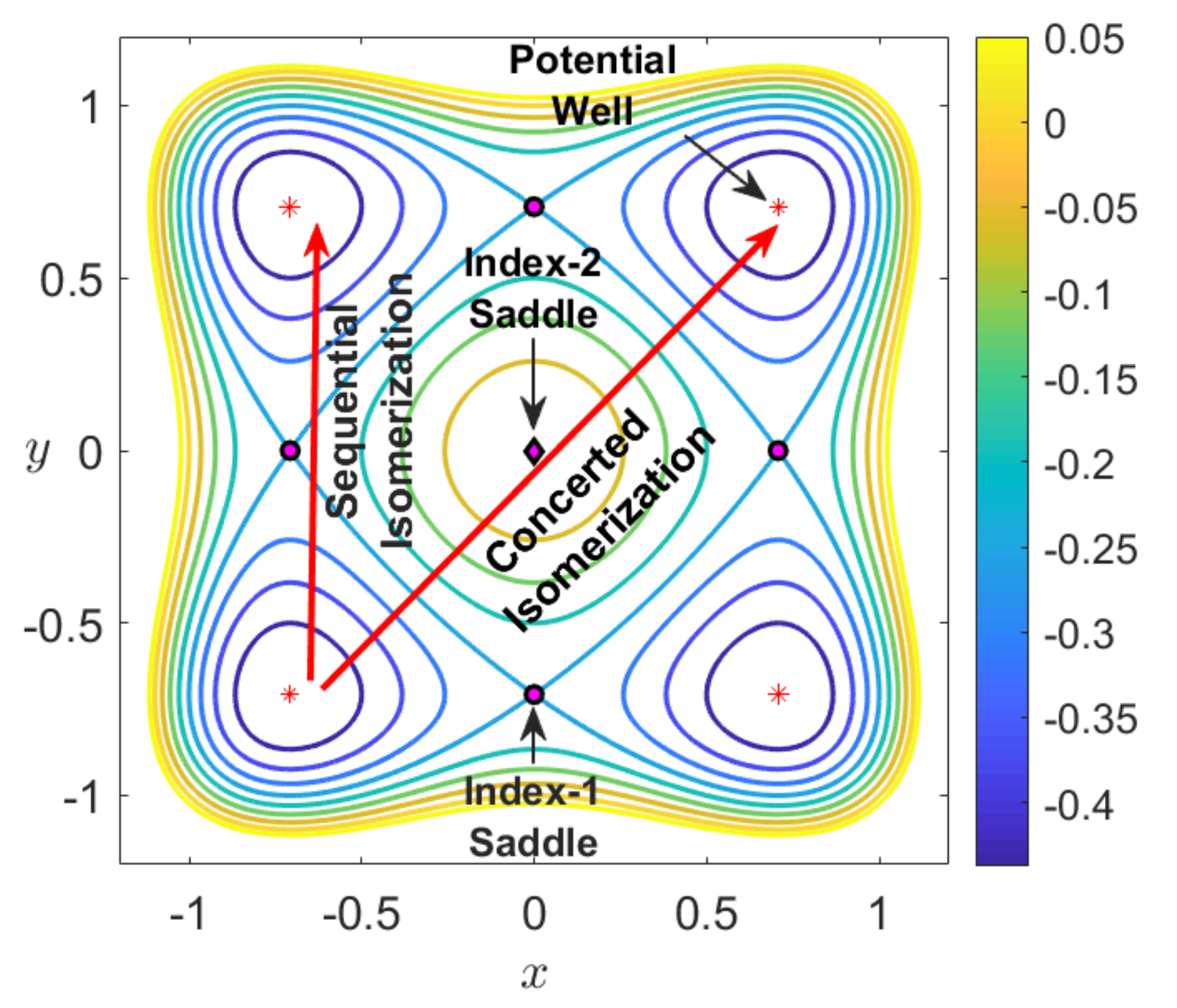}
	\end{center}
	\caption{Potential energy surface in Eq. \eqref{pes_model} for the uncoupled ($\beta = 0$) and symmetric ($\delta = 0$) Hamiltonian system.}
	\label{PES_isom_rout}
\end{figure}

In the framework of chemical reaction dynamics, each well in the PES would correspond to a stable isomer confi\-guration of the molecule under study. For the isomerization model that we consider here, the wells are interconnected by saddle critical points of the PES, which are identified with chemical transition states. Therefore, when the energy of the molecule is higher than that of a saddle point connecting two neighboring wells, then the state of the system can transit from well to well, and thus isomerization is possible. This type of isomerization is known in the literature as \textit{sequential isomerization}. Interestingly, another type of isomerization route exists for energy landscapes that exhibit an index-2 saddle in their topography, which consists of trajectories that transit from a given well to a diametrically opposing well through the ``hilltop'' of the PES. This is known as \textit{concerted isomerization}, and it is a mechanism that can occur when the energy of the system is above that of the index-2 saddle. In Fig. \ref{PES_isom_rout} we show a representation of the PES in Eq. \eqref{pes_model} together with the two possible isomerization routes that can occur in the problem for the uncoupled ($\beta = 0$) and symmetric ($\delta = 0$) Hamiltonian system.

\section{Analysis of the Uncoupled System}
\label{sec:sec2}

This section is devoted to describing the dynamics of the Hamiltonian in Eq. \eqref{hamiltonian} in the case where both DoF are uncoupled, so that the coupling strength parameter is set to $\beta = 0$. We start our analysis by focusing on the symmetric Hamiltonian with $\delta = 0$, and later we will move on to discuss the dynamical effect that symmetry-breaking perturbations of the double well potential in the $x$ DoF has on the system. 

\subsection{Dynamics of the Symmetric Problem}

Consider the symmetric and uncoupled system with energy $H_0$. Since the system is conservative, dynamics is constrained to the three-dimensional energy hypersurface:
\begin{equation}
\mathcal{S}(H_0) = \left\{ (x,y,p_x,p_y) \in \mathbb{R}^4 \; \bigg| \; H_0 = \frac{1}{2}\left(p_x^2+p_y^2\right) + x^4 - x^2 + y^4 - y^2 \right\} \;,
\end{equation} 
and the total energy of the system can be split between both DoF to yield:
\begin{equation}\label{eqsymun}
H(x,y,p_x,p_y) = H_{x}(x,p_x) + H_{y}(y,p_y) \;,
\end{equation}
so that the partitioned energy is:
\begin{equation}
H_{x}(x,p_x) = \frac{1}{2} \, p_x^2 + W(x) \quad,\quad H_{y}(y,p_y) = \frac{1}{2} \, p_y^2 + W(y) \;,
\end{equation}
where the potential in each DoF is a double-well in the form:
\begin{equation}
W(z) = z^4 - z^2 \;.
\label{1D_potSymm}
\end{equation}
Since the energy in each DOF is conserved separately, we have two independent constants of the motion in involution with each other. Therefore, the symmetric and uncoupled system is integrable, implying that all motion is regular and chaotic dynamics is not allowed. Before describing the equilibrium points of Hamilton's equations for this system, let us introduce some useful terminology to analyze later the results of this work. The Hill's region, which has its origins in Celestial Mechanics, is defined as the intersection of the energy hypersurface with configuration space, that is:
\begin{equation}
\mathcal{C}(H_0) = \left\{ (x,y,p_x,p_y) \in \mathbb{R}^4 \; \big| \; x^4 - x^2 + y^4 - y^2 \leq H_0 \right\}
\end{equation}
This region determines the energetically allowed configurations for the DoFs of the system, and all the points outside the Hill's region are energetically forbidden. The boundary of the Hill's region, $\partial \mathcal{C}(H_0)$, is known in the literature as the zero velocity curve and corresponds to phase space points for which kinetic energy is zero.

In order to determine the phase space structures that characterize isomerization dynamics, we need to look first at the equilibrium points of Hamilton's equations. A phase space point  $\mathbf{x}_e = (x_e,y_e,p_{x,e},p_{y,e})$ is an equilibrium of Eq. \eqref{ham_eqs} when it satisfies $p_{x,e} = p_{y,e} = 0$ and $\nabla V(x_e,y_e) = \mathbf{0}$, that is, it is a point located in configuration space and also a critical point of the PES. The local stability of this stationary point is determined by the eigenvalues of the Jacobian matrix obtained by linearizing Hamilton's equations in its neighborhood. In the symmetric and uncoupled system, we have 9 equilibrium points with configuration coordinates and energies given below: 
\begin{itemize}
	\item Four potential wells at the points $(\pm\sqrt{2}/2,\pm\sqrt{2}/2)$ with energies $V(\pm\sqrt{2}/2,\pm\sqrt{2}/2) = -1/2$. \\[-.5cm]
	\item Four index-1 saddles located at $(\pm\sqrt{2}/2,0)$, $(0,\pm\sqrt{2}/2)$ with energies $V(\pm\sqrt{2}/2,0) = V(0,\pm\sqrt{2}/2) = -1/4$. \\[-.5cm]
	\item One index-2 saddle at the origin with energy $V(0,0) = 0$.
\end{itemize}
Potential wells of a PES correspond to center-stability equilibrium points of the Hamiltonian, and are characterized by the fact that the Hessian matrix of the PES evaluated at these critical points, denoted by $\text{Hess}_V$, has a pair of real and positive eigenvalues. This results in the Jacobian of the linearization having two pairs of complex (and purely imaginary) eigenvalues, which yields quasiperiodic motion in their neighborhood. In the context of chemical reaction dynamics, potential wells correspond to stable isomer configurations of the molecule under study. On the other hand, index-1 saddles of the PES are identified with saddle points of the potential energy landscape, and the associated phase space equilibrium point is of saddle-center stability type. This means that $\text{Hess}_V$ has two real eigenvalues of different sign, which is equivalent to the Jacobian having a pair of real eigenvalues of opposite sign and a pair of purely imaginary eigenvalues. Geometrically, an index-1 saddle is a local maximum in one direction and a local minimum in a perpendicular direction, as seen in the normal coordinates associated to the eigenvectors. Moreover, the eigenvector pointing in the maximum direction can be taken as a local approximation to define the reaction coordinate. The role that the index-1 saddles have for isomerization in the model PES that we  consider here is to connect pairs of wells, and therefore, they control access of phase space trajectories from well to well. From the perspective of chemical reactions, index-1 saddles are identified with transition state configurations of the given molecule, and, as we will see shortly, the phase space structures they give rise to are essential for the accurate computation of chemical reaction rates. Finally, index-2 saddles of a two-dimensional PES have the geometrical shape of a local maximum, i.e. a ``hilltop'', which is characterized by the fact that the Hessian of the PES has two real and negative eigenvalues. This implies that when we linearize Hamilton's equations about this type of equilibrium point, the Jacobian yields two pairs of real eigenvalues, each pair having opposite signs. Thus, their stability is of saddle-saddle type, which has the dynamical effect of deflecting incoming trajectories in the neighborhood of an index-2 saddle at an exponential rate.

At this point, it is important to define what we mean for a trajectory to be trapped in one of the wells of the PES. From the energy landscape displayed in Fig. \ref{PES_isom_rout}, this concept can be determined by studying the change in the sign of the configuration coordinates $x$ and/or $y$. For example, if we focus on the lower-left well, we can say that a trajectory is trapped in that well when the configuration coordinates of both $x$ and $y$ DoF remain negative during its evolution. Moreover, in this setup we can identify \textit{reactive} trajectories as those that move from one well to another along their evolution. Since we are dealing in this case with a symmetric PES with respect to the origin, and there is one well in each quadrant, reaction would imply a change in the sign of one of the configuration coordinates of the trajectory. Notice that two types of reactive trajectories are possible when considering an initial condition starting on the lower-left well whose potential destination is the upper-right well. First, we could have sequential isomerization which implies transition from the lower-left well to an adjacent well through the phase space bottleneck region in the neighborhood of the index-1 saddle point that sits between both wells. This situation is only possible given that the system has enough energy to surmount the energy barrier (the energy of the index-1 saddle separating the wells). A second alternative for reaching the upper-right well from the lower-left well is that trajectories cross directly through the index-2 saddle located at the origin (the hilltop of the PES), which is known as concerted isomerization. For an illustration of these two isomerization processes refer to Fig. \ref{PES_isom_rout}.

We describe next the isomerization dynamics for the symmetric and uncoupled Hamiltonian system in terms of the phase space geometrical structures associated to the index-1 and index-2 saddles present in the model PES. We begin our discussion by fixing a value $H_0$ for the total energy of the system. As we have mentioned earlier, this energy is distributed among both DoF, so that we can write $H_0 = H_{x,0} + H_{y,0}$ where $H_{x,0} $ and $H_{y,0}$ are the energies in the $x$ and $y$ DoFs respectively. We discuss all the different cases in terms of the energy below: 
\begin{itemize}
	\item \underline{\textbf{Energy Level} $(-1/2 \leq H_0 \leq -1/4)$:} For this case the wells of the PES are isolated from each other since the energy of the system is below that of the index-1 saddles that interconnect the wells. Therefore, motion of an initial condition that starts in one of the wells will remain trapped in that well forever, displaying librational quasiperiodic motion.
	
	\item \underline{\textbf{Energy Level} $(-1/4 < H_0 \leq 0)$:} In this situation the energy is above that of any of the index-1 saddles in the PES, but below the energy of the index-2 saddle at the origin. Therefore, all potential wells are connected and isomerization can take place. This allows the transit of trajectories from well to well through the phase space bottlenecks that open in the neighborhood of the equilibrium points associated to the index-1 saddles. However, for this energy level the index-2 saddle region of the PES is energetically forbidden, and hence, only sequential isomerization is allowed. 
	
	Given that the system is completely symmetric in this case, in order to describe the phase space structures that govern isomerization dynamics and characterize the bottleneck regions in the vicinity of the index-1 saddles of the PES, we focus our analysis on the equilibrium point $(0,\sqrt{2}/2,0,0)$ associated to the upper index-1 saddle, separating the upper-left and upper-right wells. Since this stationary point has saddle-center type stability, we know from Lyapunov subcenter theorem, see e.g. \cite{weinstein1973,moser1976,rabinowitz1982}, that a family of normally hyperbolic invariant manifolds (NHIMs) bifurcates as the energy of the system is increased above that of the index-1 saddle. Recall that a NHIM is a dynamical structure exhibiting saddle-like stability in the directions transverse to it \cite{wiggins2013normally,wiggins2016}. For Hamiltonian systems with two DoF, the NHIM is an unstable periodic orbit (UPO) with the topology of $S^1$. This phase space geometrical object plays a crucial role in the determination of chemical reaction rates, as it is the scaffolding that serves to construct what is known in the chemistry literature as the dividing surface with the non-recrossing and minimal flux properties. This object, which is one of the cornerstones of Transition State Theory \cite{pechukas1981} was envisioned by E. Wigner in his works \cite{Wigner38,Wigner39} and has the role of separating phase space regions corresponding to ``reactants'' and ``products'' (in our model PES it separates two neighboring wells).
	
	We take a look first at the linearized dynamics about the equilibrium point $\mathbf{x}_e = (x_e.y_e,p_{x,e},p_{y,e}) = (0,\sqrt{2}/2,0,0)$. The Jacobian of Hamilton's equations in Eq. \eqref{ham_eqs} has eigenvalues $\lambda_{1,2} = \pm \lambda = \pm \sqrt{2}$ and $\omega_{1,2} = \pm \, \omega = \pm i$. The eigenvectors associated with the real eigenvalues span what is known as the saddle space, and correspond to the tangent directions of the unstable and stable manifolds of the NHIM at the index-1 saddle. They are given by the expression $\mathbf{u}_{\pm} = \left(1,0,\pm\sqrt{2},0\right)$. The eigenvectors associated with the complex eigenvalues span the center subspace of the NHIM at the index-1 saddle and have the form $\mathbf{w}_{\pm} = \left(0,1,0,\pm  i\right)$ Therefore, we can write the solution to the linearized system at the index-1 saddle as:
	\begin{equation}
	\mathbf{x}(t) = \mathbf{x}_e + C_1 \, e^{\lambda t} \mathbf{u}_{+} + C_2 \, e^{-\lambda t} \mathbf{u}_{-} + 2 \, \text{Re}\left(\eta \, e^{i\omega t} \mathbf{w}_{+}\right) \,,
	\label{geneq_lin_ham}
	\end{equation}
	where $C_1,C_2 \in \mathbb{R}$ and $\eta = \eta_1 + \eta_2 \, i \in \mathbb{C}$ are constants determined from initial conditions. The linearized Hamiltonian has the form:
	\begin{equation}
	\widetilde{H} = \dfrac{1}{2} \left(p_x^2+p_y^2\right) -\dfrac{1}{4} - X^2 +2Y^2 = -\dfrac{1}{4} + \dfrac{1}{2}\left(p_x^2 - 2X^2\right) + \dfrac{1}{2}\left(p_y^2 + 4Y^2\right) \,,
	\label{lin_ham}
	\end{equation}
	where $\left(X,Y\right) = \left(x,y-\sqrt{2}/2\right)$ are the configuration coordinates about the equilibrium point. In this setup we can define the isoenergetic dividing surface as the intersection of the energy surface of the linearized system with the slice $X = 0$, so that:
	\begin{equation}
	\mathcal{D}\left(\widetilde{H}_0\right) = \left\{\left(X,Y,p_x,p_y\right) \in \mathbb{R}^4 \; \bigg| \; \widetilde{H}_0 + \dfrac{1}{4} = \dfrac{1}{2} \, p_x^2 + \dfrac{1}{2}\left(p_y^2 + 4Y^2\right) \;,\; X = 0 \right\} \,,
	\end{equation}
	which has the topology of a sphere, $S^2$, in $\mathbb{R}^3$. The dividing surface can be split into two halves, known as the forward and backward dividing surface respectively with the form:
	\begin{equation}
	\begin{split}
	\mathcal{D}_{f}(\widetilde{H}_0) &= \left\{\left(X,Y,p_x,p_y\right) \in \mathbb{R}^4 \; \bigg| \; \widetilde{H}_0 + \dfrac{1}{4} = \dfrac{1}{2} \, p_x^2 + \dfrac{1}{2}\left(p_y^2 + 4Y^2\right) \;,\; p_x > 0 \;,\; X = 0 \right\} \\[.2cm]
	\mathcal{D}_{b}(\widetilde{H}_0) &= \left\{\left(X,Y,p_x,p_y\right) \in \mathbb{R}^4 \; \bigg| \; \widetilde{H}_0 + \dfrac{1}{4} = \dfrac{1}{2} \, p_x^2 + \dfrac{1}{2}\left(p_y^2 + 4Y^2\right) \;,\; p_x < 0 \;,\; X = 0 \right\} 
	\end{split}
	\,,
	\end{equation} 
	and all trajectory that evolve from the upper-left well to the upper-right well (reactive trajectories) cross the forward/backward DS along their forward/backward evolution respectively. The two hemispheres of the DS meet at the equator, which is a NHIM, or an UPO, given by:
	\begin{equation}
	\mathcal{N}(\widetilde{H}_0) = \left\{\left(X,Y,p_y,p_y\right) \in \mathbb{R}^4 \; \bigg| \; 2\widetilde{H}_0 + \dfrac{1}{2} = p_y^2 + 4Y^2 \;,\; X = p_x = 0 \right\} \;,
	\end{equation}
	which is an ellipse with semiaxis $\dfrac{\sqrt{4\widetilde{H}_0 + 1}}{2\sqrt{2}}$ and $\dfrac{\sqrt{4\widetilde{H}_0 + 1}}{\sqrt{2}}$. The NHIM has stable and unstable manifolds:
	\begin{equation}
	\begin{split}
	\mathcal{W}^{u}(\widetilde{H}_0) &= \left\{\left(X,Y,p_x,p_y\right) \in \mathbb{R}^4 \; \bigg| \; 2\widetilde{H}_0 + \dfrac{1}{2} = p_y^2 + 4Y^2 \;,\; p_x = \sqrt{2}X \right\} \\[.2cm]
	\mathcal{W}^{s}(\widetilde{H}_0) &= \left\{\left(X,Y,p_x,p_y\right) \in \mathbb{R}^4 \; \bigg| \; 2\widetilde{H}_0 + \dfrac{1}{2} = p_y^2 + 4Y^2 \;,\; p_x = -\sqrt{2}X \right\}
	\end{split}
	\;,
	\end{equation}
	which are known in the literature as \textit{spherical cylinders} due to the fact that their topology is of the form $S^1 \times \mathbb{R}$. These two-dimensional tube manifolds partition the three-dimensional energy surface into reactive and non-reactive regions. Initial conditions that start on the stable/unstable manifold evolve asymptotically to the NHIM in forward/backward time. Moreover, trajectories that lie inside the stable/unstable spherical cylinders will cross the phase space bottleneck  in forward/backward time and thus give rise to reaction. With the purpose of developing a complete understanding and providing a clear visualization on how these phase space structures characterize transport, i.e. reaction, across the phase space bottleneck in the neighborhood of the equilibrium point $\mathbf{x}_e = (0,\sqrt{2}/2,0,0)$, we illustrate in Fig. \ref{linear_phasePort} the dynamical behavior of the linearized Hamiltonian system.
	
		\begin{figure}[!h]
		\begin{center}
			A)\includegraphics[scale=0.22]{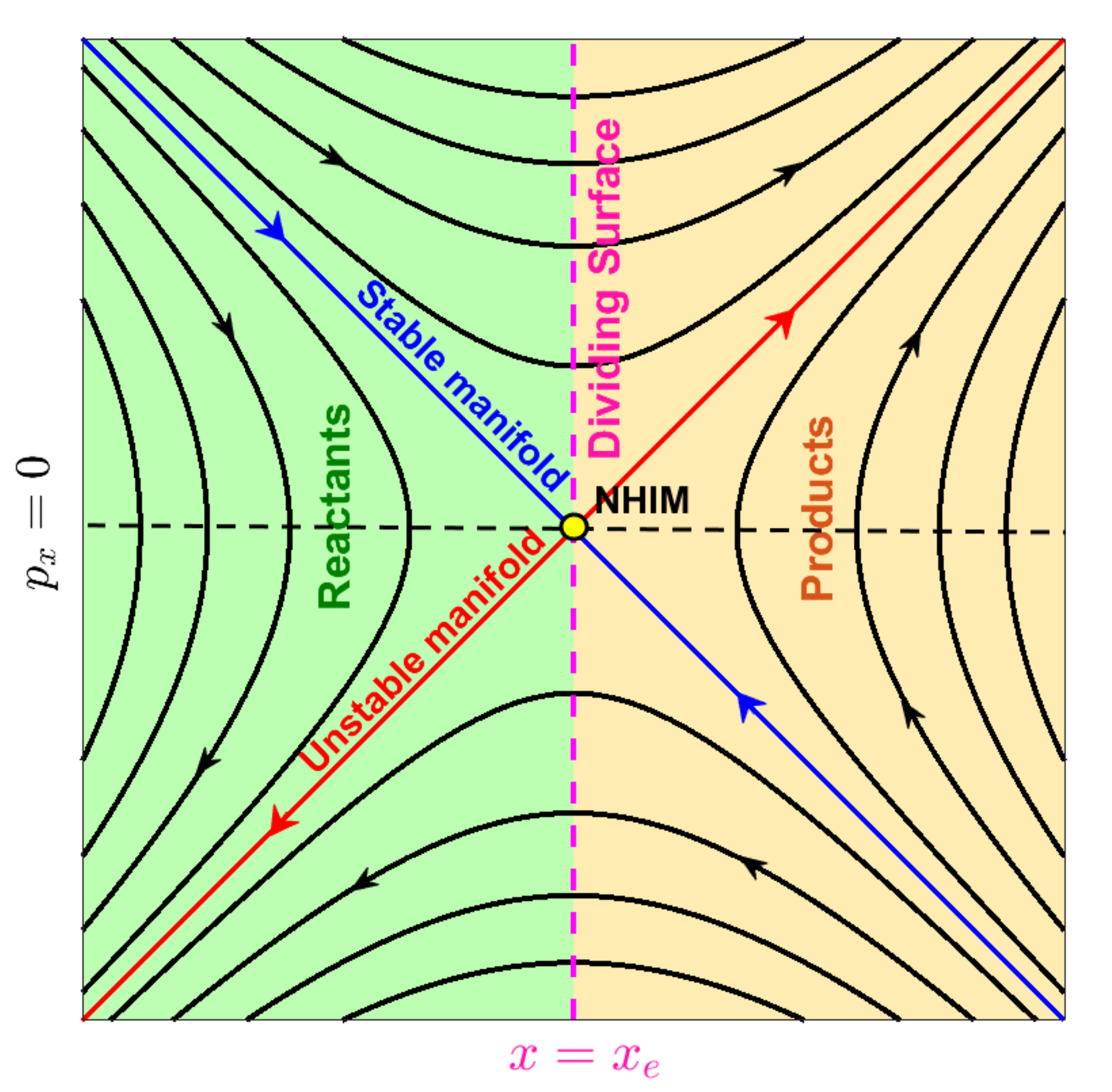}
			B)\includegraphics[scale=0.22]{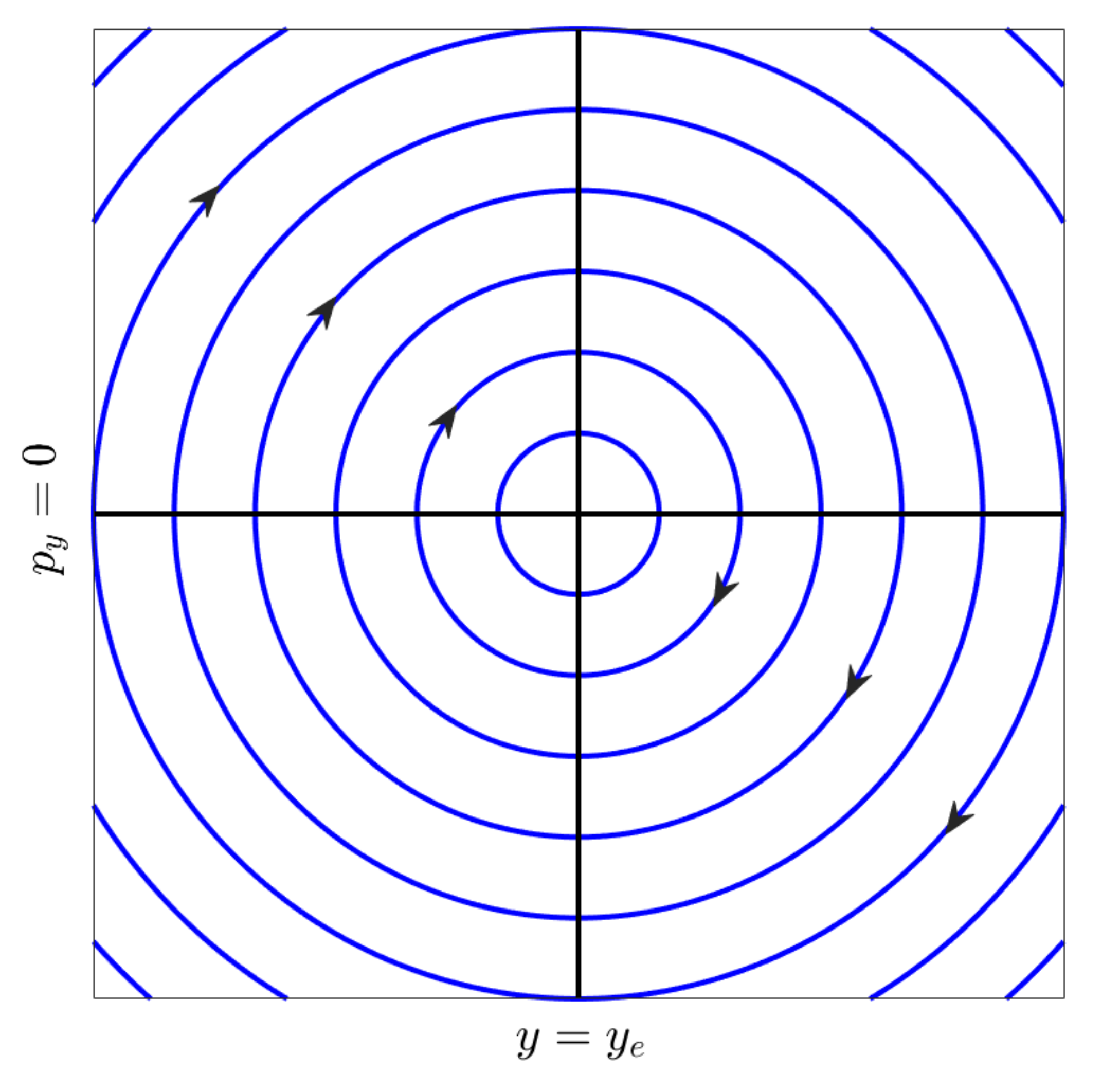}
			C)\includegraphics[scale=0.24]{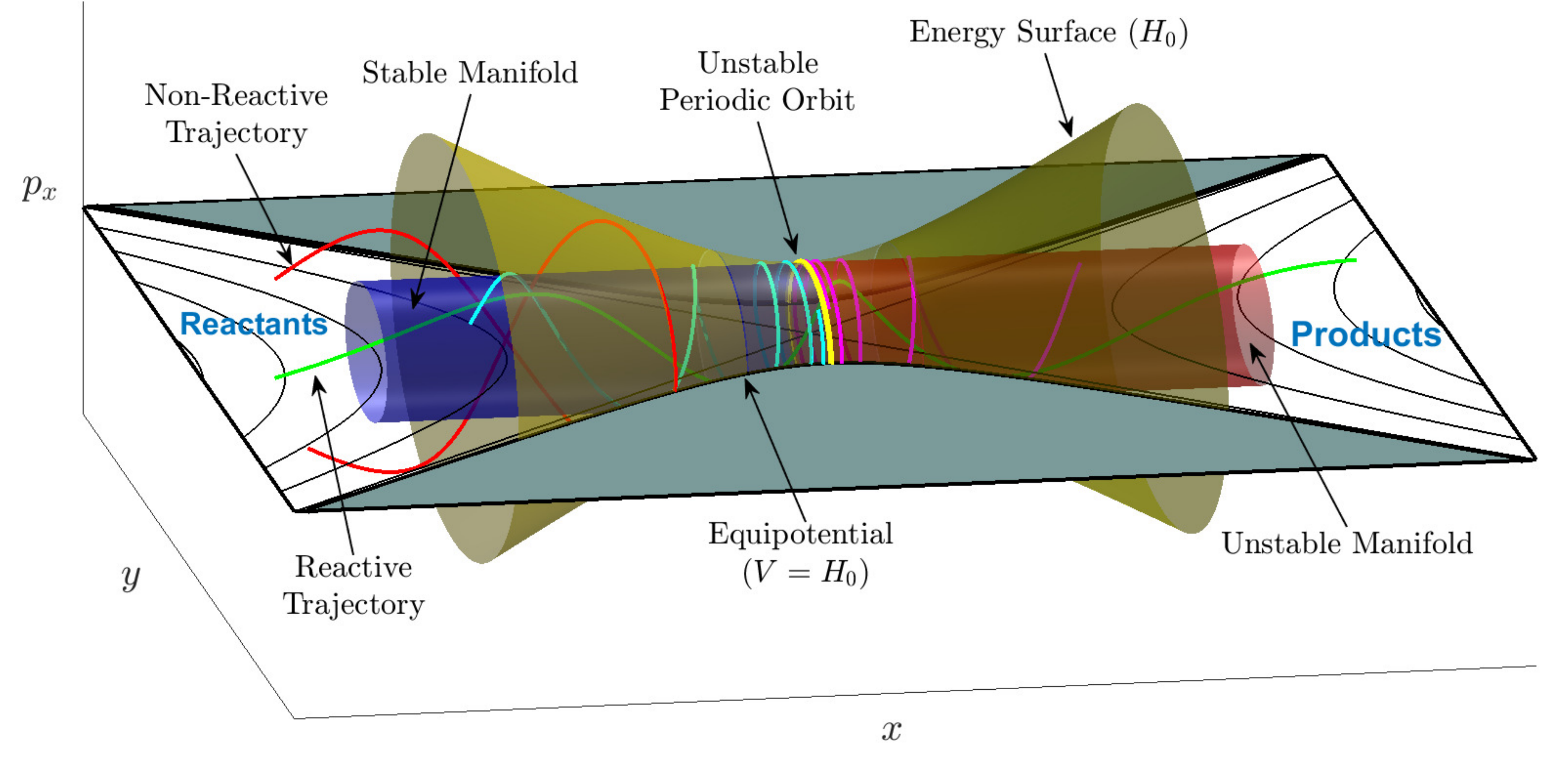}
		\end{center}
		\caption{Phase space structures of the linearized Hamiltonian given in Eq. \eqref{lin_ham} in the neighborhood of the index-1 saddle $\mathbf{x}_e = (0,\sqrt{2}/2,0,0)$. Panel A) depicts the saddle space, B) is the center space, and C) represents the phase space bottleneck in the vicinity of the index-1 saddle that allows transport, i.e. reaction, from reactants to products.}
		\label{linear_phasePort}
	\end{figure}
	
	Once the dynamical concepts have been introduced for the linearized dynamics, we move on to describe the phase space structures for the nonlinear system. The phase space dividing surface separating the upper-left from the upper-right well regions of the PES is also defined in this situation by intersecting the energy hypersurface with the slice $x = 0$, that is:
	\begin{equation}
	\mathcal{D}\left(H_0\right) = \mathcal{S}\left(H_0\right) \cap \lbrace x = 0 \rbrace = \left\{\left(x,y,p_x,p_y\right) \in \mathbb{R}^4 \; \bigg| \; H_0 = \frac{1}{2}\left(p_x^2+p_y^2\right) + y^4 - y^2 \;,\; x = 0 \right\} \;,
	\label{ds_sym_unc}
	\end{equation}
	which is a non-invariant surface with the local non-recrossing property. Moreover, it has the topology of a sphere $S^2$ with two hemispheres, known as the forward and backward dividing surface:
	\begin{equation}
	\begin{split}
	\mathcal{D}_{f}(H_0) &= \left\{\left(x,y,p_x,p_y\right) \in \mathbb{R}^4 \; \bigg| \; H_0 = \frac{1}{2}\left(p_x^2+p_y^2\right) + y^4 - y^2 \;,\; x = 0 \;,\; p_x > 0 \right\} \\[.2cm]
	\mathcal{D}_{b}(H_0) &= \left\{\left(x,y,p_x,p_y\right) \in \mathbb{R}^4 \; \bigg| \; H_0 = \frac{1}{2}\left(p_x^2+p_y^2\right) + y^4 - y^2 \;,\; x = 0 \;,\; p_x < 0 \right\}
	\end{split}
	\;.
	\end{equation}
	The two hemispheres meet at the equator, which is an UPO in the form:
	\begin{equation}
	\mathcal{N}(H_0) = \left\{\left(x,y,p_y,p_y\right) \in \mathbb{R}^4 \; \bigg| \; H_0 = \frac{p_y^2}{2} + y^4 - y^2 \;,\; x = p_x = 0 \right\} \;,
	\end{equation}
	with the topology of a circle $S^1$. The NHIM has stable and unstable manifolds:
	\begin{equation}
	\mathcal{W}^{u}(H_0) = \mathcal{W}^{s}(H_0) = \left\{\left(x,y,p_y,p_y\right) \in \mathbb{R}^4 \; \bigg| \; H_y = \frac{p_y^2}{2} + y^4 - y^2 = H_0 \;,\; H_x = \frac{p_x^2}{2} + x^4 - x^2 = 0 \right\} \;,
	\end{equation}
	which topologically have the structure of $S^1 \times \mathbb{R}$, representing \textit{tube} or \textit{cylindrical manifolds}. Observe that they have a homoclinic structure in phase space. All these phase space structures, depicted in Fig. \ref{phasePort_1DoF_symm}, characterize the bottleneck region connecting the upper-left and upper-right wells of the PES and are responsible for the reaction mechanisms that take place in the phase space of the Hamiltonian system.
	
	\begin{figure}[!h]
		\begin{center}
			A)\includegraphics[scale=0.23]{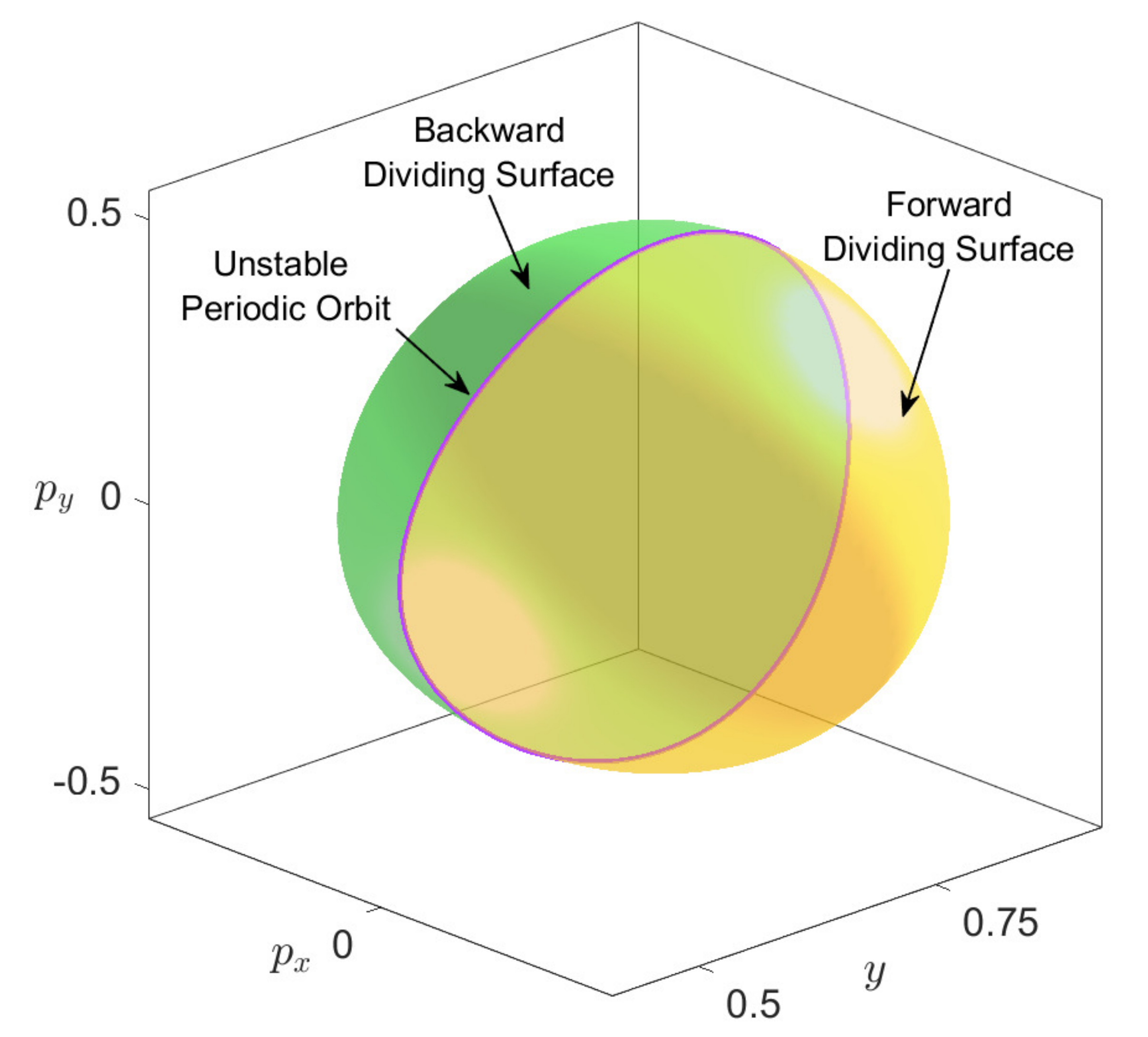} 
			B)\includegraphics[scale=0.3]{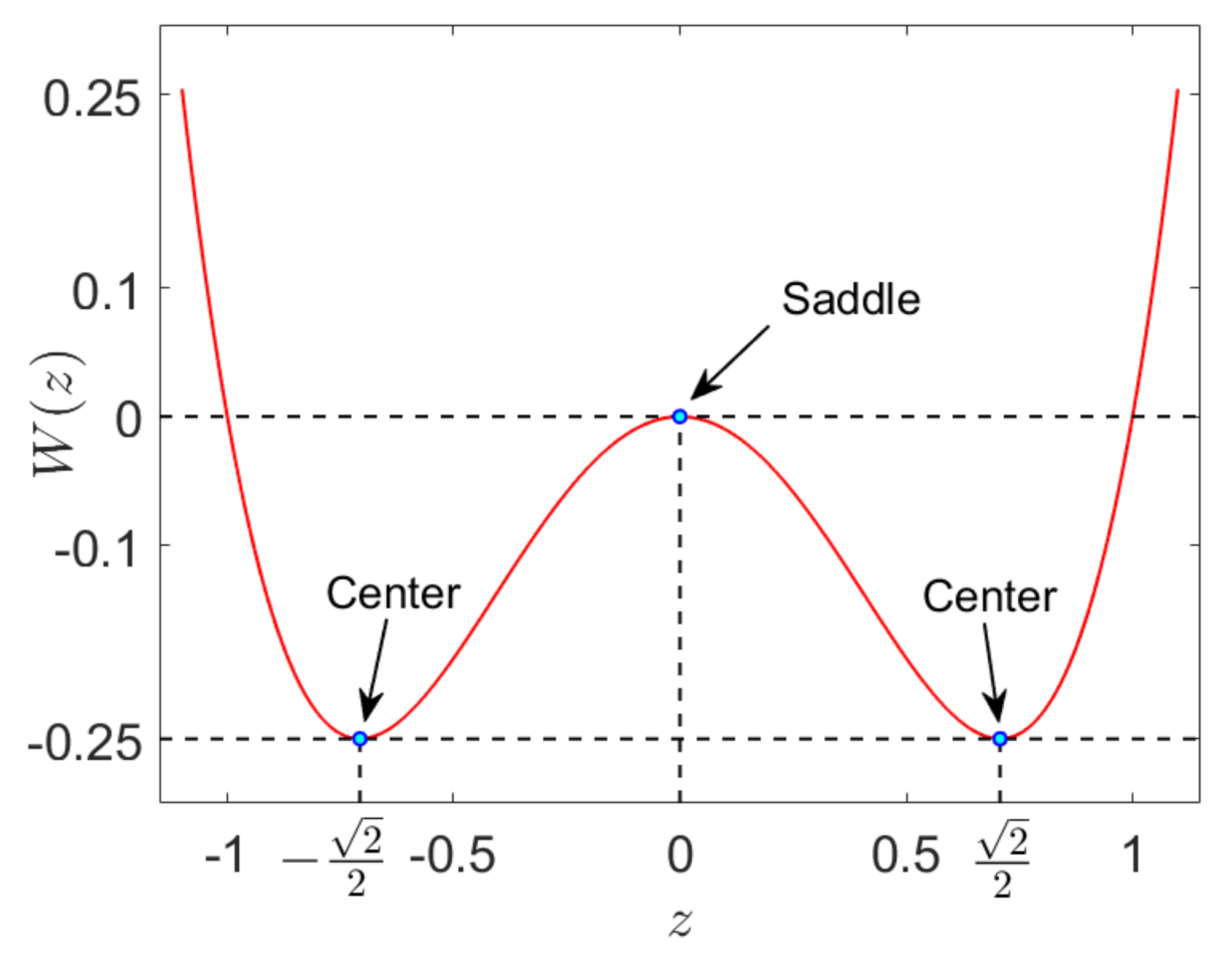}\\
			C)\includegraphics[scale=0.29]{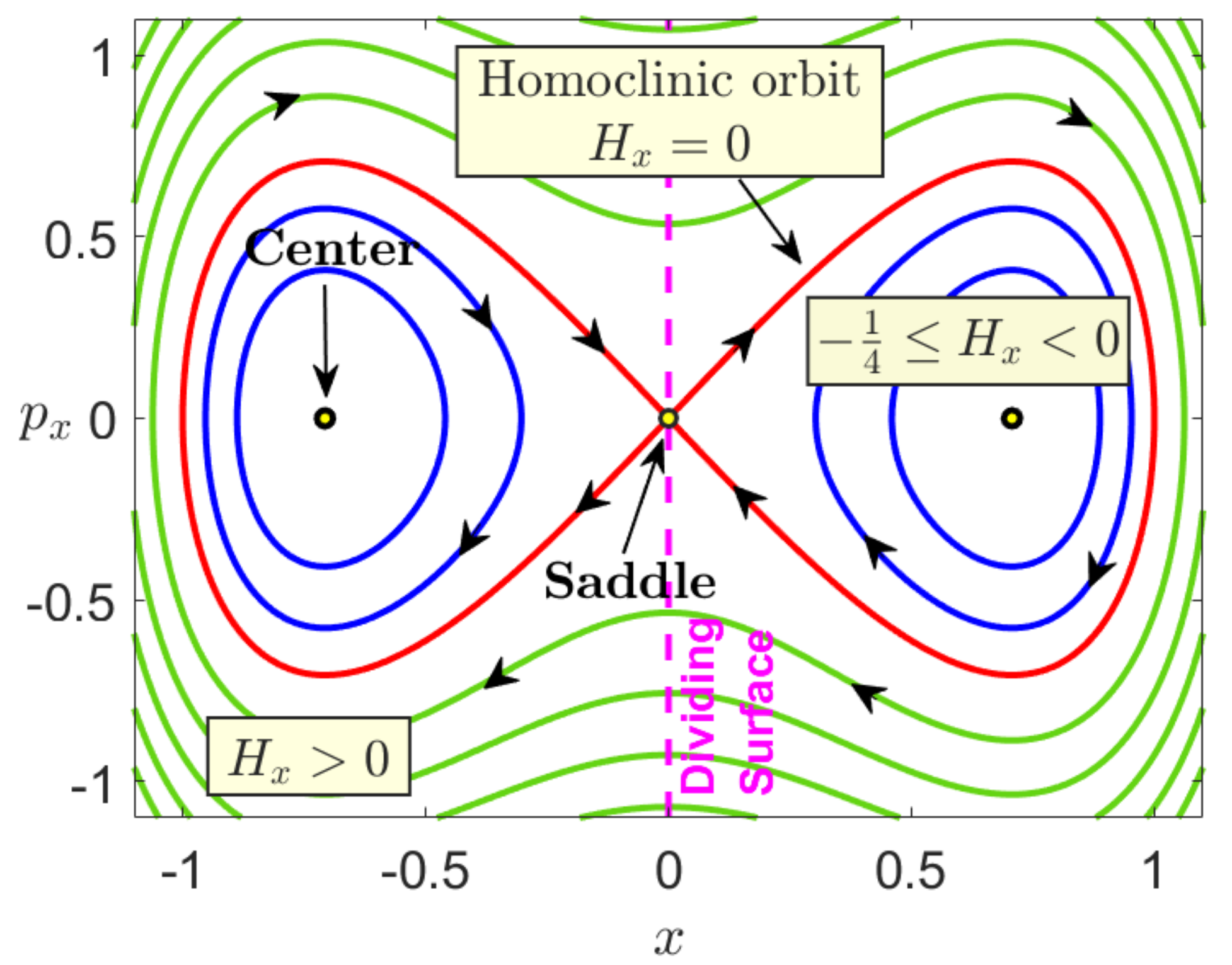}
			D)\includegraphics[scale=0.29]{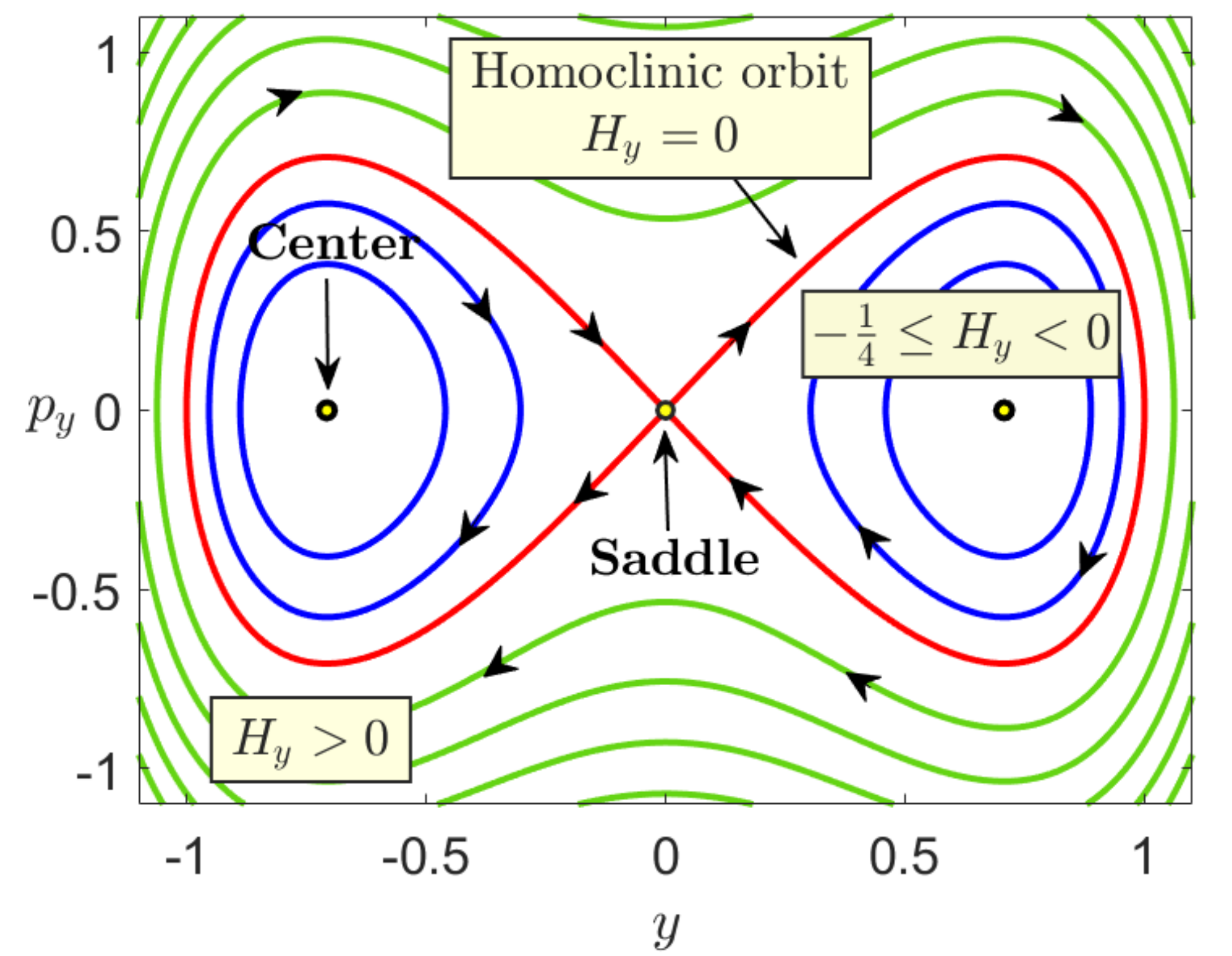}
		\end{center}
		\caption{A) Dividing surface described in Eq. \eqref{ds_sym_unc} for the symmetric and uncoupled Hamiltonian system with energy $H_0 = -0.15$ in the neighborhood of the equilibrium point $\mathbf{x}_e = (0,\sqrt{2}/2,0,0)$. B) Symmetric double well potential given in Eq. \eqref{1D_potSymm}. C) and D) depict the phase portraits in the $x-p_x$ and $y-p_y$ planes respectively. We have marked with a magenta line the dividing surface $x = 0$ that separates the upper-left and upper-right wells of the PES.}
		\label{phasePort_1DoF_symm}
	\end{figure}

	We describe now the dynamics of the system by applying the method of Lagrangian descriptors (LDs) to reveal the geometrical template of phase space structures governing transport. In particular, throughout this work we will use the $p$-norm definition of LDs with $p = 1/2$. A detailed introduction about the use of this technique can be found in Appendix \ref{sec:appA}. For our analysis of the system, we fix a total energy of $H_0 = -0.2$ and consider a phase space slice that goes through the lower index-1 saddle of the PES:
	\begin{equation}
	\mathcal{P}_1 = \left\{ (x,y,p_x,p_y) \in \mathbb{R}^4 \;\big|\; y = -1/\sqrt{2} \; ,\; p_y > 0 \right\} \;,
	\label{psos1}
	\end{equation}	
	and another Poincar\'e surface of section (PSOS) that coincides with the configuration plane:
	\begin{equation}
	\mathcal{P}_2 = \left\{ (x,y,p_x,p_y) \in \mathbb{R}^4 \;\big|\; p_x = 0 \; ,\; p_y > 0 \right\} \;.
	\label{psos2}
	\end{equation} 
	When we compute LDs with an integration time of $\tau = 5$ on the two phase space slices defined above, we can clearly see that the method nicely captures the intersection of the relevant invariant manifolds that determine isomerization dynamics with the surfaces of section. An important conclusion that can be drawn from the success of this technique for unveiling the geometry of phase space is that by looking at the LD output and having a detailed knowledge and understanding about the phase space structures present in the problem, one can easily and systematically determine the dynamical fate of any initial condition of the Hamiltonian system.
	
	For instance, in Fig. \ref{LD_sym_h_neg} A) we depict the output of LDs calculated on the slice $\mathcal{P}_1$. Three distinct phase space regions are visible, separated by curves where the LD scalar field is non-differentiable, and which correspond to the invariant stable and unstable manifolds of the UPOs associated to the index-1 saddles of the PES. The cross located at the point $(x,p_x) = (0,0)$ represents the intersection of the UPO corresponding to the bottom $(\mathcal{B})$ index-1 saddle with $\mathcal{P}_1$, and the phase portrait $x-p_x$ has a saddle-like structure because the exponential growth and decay eigendirections for the bottom index-1 saddle lie in this plane. Moreover, the homoclinic orbits that emerge from the saddle point $(x,p_x) = (0,0)$ represent the stable and unstable spherical cylinders that are born from the UPO of the bottom index-1 saddle. These manifolds, which are coincident, extend parallel to the $x$ axis until they reach the boundary of the energy hypersurface, where they bounce back to return to the UPO. Interestingly, Fig. \ref{LD_sym_h_neg} A) also reveals a cross-section of the stable and unstable manifolds associated to the left $(\mathcal{L})$ and right $(\mathcal{R})$ index-1 saddles, which are located inside the homoclinic orbits. At first, this fact seems to indicate that the tube manifolds of the left and right index-1 saddles are inside those of the bottom index-1 saddle, but this is misleading. One has to be very careful when interpreting the phase space geometry, since the interior of the tube manifolds corresponding to the bottom index-1 saddle is in fact the region outside the homoclinic orbit. This can be easily understood by taking a look at the phase portrait shown in Fig. \ref{phasePort_1DoF_symm} C). Furthermore, the circular-shaped cross-sections obtained from the intersection of the stable and unstable invariant manifolds of the left and right index-1 saddles with the $\mathcal{P}_1$ slice are known in the chemistry literature as \textit{reactive islands}. The reason for this name is that all initial conditions chosen inside the region delimited by these curves will be transported along their evolution through the phase space bottleneck that exists in the neighborhood of the UPO, from which the invariant manifolds that give rise to this reactive island structure are born. Therefore, all initial conditions inside a reactive island are classified as reactive, since they will move between neighboring wells through the index-1 saddle of the PES that interconnects both wells. On the other hand, if an initial condition is selected outside a reactive island, it will be trapped forever in the well region where its evolution started. 
	
	In order to validate the dynamical behavior discussed above, we have chosen three initial conditions in the $\mathcal{P}_1$ slice that are presented as red, cyan and magenta squares in Fig. \ref{LD_sym_h_neg} A), and we have evolved them in forward time. We also show their projections onto configuration space in Fig. \ref{LD_sym_h_neg} B) superimposed with LDs calculated using $\tau = 5$ on the $\mathcal{P}_2$ PSOS. First, the magenta initial condition is initially located in the lower-right well and, since it is inside the reactive island of the right index-1 saddle, it goes through the bottleneck and ends up in the upper-right well. If we extend the time of evolution further, it will move back and forth between both wells. The same type of behavior is observed for the red initial condition, that evolves between the lower-left ad lower-right wells of the PES because it is inside the spherical cylinders of the bottom index-1 saddle. On the other hand, the cyan initial condition is outside the tube manifolds of the UPOs of the left and bottom index-1 saddles, so that it remains trapped forever in the lower-left well region. We provide a three-dimensional visualization of the phase space structures and the resulting dynamics they give rise to inside the energy hypersurface in Fig. \ref{LD_sym_h_neg} C). From the observed dynamical behavior, it is important to remark that only sequential isomerization between neighboring wells is allowed in this situation. Moreover, since the tube manifolds of different UPOs do not intersect, i.e. there are no heteroclinic connections, an initial condition that starts in the lower-left well region cannot reach the upper-right well of the PES along its evolution by moving from well to well. The only two options available for its dynamical behavior are that it remains trapped forever in the lower-left well, or that it moves back and forth between the lower-left and lower-right (or upper-left) well if it is located inside the corresponding reactive island.  
	
	\begin{figure}[!h]
		\begin{center}
			A)\includegraphics[scale=0.21]{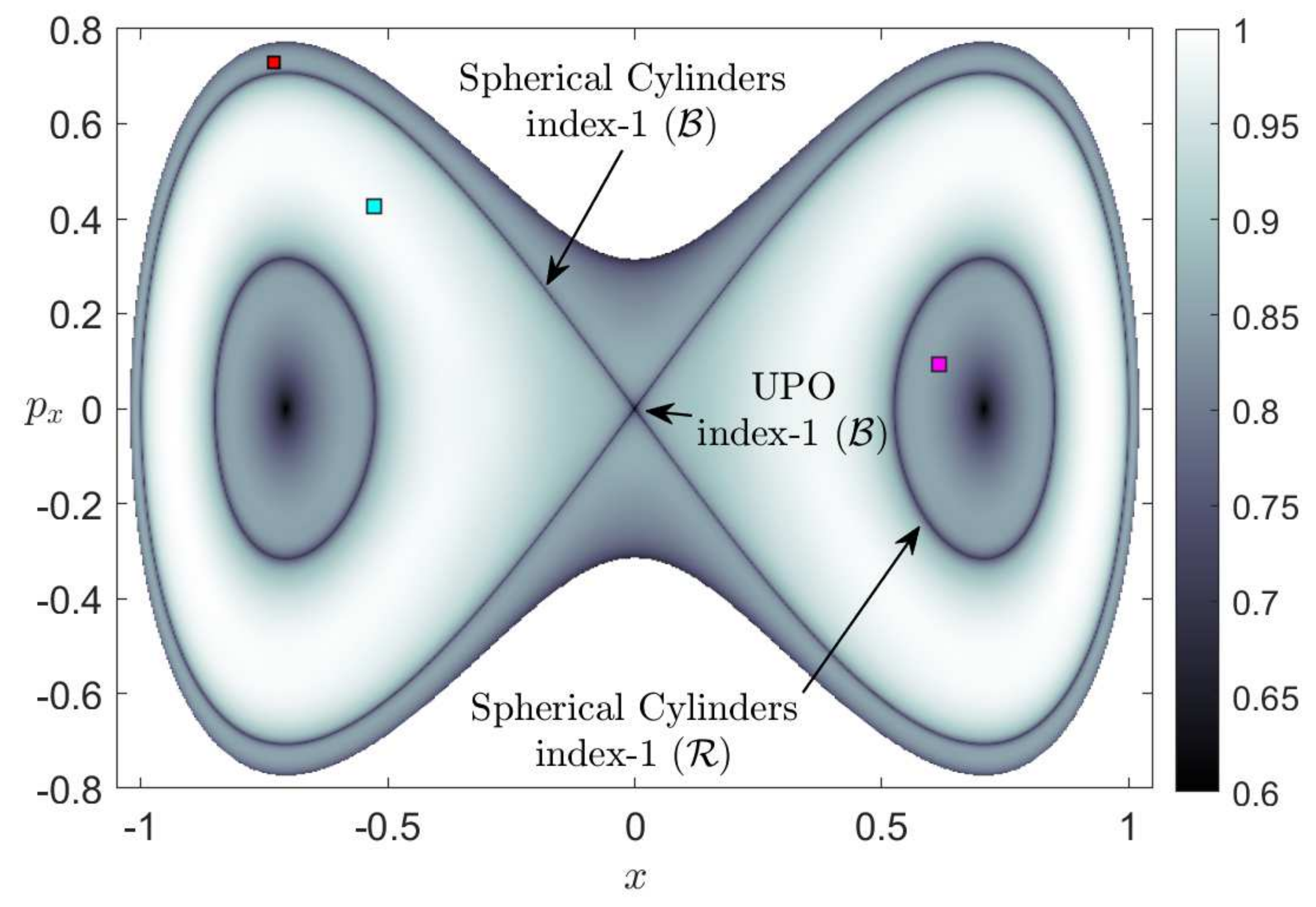}
			B)\includegraphics[scale=0.205]{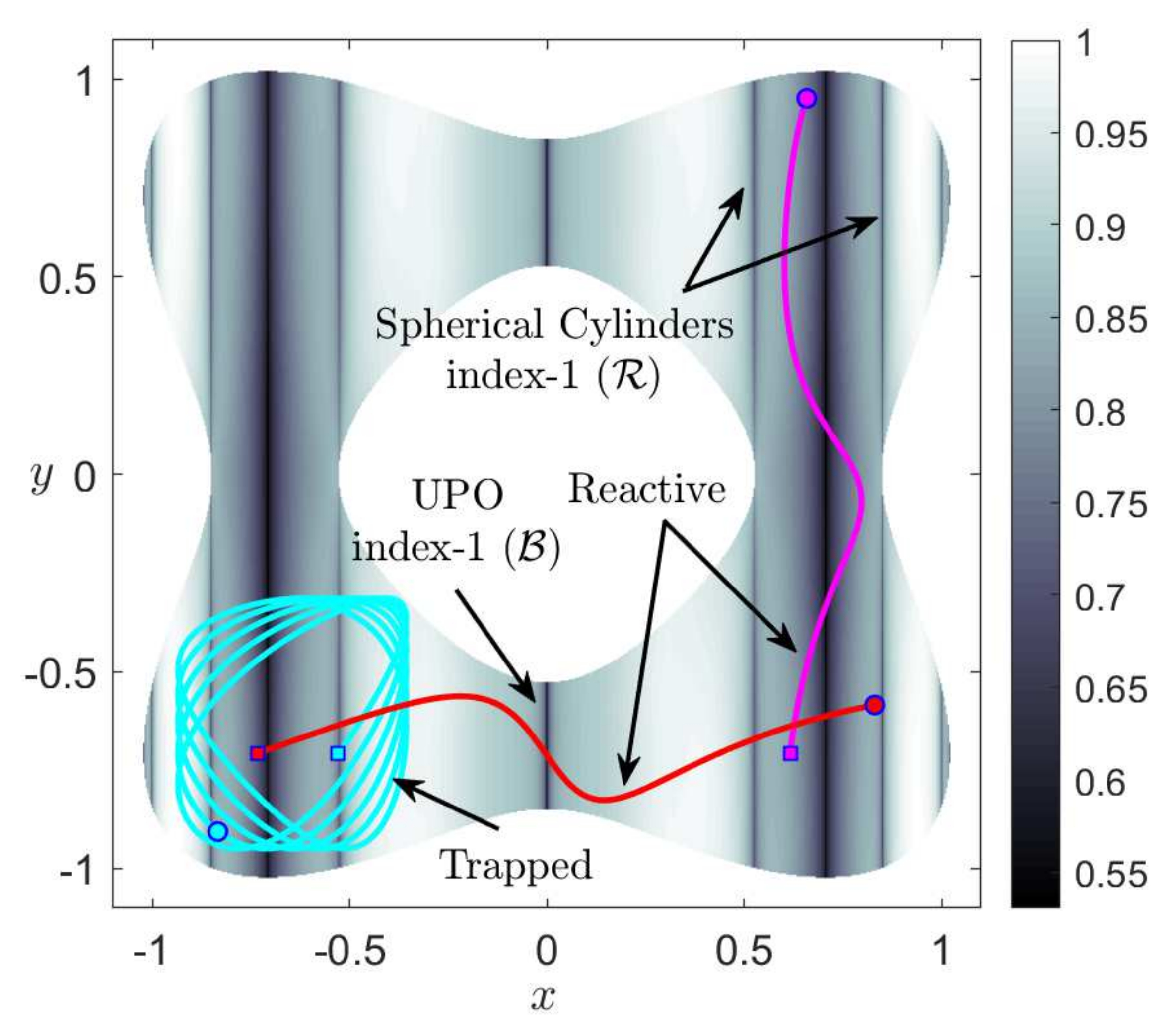} \\[.3cm]
			C)\includegraphics[scale=0.26]{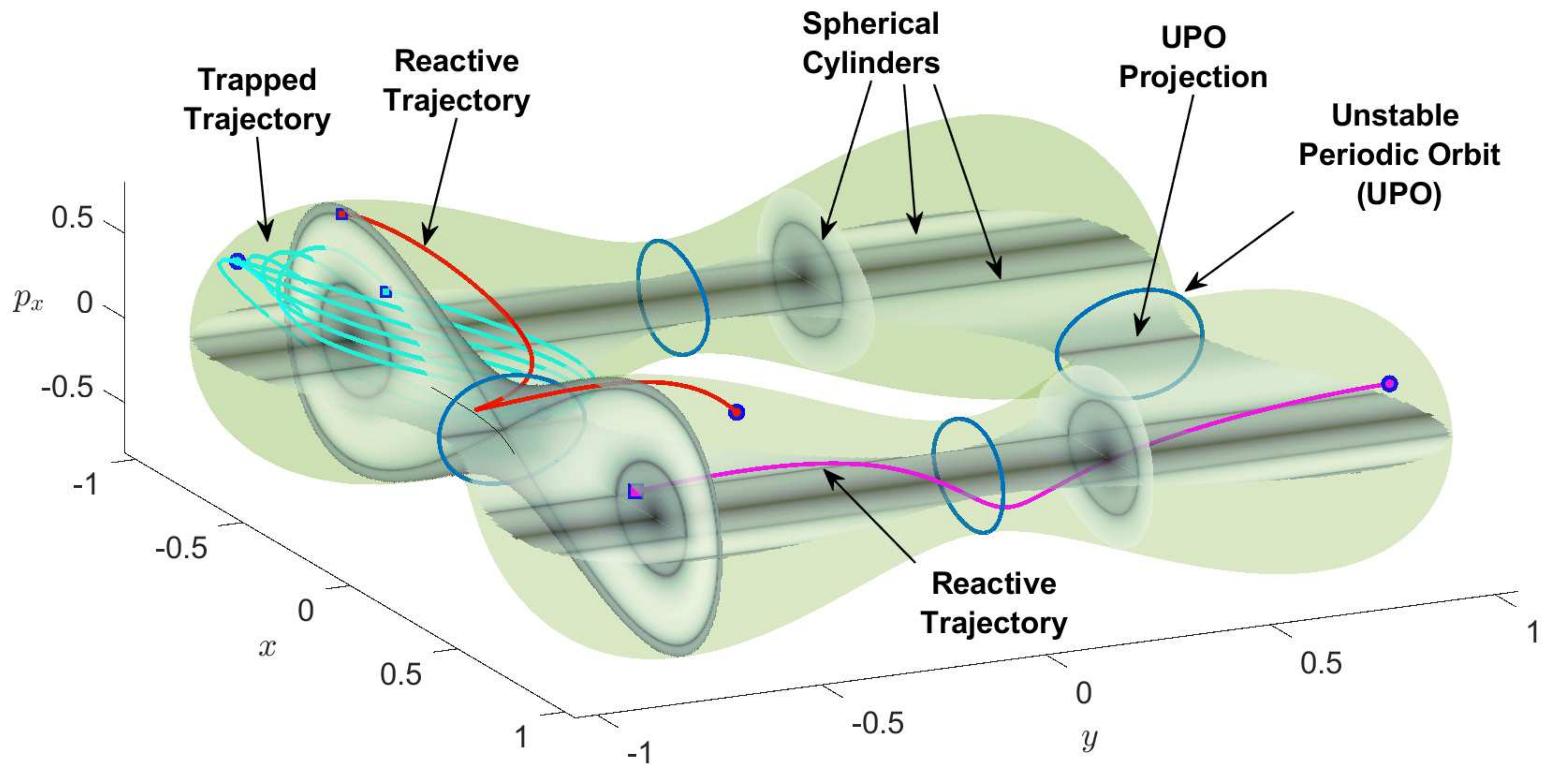}
		\end{center}
		\caption{Phase space structures and evolution of initial conditions at an energy  $H_0 = -0.2$ for the symmetric and uncoupled Hamiltonian. A) LDs calculated using $\tau = 5$ on the phase space slice in Eq. \eqref{psos1}. B) Superposition of LDs calculated using $\tau = 5$ on the PSOS in Eq. \eqref{psos2}  with the dynamical evolution of the initial conditions selected in panel A. C) Visualization of the phase space dynamics in the three-dimensional energy hypersurface.}
		\label{LD_sym_h_neg}
	\end{figure}
	
	All the dynamical behavior that we have explained so far can also be understood by looking at the phase portraits in Fig. \ref{phasePort_1DoF_symm} C) and D). Given that the system is separable, so that the total energy $H_0$ can be naturally partitioned between each DoF, that is, $H_0 = H_{x,0} + H_{y,0}$, we discuss the following cases:

\begin{enumerate}
	\item \underline{$H_{x,0} \, , \, H_{y,0} < 0$}: In this situation, these energy levels correspond to trajectories lying inside the homoclinic curves, and therefore we know that neither $x$ nor $y$ will change sign. This means that trajectories will be trapped in the lower-left well region of the PES.
	
	\item \underline{$H_{x,0} < 0 \, , \, H_{y,0} > 0$}: Since the energy in the $y$ DoF is positive, we are outside the separatrix in the $y-p_y$ plane, and thus the trajectory is periodic in $y$, and the sign of $y$ changes along the evolution. Moreover, since the energy in $x$ is negative, this trajectory is inside the separatrix in the $x-p_x$ plane. Thus, $x$ does not change sign along the evolution. Consequently, we have trajectories that move back and forth between the lower-left and upper-left wells through the bottleneck associated to the left index-1.
	
	\item \underline{$H_{x,0} > 0 \, , \, H_{y,0} < 0$}: This case is analogous to the previous one, but the roles of the $x$ and $y$ DoF have been interchanged. Hence, we will have trajectories that move back and forth between the lower-left and lower-right wells through the phase space bottleneck corresponding to the bottom index-1.
	
	\item \underline{$H_{x,0} = 0 \, , \, H_{y,0} < 0$}: For this case, the energy in the $x$ DoF is zero, and this implies that we are on the separatrix in the $x-p_x$ plane, where the trajectory does not change sign in $x$. Due to the fact that the energy in $y$ is negative, we are inside the separatrix in the $y-p_y$ plane and $y$ does not change sign along the trajectory evolution. Therefore, we have trajectories evolving on the spherical cylinders of the bottom index-1 that asymptotically approach the UPO.
	
	\item \underline{$H_{x,0} < 0 \, , \, H_{y,0} = 0$}: This partition of energies is similar to that discussed in the previous case, but with the roles of $x$ and $y$  interchanged. Therefore, initial conditions that start in the lower-left well region of the PES will evolve on the spherical cylinder of the left index-1, asymptotically approaching the UPO.
\end{enumerate}

\item \underline{\textbf{Energy Level} $(H_0 = 0)$:} We take a look now at the situation where the energy of the system is exactly that of the index-2 saddle at the origin. At this critical energy value, the UPOs of the four index-1 saddles that exist in the PES meet at the origin. This is a consequence of the growth in the size of the UPOs and their corresponding spherical cylinders as the system's energy is raised towards zero from below. As a result, the phase space region associated to initial conditions that are trapped in one of the potential wells shrinks and disappears at the zero energy level. This is nicely captured by LDs in Fig. \ref{LD_sym_h_zero}, where we see that the homoclinic trajectories in the form of a figure eight, which are born from the UPO of the bottom index-1 saddle, partition the phase space into two regions with distinct dynamical behavior. Moreover, the left/right homoclinic trajectory also represents the spherical cylinders of the left/right index-1 saddle of the PES. Notice that the region that led to well trapping in the negative energy case discussed earlier no longer exists, see Fig. \ref{LD_sym_h_neg} A), since the spherical cylinders of the index-1 saddles completely fill the energetically accessible phase space. Initial conditions located outside the figure eight shape will evolve back and forth between the lower-left and lower-right wells, and those inside the left homoclinic trajectory move between the lower-left and upper-left wells.

\begin{figure}[!h]
	\begin{center}
		\includegraphics[scale=0.36]{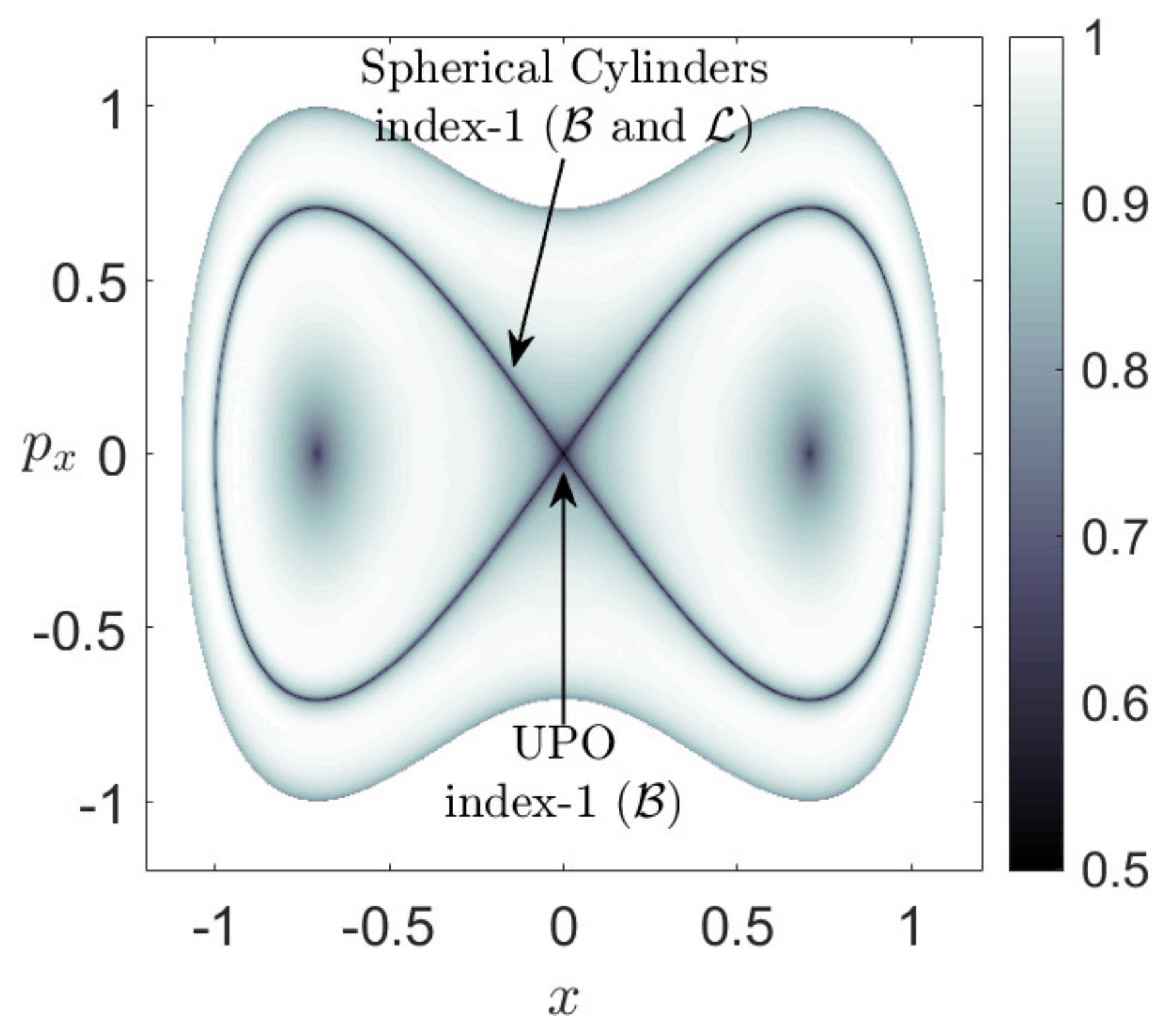}
	\end{center}
	\caption{Phase space structures at an energy $H_0 = 0$ for the symmetric and uncoupled Hamiltonian, as revealed by Lagrangian descriptors using $\tau = 5$ on the PSOS in Eq. \eqref{psos1}.}
	\label{LD_sym_h_zero}
\end{figure}

\item \underline{\textbf{Energy Level} $(H_0 > 0)$:} We move on to explore the dynamics in the case where the total energy of the system is positive, that is, above that of the index-2 saddle located at the origin. In this situation, the UPOs associated to the left and right index-1 saddles that existed in the system for non-positive energies merge, forming a larger UPO related to the index-2 saddle. This phenomenon occurs also for the UPOs of the top and bottom index-1 saddles. The result is that two UPOs are responsible for characterizing the dynamics induced by the index-2 saddle in the phase space. We can easily understand the emergence of these geometrical structures by looking at the phase portraits depicted in Fig. \ref{phasePort_1DoF_symm} C) and D). Since the DoF are uncoupled, any phase space trajectory with energy $H_0$ can be constructed by performing a cartesian product between a trajectory in the $x-p_x$ phase plane with energy $H_{x,0}$, and another trajectory in the $y-p_y$ space with energy $H_{y,0}$. In this way, the UPOs associated to the index-2 saddle at the origin are described by the sets:	\begin{equation}
\begin{split}
\mathcal{U}_1 &= \left\{ (x,y,p_x,p_y) \in \mathbb{R}^4 \;\big|\; H_0 = H_{x,0} = \frac{1}{2} \, p_x^2 + x^4 - x^2 \; , \; y = p_y = 0 \right\} \\[.1cm]
\mathcal{U}_2 &= \left\{ (x,y,p_x,p_y) \in \mathbb{R}^4 \;\big|\; x = p_x = 0 \; ,\; H_0 = H_{y,0} = \frac{1}{2} \, p_y^2 + y^4 - y^2 \right\}
\end{split}
\;,
\label{upos_index2}
\end{equation}
which are topologically equivalent to $S^1$, and they have stable and unstable spherical cylinders given by:
\begin{equation}
\begin{split}
\mathcal{W}^{u}_1 = \mathcal{W}^{s}_1 &= \left\{ (x,y,p_x,p_y) \in \mathbb{R}^4 \;\big|\; H_{x,0} = \frac{1}{2} \, p_x^2 + x^4 - x^2 \; , \; H_{y,0} = \frac{1}{2} \, p_y^2 + y^4 - y^2 = 0 \right\} \\[.1cm]
\mathcal{W}^{u}_2 = \mathcal{W}^{s}_2 &= \left\{ (x,y,p_x,p_y) \in \mathbb{R}^4 \;\big|\; H_{x,0} = \frac{1}{2} \, p_x^2 + x^4 - x^2 = 0 \; , \; H_{y,0} = \frac{1}{2} \, p_y^2 + y^4 - y^2  \right\}
\end{split}
\;,
\label{sphcyl_upos_index2}
\end{equation}
In terms of the energy distribution among the DoF of the system, we discuss the following scenarios:
\begin{enumerate}
	\item \underline{$H_{x,0} \, , \, H_{y,0} > 0$:} An initial condition with this energy partition is located outside the separatrices in the $x-p_x$ and $y-p_y$ phase portraits, which means that it is inside the spherical cylinders of both UPOs associated to the index-2 saddle, and therefore, both coordinates $x$ and $y$ will change sign along the trajectory evolution. The motion will be quasiperiodic and the trajectory goes over the index-2 saddle at the origin at some point, visiting any of the four wells in the PES. Consequently, this case results in concerted isomerization. 	
	
	\item \underline{$H_{x,0} > 0 \, , \, H_{y,0} < 0$:} For this case, the initial condition is outside the homoclinic trajectory in the $x-p_x$ plane, and thus the trajectory is periodic in $x$, and changes sign in $x$ along its motion. Also, since the energy in $y$ is negative, this indicates that we are inside the separatrix in the $y-p_y$ plane, and $y$ does not change sign. This dynamical behavior is a consequence of the initial condition being inside the spherical manifolds of the $\mathcal{U}_2$ unstable periodic orbit and outside those of $\mathcal{U}_1$. Moreover, if the initial condition is in the phase space region inside the left (right) homoclinic trajectory in the $y-p_y$ plane, the trajectory will evolve back and forth between the lower-left and lower-right wells (upper-left and upper-right wells). In this context, these trajectories follow sequential isomerization between neighboring wells. 
	
	\item \underline{$H_{x,0} < 0 \, , \, H_{y,0} > 0$:} This situation is similar to the one we have just described in the previous case but with the roles of the $x$ and $y$ DoF interchanged. Initial conditions with this energy configuration will follow sequential isomerization routes and evolve back and forth between the lower-left and  upper-left wells (lower-right and upper-right wells), depending on whether the initial condition lies in the phase space region inside the left (right) homoclinic trajectory in the $x-p_x$ plane. In this energy distribution, trajectories of the system undergo sequential isomerization.
	
	\item \underline{$H_{x,0} = 0 \, , \, H_{y,0} > 0$:} Since the energy in the $x$ DoF is zero, this means that we are on one of the homoclinic trajectories in the $x-p_x$ plane, and thus the trajectory will not change sign in $x$. Moreover, since the energy in $y$ is positive, this yields that we are outside the separatrix in the $y-p_y$ plane, implying that $y$ changes sign along the trajectory evolution. Consequently, an initial condition starting on the left (right) homoclinic trajectory in the $x-p_x$ plane will evolve periodically in $y$ back and forth from the lower-left to the upper-left well (lower-right to upper-right well), without going over the index-2, but approaching asymptotically the UPO $\mathcal{U}_2$ in forward and backward time. This situation corresponds to sequential isomerization.
	
	\item \underline{$H_{x,0} > 0 \, , \, H_{y,0} = 0$:} The dynamical behavior of the system in this energy configuration is analogous to the previous case, but with the roles of the $x$ and $y$ DoF interchanged. Therefore, an initial condition starting on the left (right) homoclinic trajectory in the $y-p_y$ plane will evolve periodically in $x$ back and forth from the lower-left to the lower-right well (upper-left to  upper-right well), without going over the index-2, but approaching asymptotically the UPO $\mathcal{U}_1$ in forward and backward time. This situation corresponds to sequential isomerization. 
\end{enumerate}

\end{itemize}

\begin{figure}[!h]
	\begin{center}
		A)\includegraphics[scale=0.315]{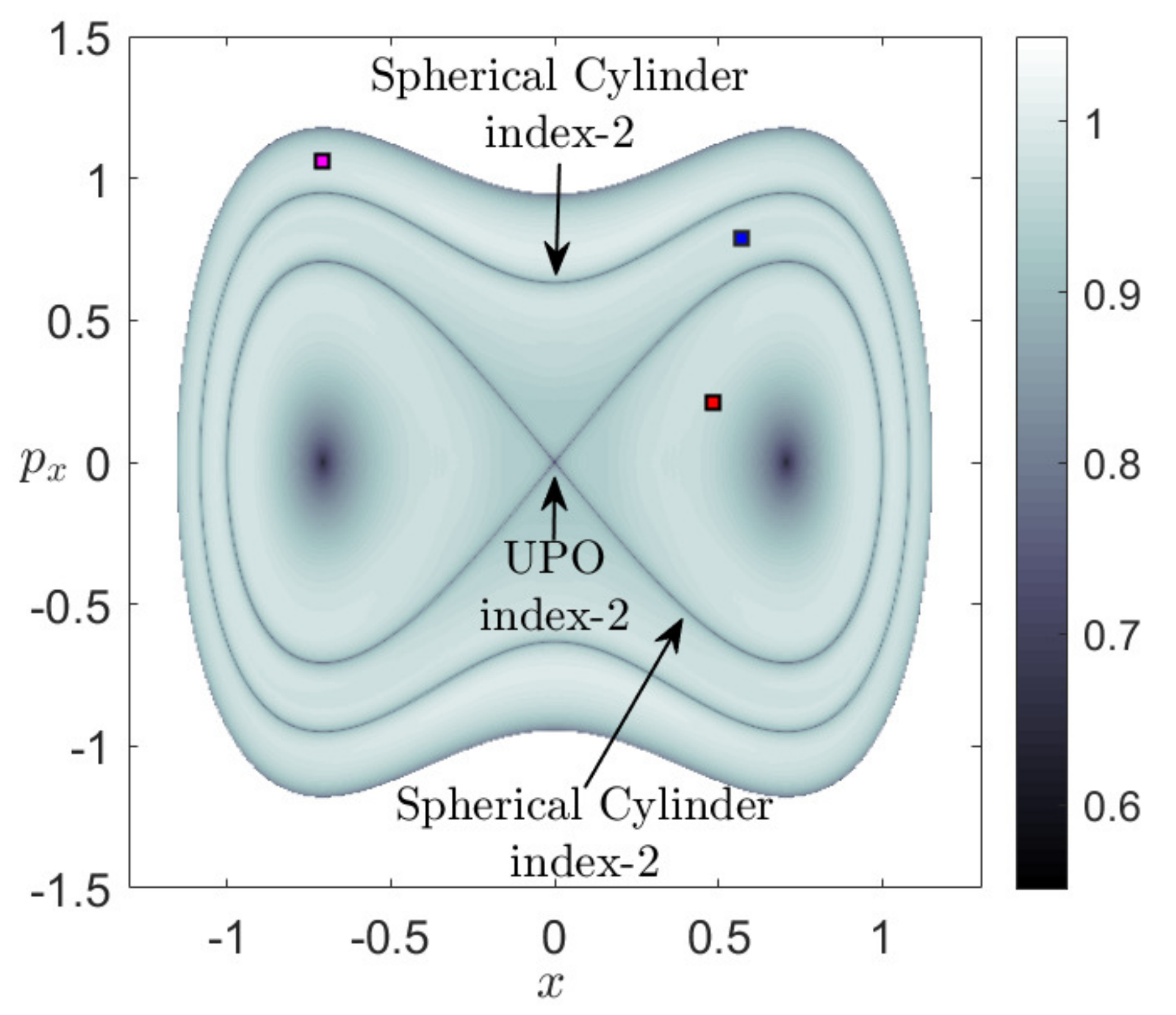}
		B)\includegraphics[scale=0.28]{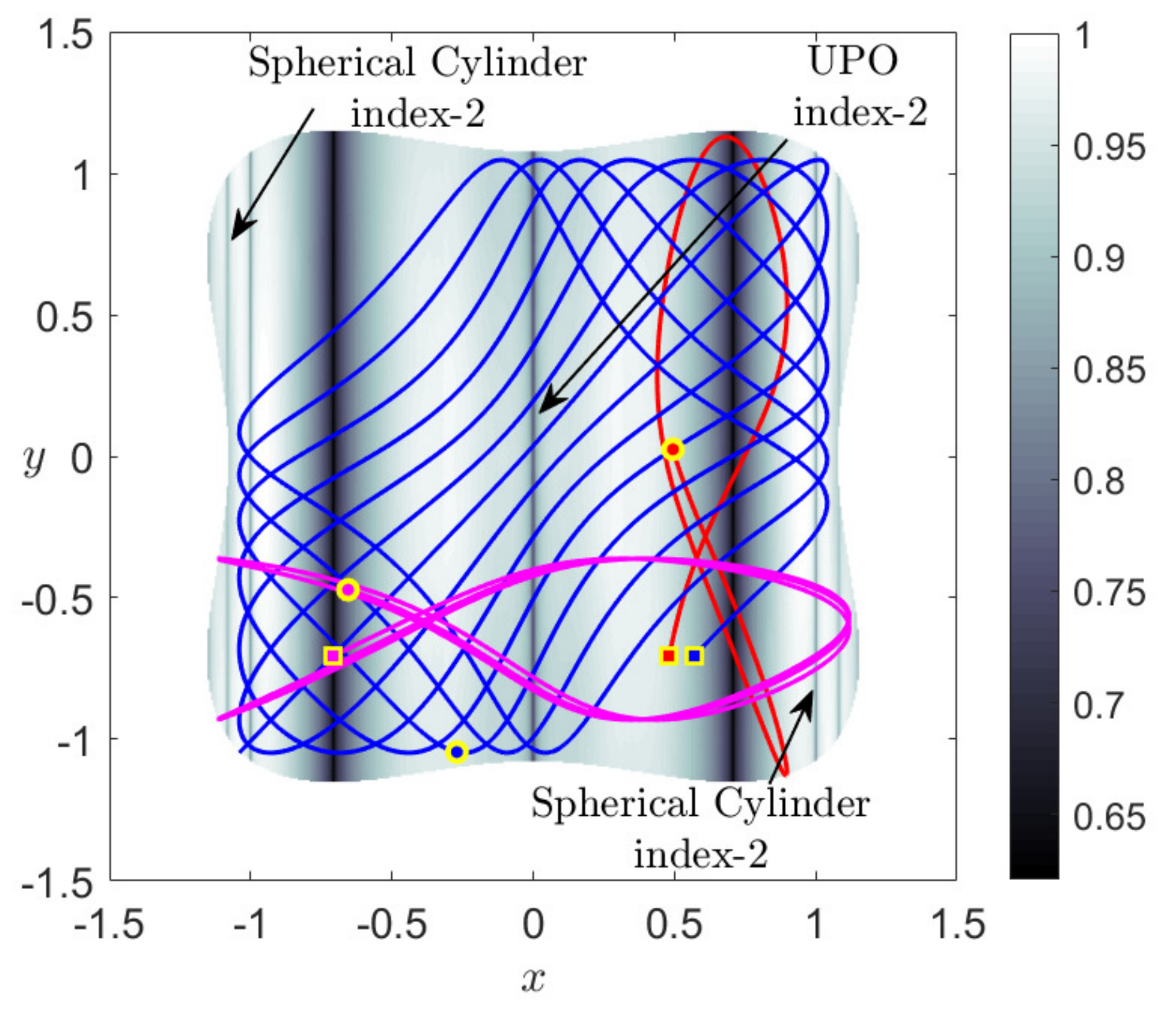} \\[.3cm]
		C)\includegraphics[scale=0.32]{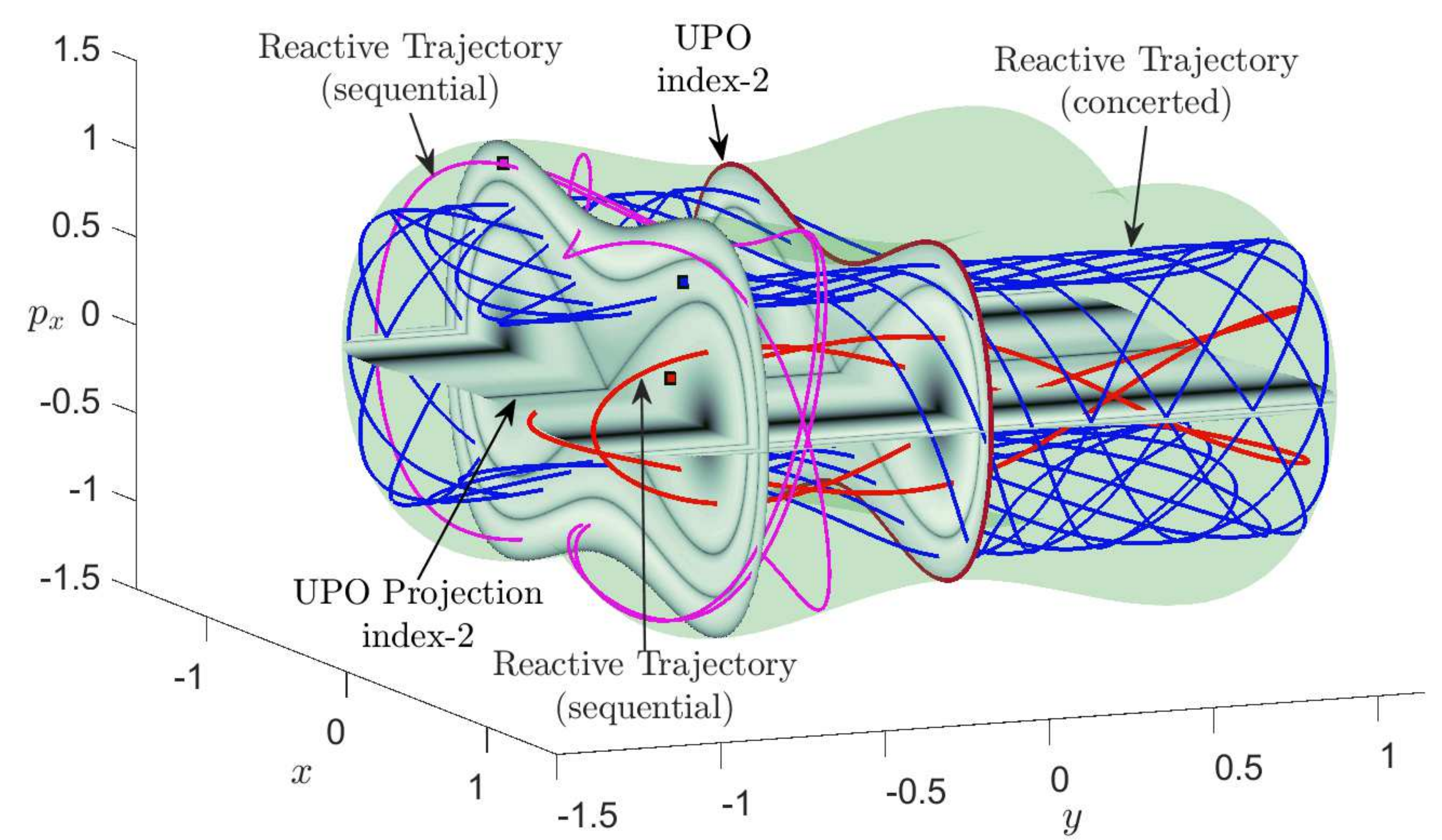}
	\end{center}
	\caption{Phase space structures and evolution of initial conditions at an energy  $H_0 = 0.2$ for the symmetric uncoupled Hamiltonian. A) LDs calculated using $\tau = 5$ on the phase space slice in Eq. \eqref{psos1}. B) Superposition of LDs calculated using $\tau = 5$ on the PSOS in Eq. \eqref{psos2} with the dynamical evolution of the initial conditions selected in panel A. C) Visualization of the phase space dynamics in the three-dimensional energy hypersurface.}
	\label{LD_sym_h_pos}
\end{figure}

We finish this section by applying Lagrangian descriptors in order to reveal the phase space structures associated to the index-2 saddle at the origin. To do so, we calculate LDs using an integration time of $\tau = 5$ on the PSOS in Eq. \eqref{psos1} and \eqref{psos2}. In Fig. \ref{LD_sym_h_pos} A) we demonstrate that LDs succeeds in highlighting three dynamically distinct phase space regions, separated by the intersections of the spherical cylinders of the index-2 saddle with this SOS. With the purpose of validating the discussion carried out previously in terms of the partition of the total energy of the system among the $x$ and $y$ DoF, we probe the dynamical fate of trajectories by choosing three initial conditions, one in each region. We evolve them in forward time and plot their projections onto configuration space in Fig. \ref{LD_sym_h_pos} B), superimposed with the output of LDs in that phase space slice. Moreover, we depict their evolution in the three-dimensional energy hypersurface to help with the visualization of the geometrical phase space structures that characterize the ismerization dynamics. Both the red and magenta initial conditions display dynamical behavior corresponding to sequential isomerization, since they start their evolution inside the stable and unstable manifolds of the unstable periodic orbits $\mathcal{U}_1$ and $\mathcal{U}_2$ respectively. In contrast, the blue initial condition is inside the spherical cylinders of both UPOs associated to the index-2 saddle, and therefore exhibits concerted isomerization and thus the trajectory it gives rise to goes over the index-2 saddle at the origin at some point along its evolution.

\subsection{Dynamics of the Asymmetric Problem}

\begin{figure}[!h]
	\begin{center}
		A)\includegraphics[scale=0.3]{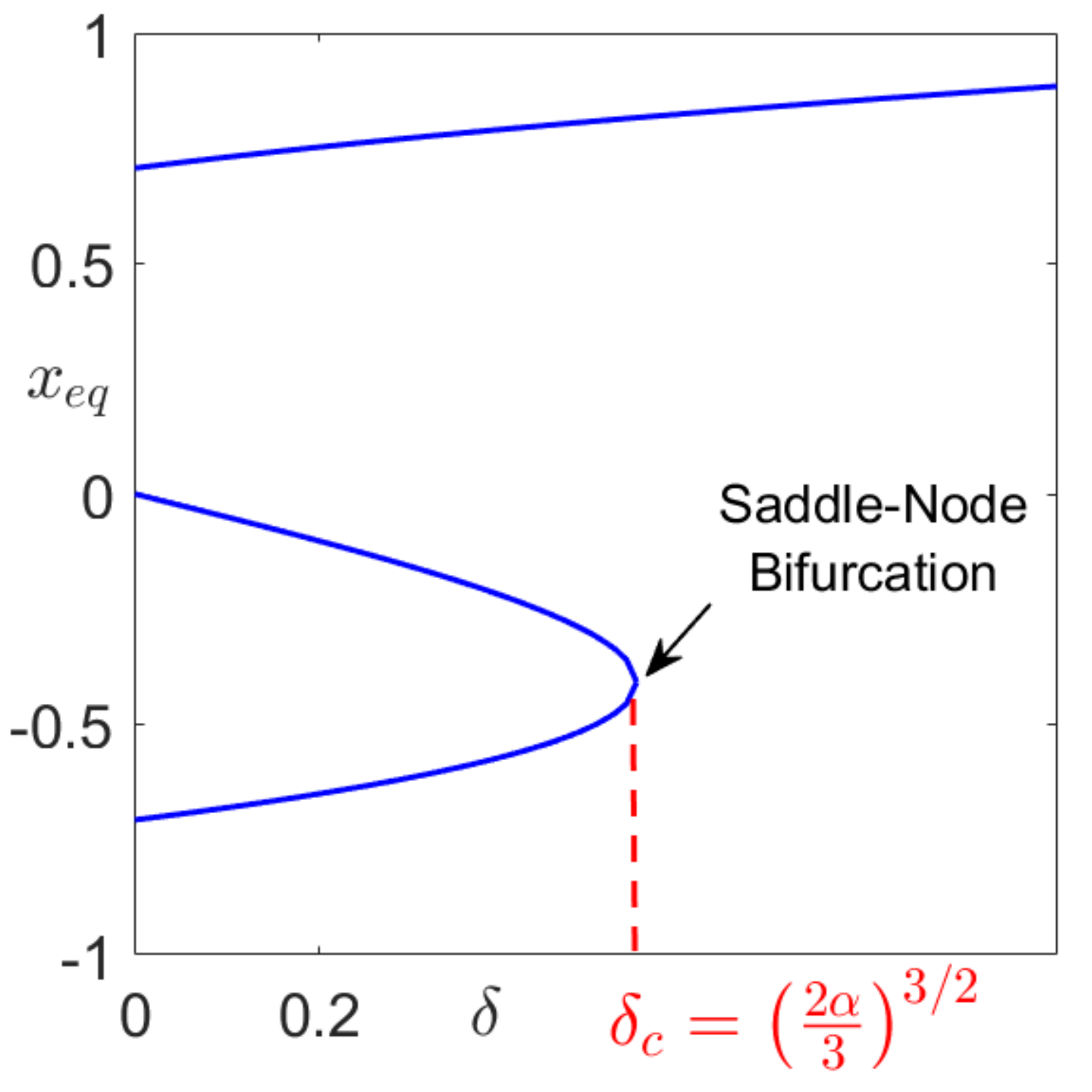}
		B)\includegraphics[scale=0.29]{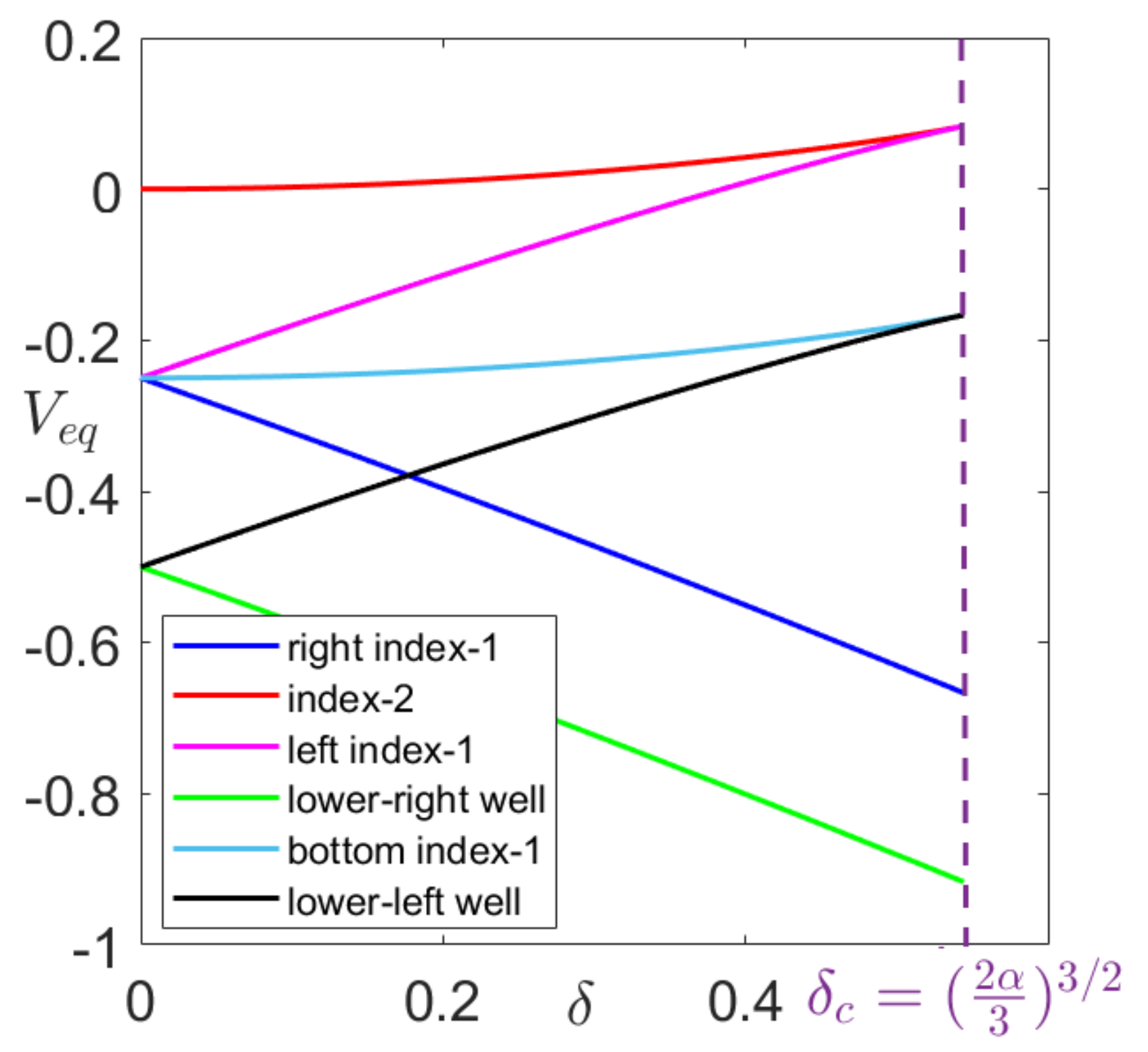}
	\end{center}
	\caption{A) Bifurcation diagram describing the $x$ coordinate location of the critical points of the PES given by Eq. \eqref{pes_model} in terms of the asymmetry parameter $\delta$. B) Potential energy of the critical points of the PES as a function of the asymmetry in the system.}
	\label{bif_diag_delta}
\end{figure}

In this section we explore the influence that breaking the symmetry in the $x$ DoF of the Hamiltonian system has on the phase space structures governing isomerization dynamics on the PES described in Eq. \eqref{pes_model}. We focus on the case where the DoF of the system are uncoupled, that is $\beta = 0$. Recall that the symmetric ($\delta = 0$) PES discussed above has nine critical pints: four index-1 saddles, four potential wells, and one index-2 saddle at the origin. The effect of increasing the value of the asymmetry parameter $\delta$ above zero is that the critical points of the symmetric system that originally lie on the lines $x = -1/\sqrt{2}$ and $x = 0$ will approach each other until they collide, and three simultaneous saddle-node bifurcations occur at the critical value of the asymmetry parameter $\delta_c(\alpha)  = (2\alpha/3)^{3/2}$. Remember that throughout this work we have fixed $\alpha = 1$, so that in the case studied here $\delta_c = (2/3)^{3/2}$. The upper index-1 saddle and the upper-left potential well merge for this critical value, and this also happens for the index-2 saddle and the left index-1 saddle, and for the bottom index-1 saddle and the lower-left potential well. This situation is illustrated in the bifurcation diagram included in Fig. \ref{bif_diag_delta} A), and the evolution of the energies of the critical points of the PES as a function of the asymmetry is also shown in Fig. \ref{bif_diag_delta} B). Interestingly, a relevant consequence that can be drawn from breaking the symmetry of the system is that the potential energy barrier of the right index-1 saddle connecting the lower-right and upper-right wells decreases, while that of the left index-1 saddle increases. This makes the sequential isomerization route between the wells on the right side of the PES energetically favorable compared to that connecting the wells on the left part of the PES. What we mean by this is that, for a given energy level of the system, the cross-sectional area of the phase space bottleneck in the neighborhood of the right index-1 saddle is larger than that related to the left index-1 saddle, and hence the flux of reactive trajectories in that region is higher. As a result, the system will end up forming more stable isomers of the types corresponding to the lower and upper right wells.

\begin{figure}[!h]
	\begin{center}
		A)\includegraphics[scale=0.16]{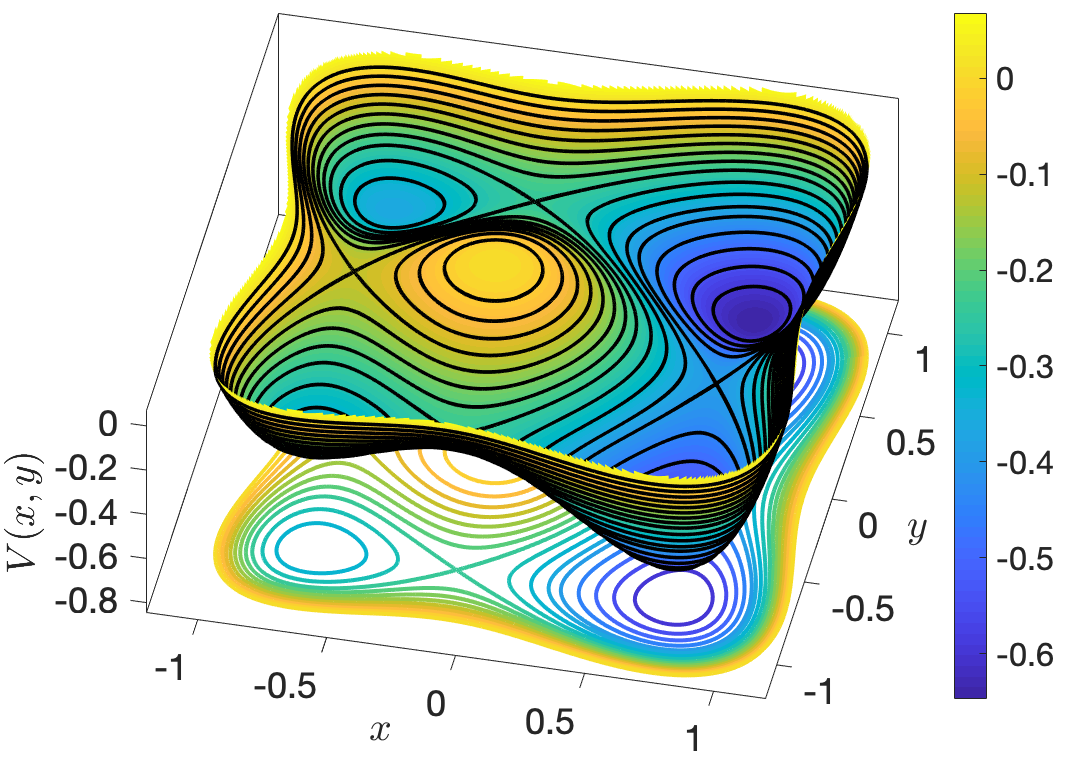}
		B)\includegraphics[scale=0.15]{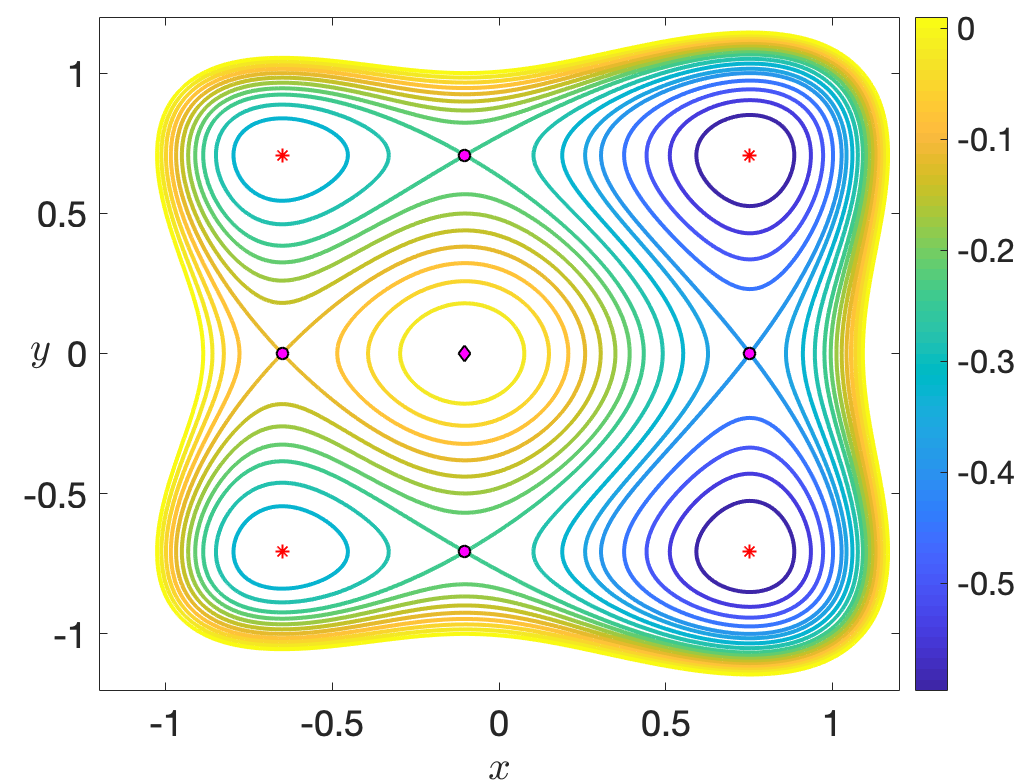}
		C)\includegraphics[scale=0.25]{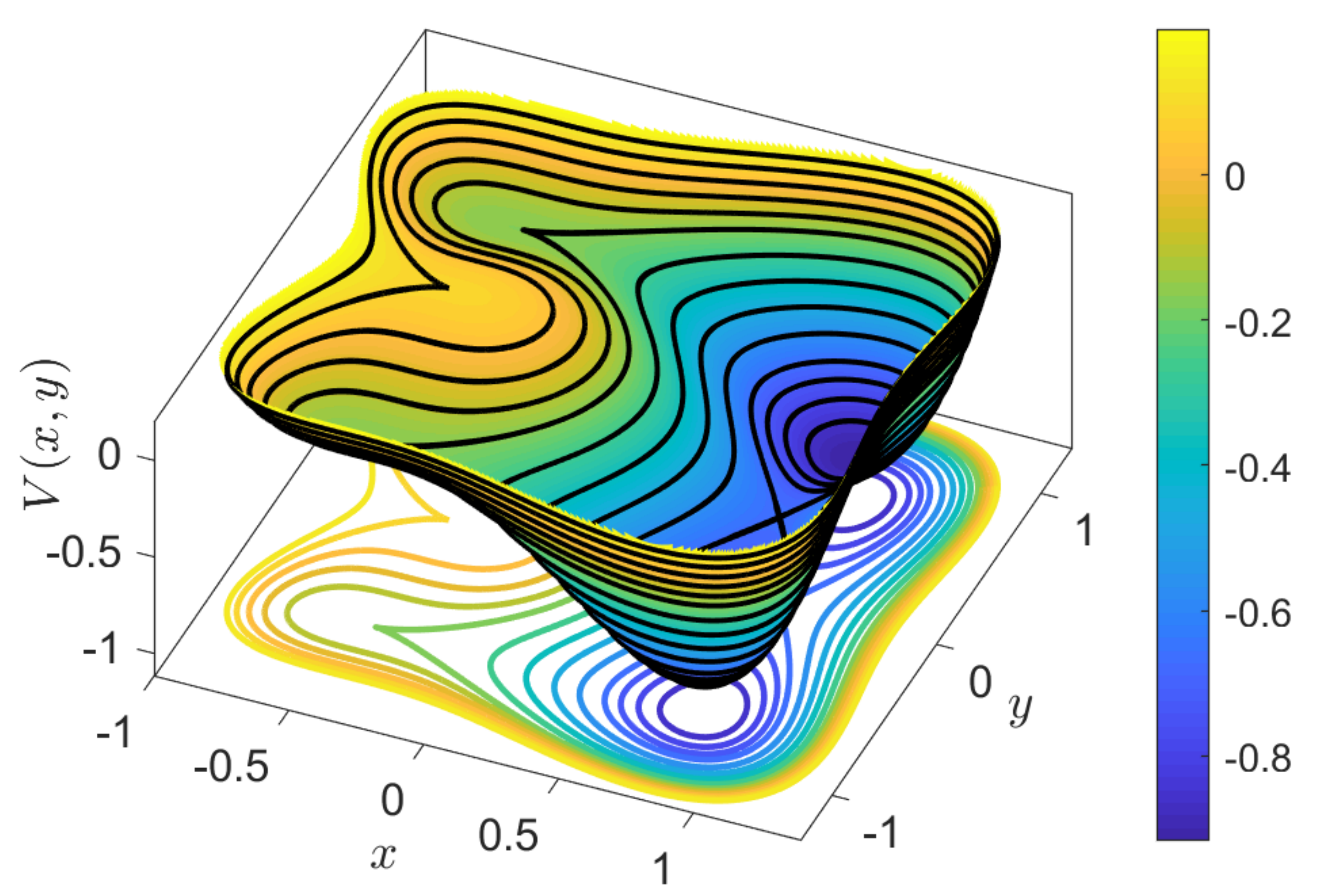}
		D)\includegraphics[scale=0.24]{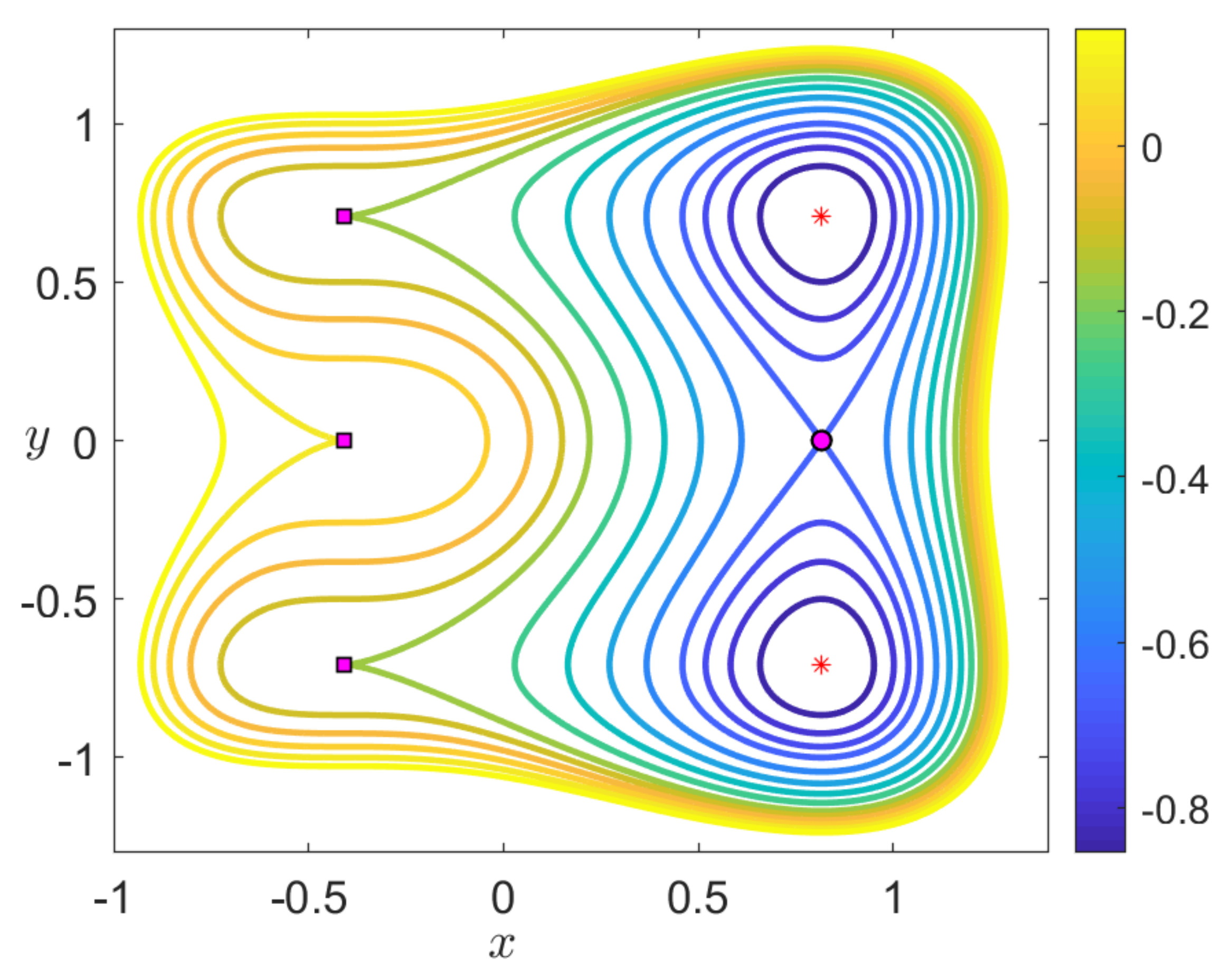}
		E)\includegraphics[scale=0.26]{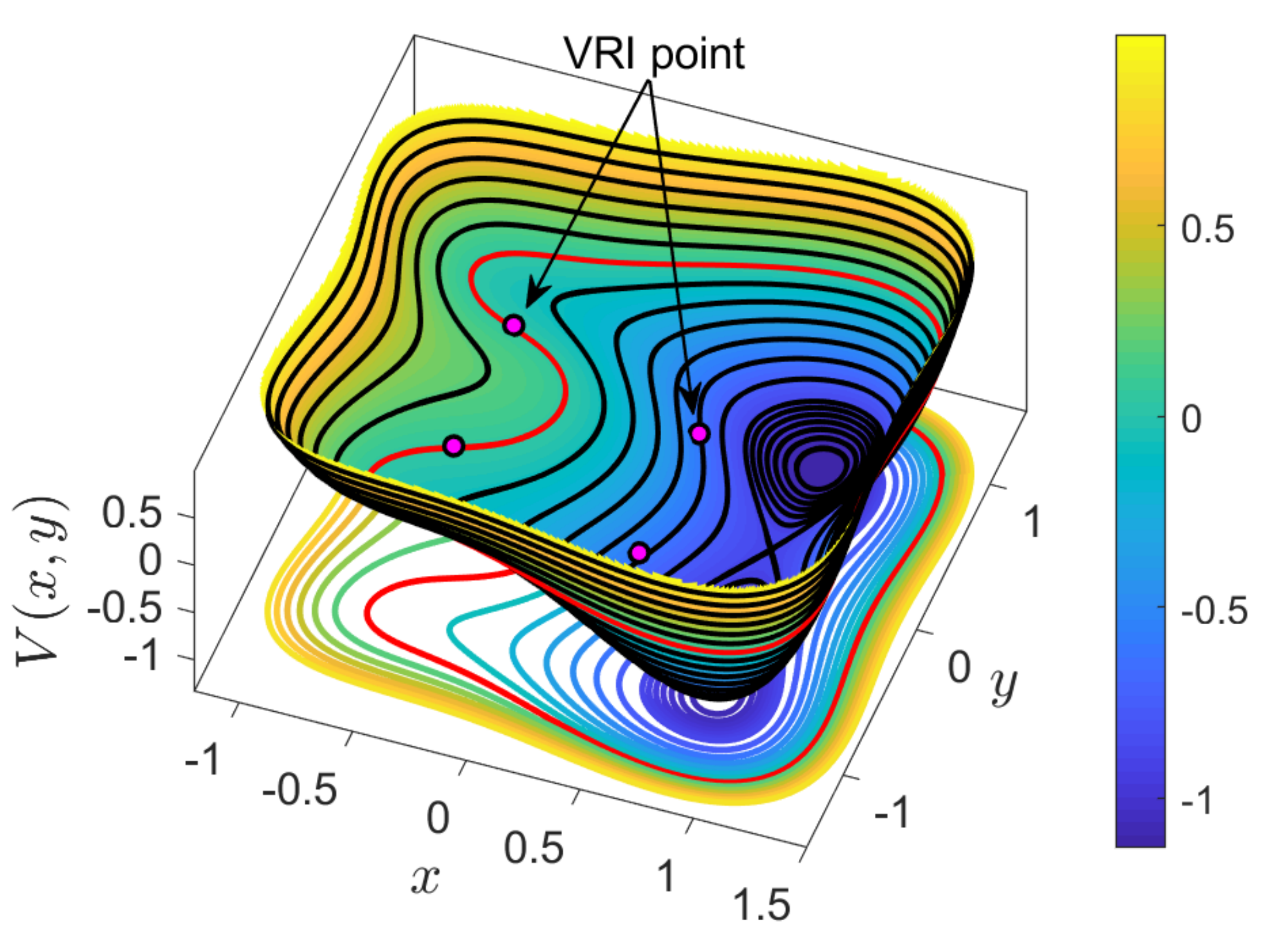}
		F)\includegraphics[scale=0.24]{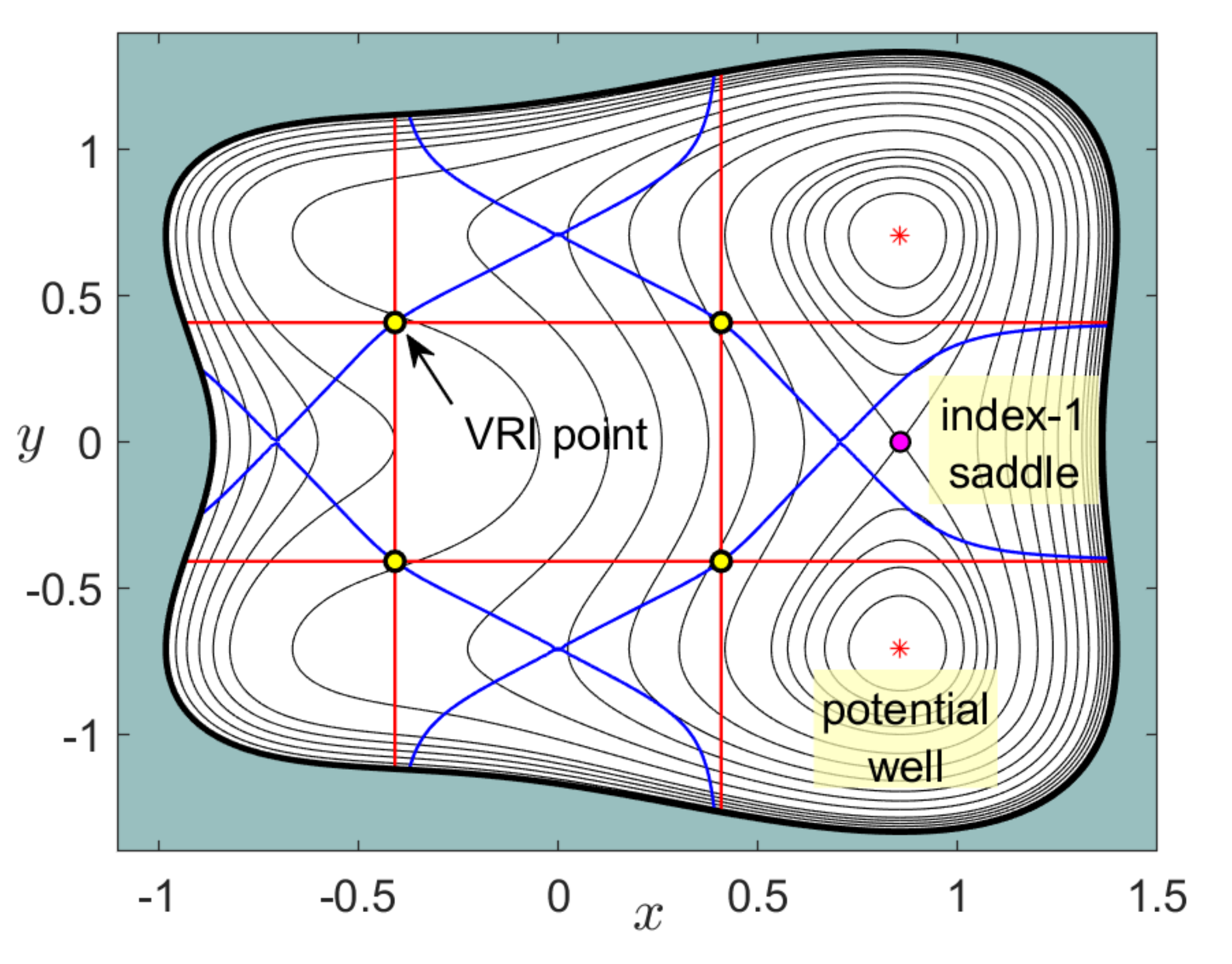}
	\end{center}
	\caption{Potential energy surface landscape (left column) and equipotential curves in configuration space (right column) for the asymmetric and uncoupled potential in Eq. \eqref{pes_model} for different values of the asymmetry parameter $\delta$. A) and B) correspond to $\delta = 0.2 < \delta_c$; C) and D) are for the critical value $\delta = \delta_c = (2/3)^{3/2}$; E) and F) represent the case $\delta = 0.8 > \delta_c$, where we have marked the locations of VRI points. In F) we depict in grey, for a given energy level of the system, the energetically forbidden realm. The red and blue curves included in panel F) represent the points of the PES that satisfy the first and second conditions for the location of VRI points described in Eq. \eqref{vri_conds}.}
	\label{asym_pot_dif_delta}
\end{figure}

In order to understand the dynamical behavior of the Hamiltonian system as the asymmetry in the $x$ DoF is varied, we will take a look at three different values of the asymmetry parameter: $\delta = 0.2, \; \delta_c,\;  0.8$. In Fig. \ref{asym_pot_dif_delta} we present how the PES landscape and its equipotential curves in configuration space change for these cases. We have marked the index-1 saddles, index-2 saddle and potential wells as circles, diamonds and asterisks respectively. Notice that in the case where $\delta = 0.8$, which is above the critical value $\delta_c$ of the asymmetry, we have also marked the location of valley-ridge inflection (VRI) points. We have done so because our analysis reveals the possible existence of phase space structures associated to them that contribute to the system dynamics. This type of configuration space points are not critical points of the PES, but they satisfy the following conditions:
\begin{equation}
\begin{cases}
\det (Hess_{V}) = 0 \\[.1cm]
\left(\nabla V\right)^{T} adj[Hess_{V}] \; \nabla V = 0
\end{cases} \;,
\label{vri_conds}
\end{equation}
where $adj[Hess_{V}]$ represents the adjugate matrix of the Hessian of the PES. These points characterize locations of the PES whose Gaussian curvature is zero, and for which there exists an eigenvector of the Hessian matrix, with eigenvalue zero, perpendicular to the gradient of the PES. For a discussion regarding the dynamical significance of VRI points for the reaction dynamics on a model PES with two DoF, see e.g. \cite{wiggins_vri} and references therein. 

\begin{figure}[!h]
	\begin{center}
		A)\includegraphics[scale=0.32]{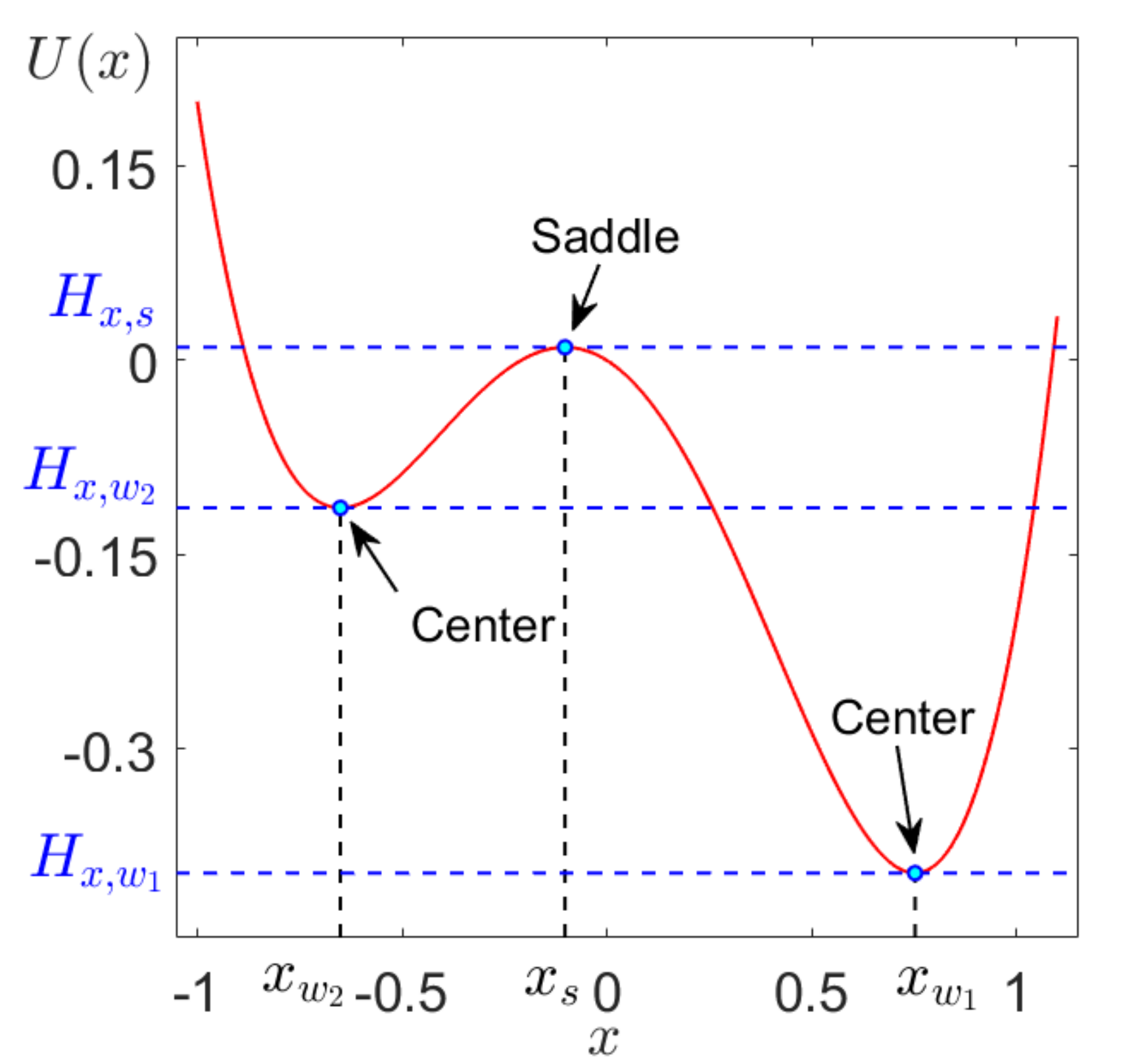}
		B)\includegraphics[scale=0.21]{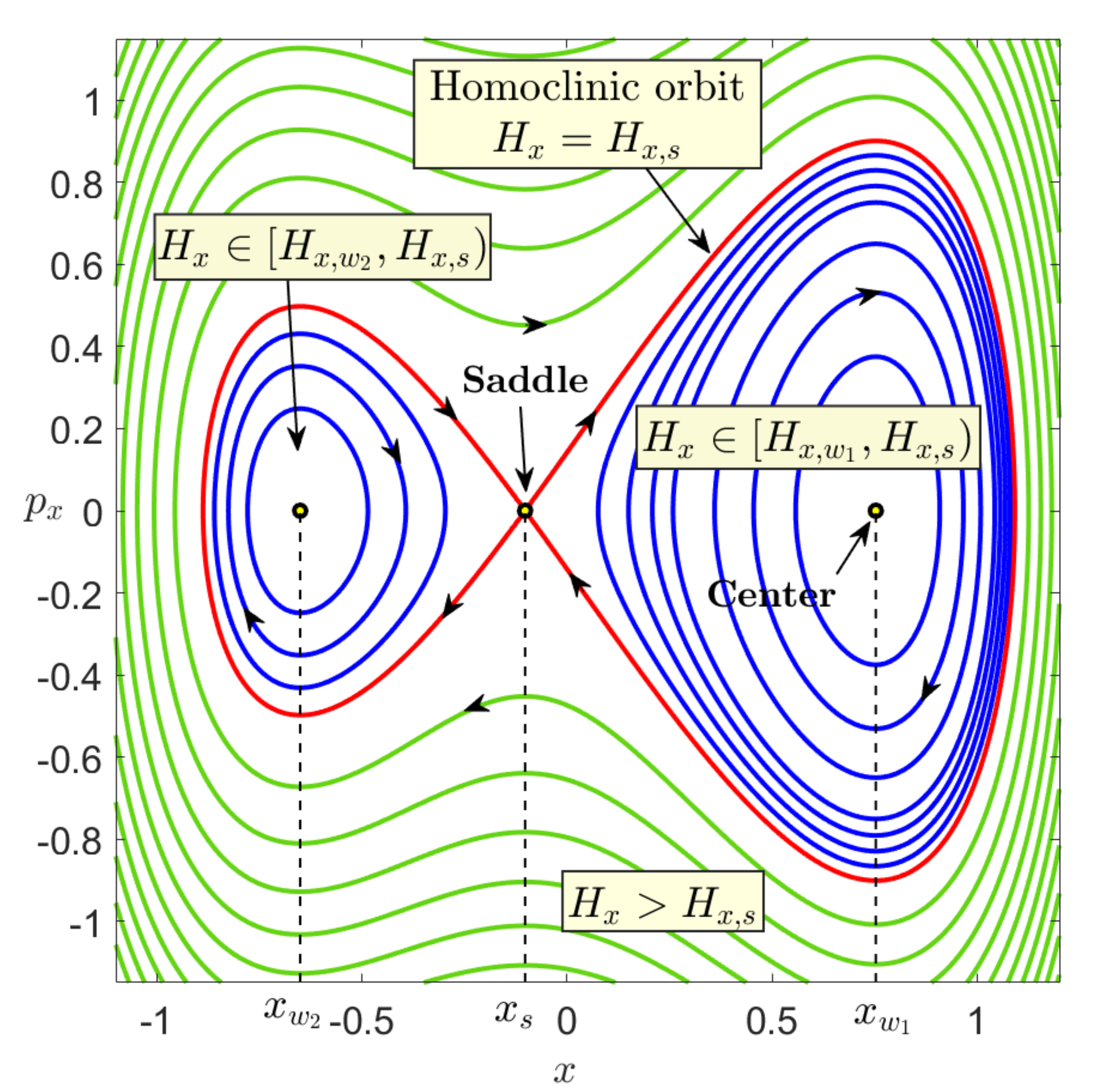} \\
		C)\includegraphics[scale=0.33]{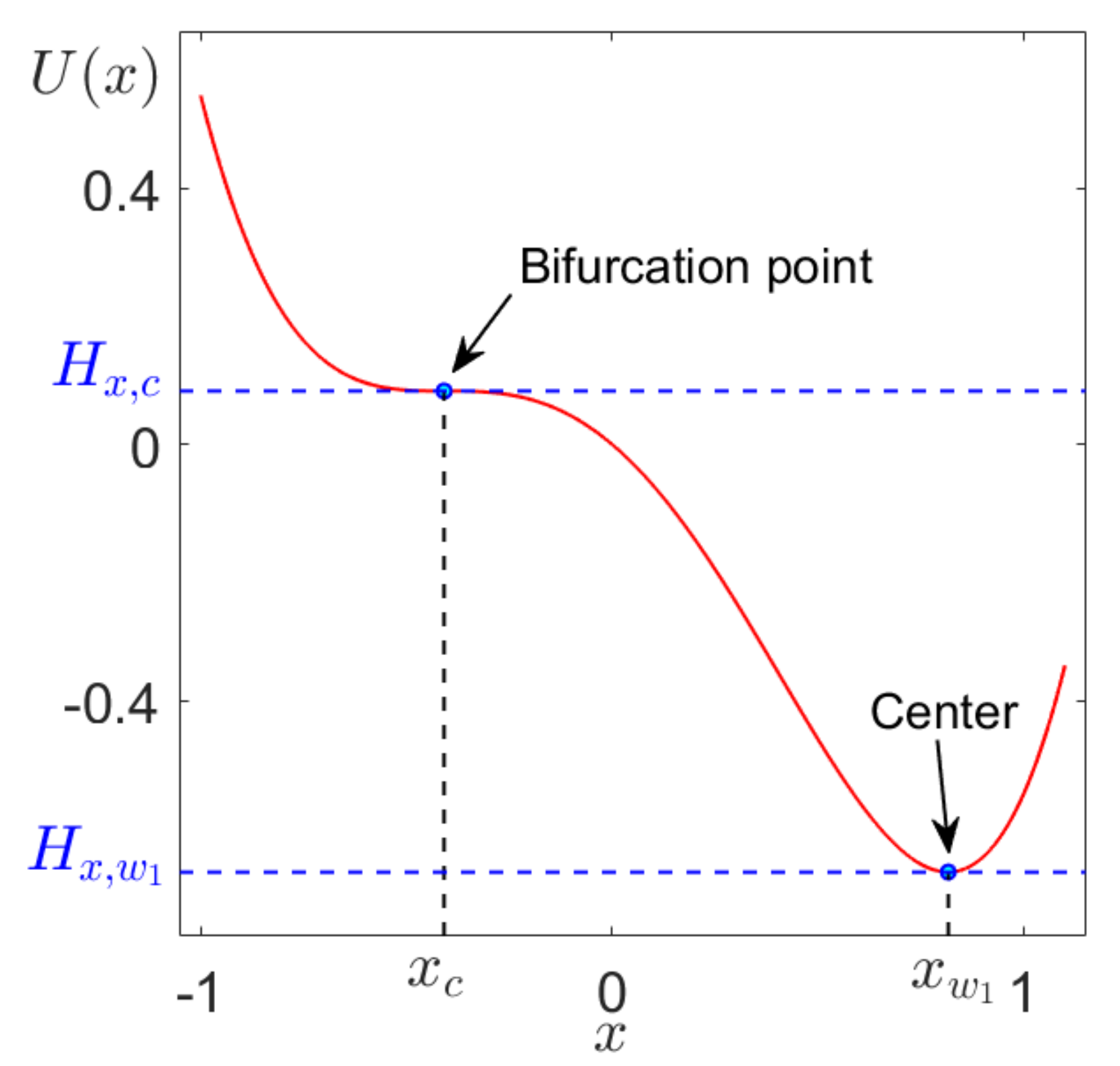} 
		D)\includegraphics[scale=0.21]{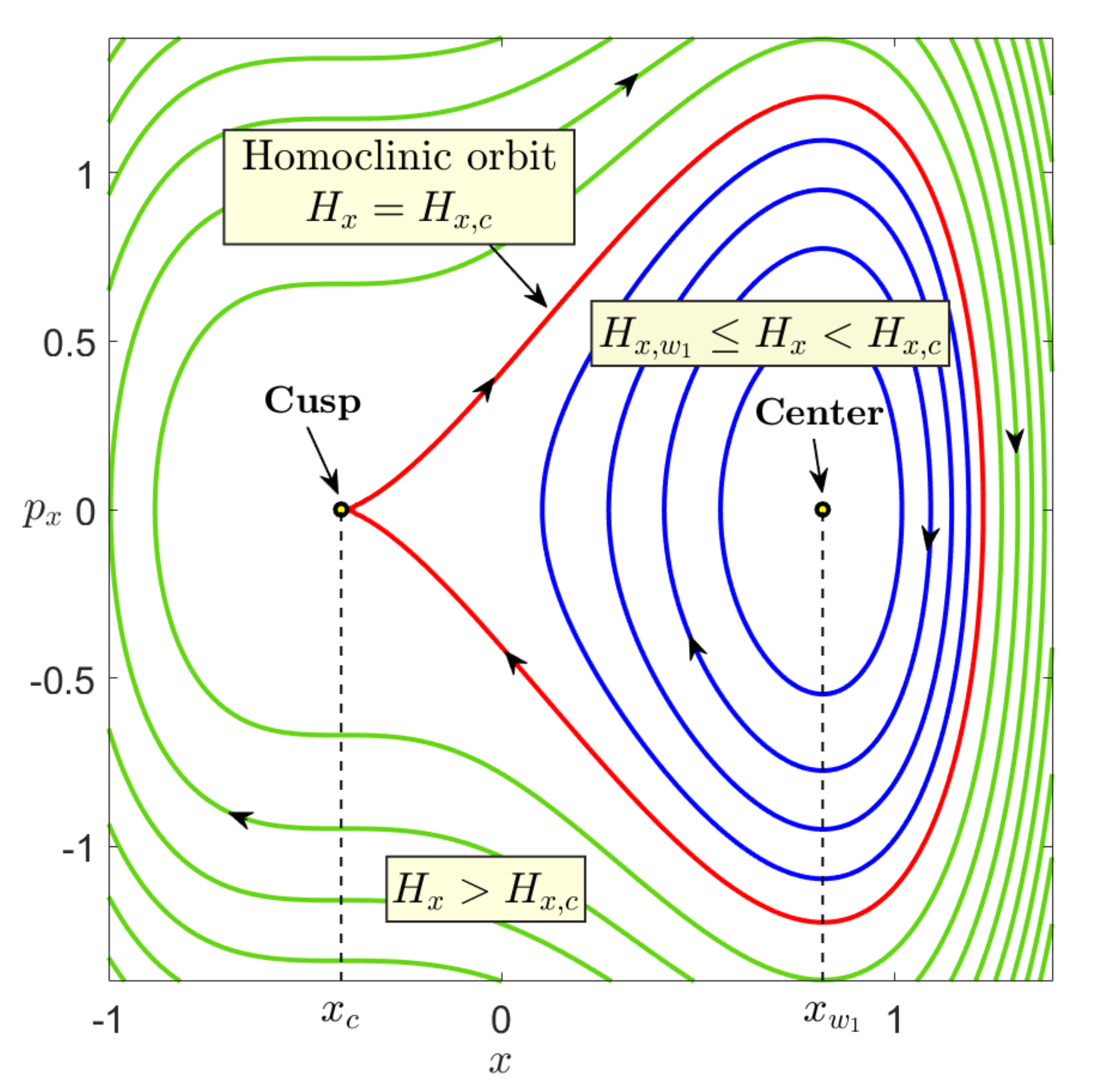} \\
		E)\includegraphics[scale=0.33]{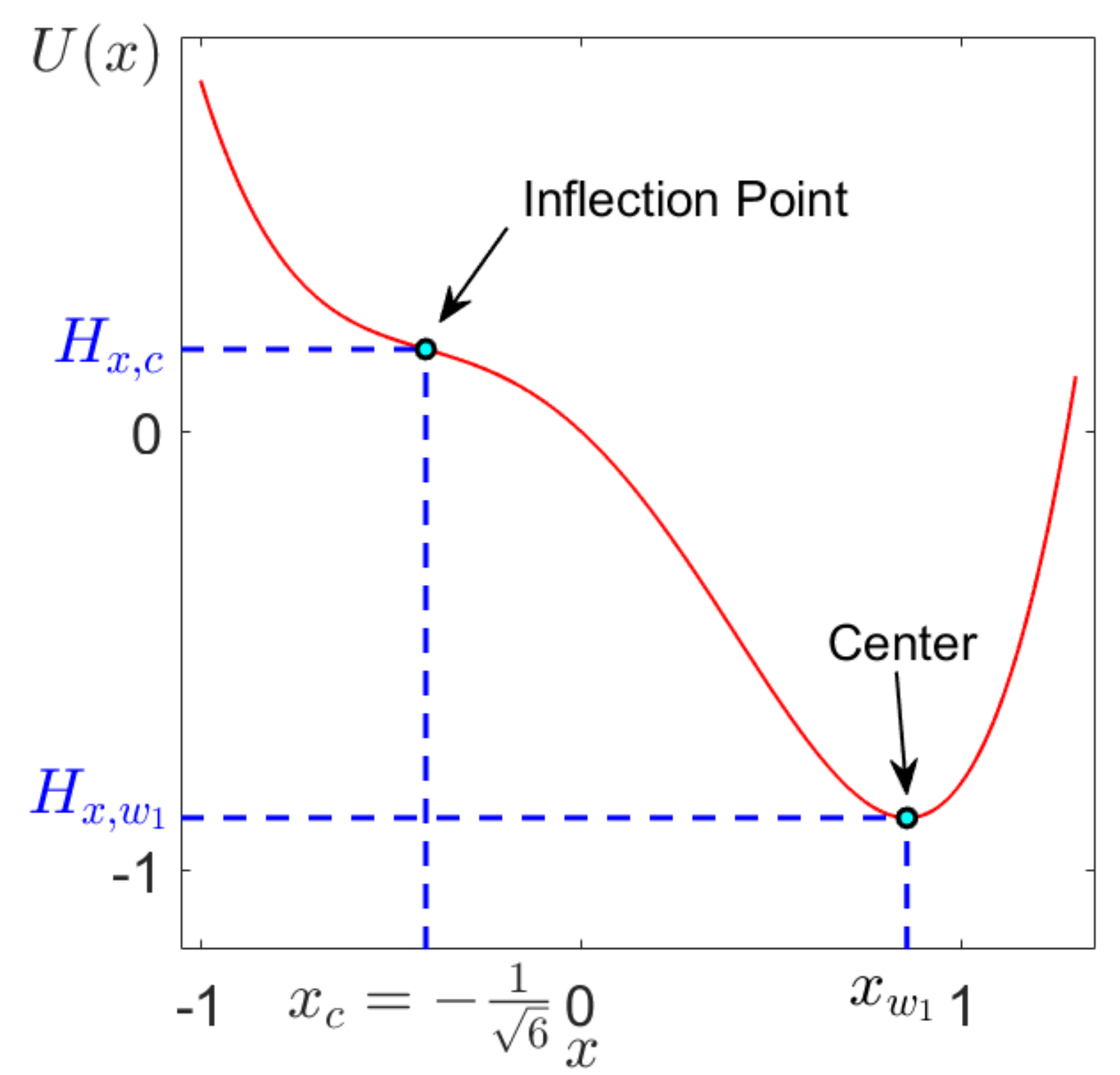} 
		F)\includegraphics[scale=0.21]{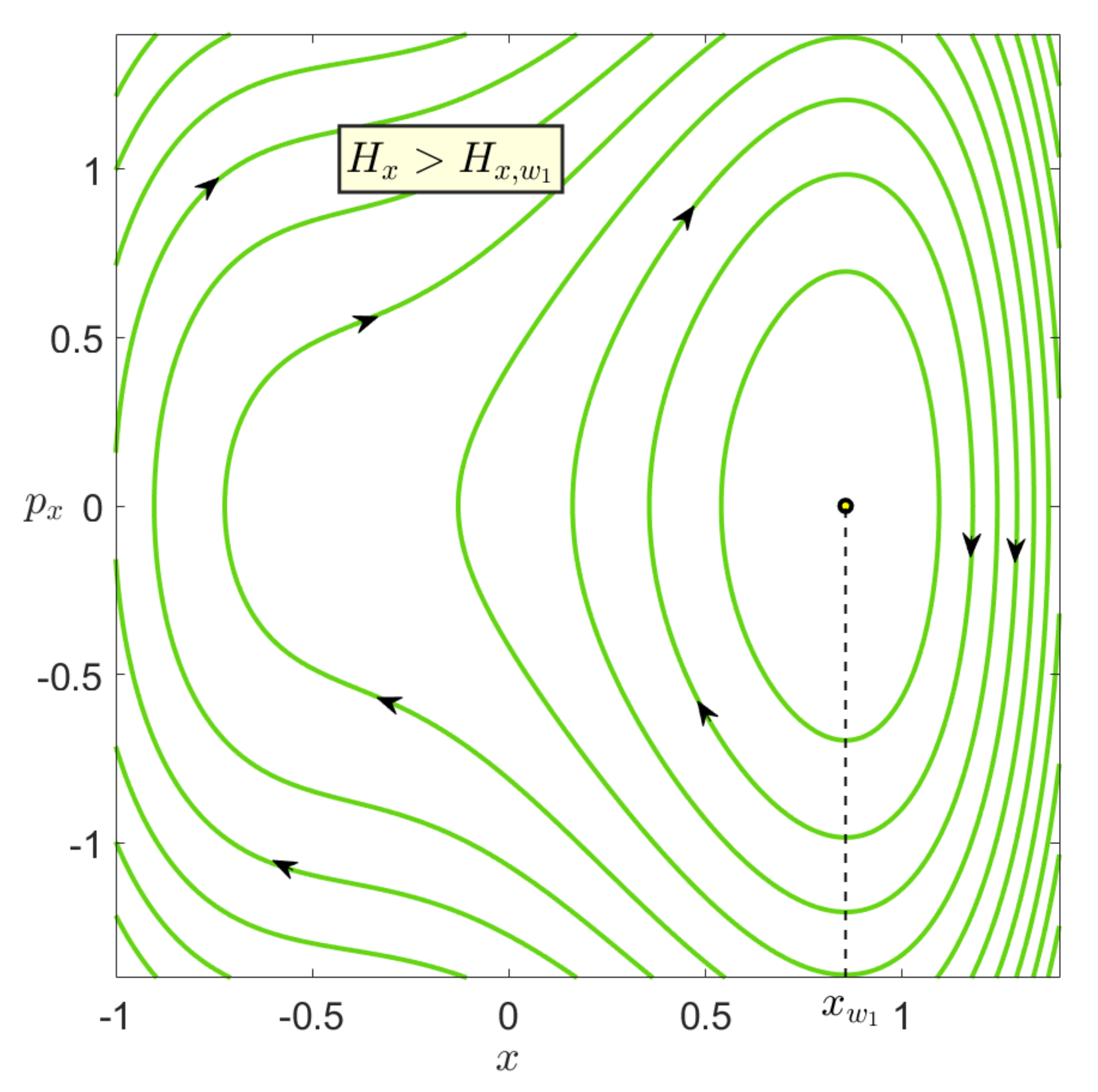}
	\end{center}
	\caption{Potential energy for the $x$ DoF (left column) and associated $x-p_x$ phase portrait (right column) for the Hamiltonian $H_x$ in Eq. \eqref{energy_asym} using different values of the asymmetry parameter $\delta$. A) and B) correspond to $\delta = 0.2 < \delta_{c}$; C) and D) are for the critical value $\delta = \delta_c$ at which a saddle-node bifurcation occurs in the $x$ direction; E) and F) represent the case $\delta = 0.8 > \delta_{c}$.}
	\label{asym_pot}
\end{figure}

We begin our analysis by looking at the case $\delta = 0.2 < \delta_c$, for which the PES has has nine critical points, as we illustrate in Fig. \ref{asym_pot_dif_delta} A)-B). Consider that the uncoupled system has a total energy $H_0$, and that it is naturally distributed among the DoF in the form $H_0 = H_{x,0} + H_{y,0}$. This allows us to write:
\begin{equation}
H_{x}(x,p_x) = \frac{1}{2} \, p_x^2 + U(x) = H_{x,0} \quad,\quad H_{y}(y,p_y) = \frac{1}{2} \, p_y^2 + W(y) = H_{y,0} \;,
\label{energy_asym}
\end{equation}
where the potential energy for the $y$ DoF, $W(y)$, has the shape of a symmetric double well described in Eq. \eqref{1D_potSymm}, and the potential energy function in the $x$ DoF is an asymmetric double-well potential in the form:
\begin{equation}
U(x) = x^4 - x^2 - \delta x \;,
\label{potasymun}
\end{equation}
characterized by the asymmetry parameter $\delta$. In Fig. \ref{asym_pot} we depict the potential $U(x)$ for different values of the asymmetry together with the corresponding dynamics it gives rise to in the $x-p_x$ phase portrait. 

Given that the energy of the Hamiltonian system is conserved, phase space dynamics takes place in the three-dimensional energy hypersurface:
\begin{equation}
\mathcal{S}(H_0) = \left\{ (x,y,p_x,p_y) \in \mathbb{R}^4 \; \bigg| \; H_0 = \frac{1}{2}\left(p_x^2+p_y^2\right) + x^4 - x^2 - \delta x + y^4 - y^2 \right\} \;.
\label{energy_surf_asym}
\end{equation}
and motion is regular because the system is fully integrable due to energy conservation in each DoF. Before describing the system dynamics, we illustrate how the phase space structures that determine transport between different potential well regions of the PES can be constructed. In particular, we take a look at the geometrical structures associated to the upper index-1 saddle that are responsible for the transport mechanisms between the upper-left and upper-right wells. The dividing surface separating both wells is defined by intersecting the energy surface with the slice $x = x_s$, and the value of $x_s$ is that of $x_2$ in Eq. \eqref{roots_cubic_asymm}, that is:
\begin{equation}
\mathcal{D}\left(H_0\right) = \mathcal{S}\left(H_0\right) \cap \lbrace x = x_{s} \rbrace = \left\{(x,y,p_x,p_y) \in \mathbb{R}^4 \; \bigg| \; H_0 + \frac{x_s}{2} \left(x_s + \frac{3\delta}{2}\right) = \frac{1}{2}\left(p_x^2+p_y^2\right) + y^4 - y^2 \right\} \;,
\label{ds_asym_unc}
\end{equation}
where we have used that the potential energy $H_{x,s} = U(x_s) = x_s^4 - x_s^2 - \delta x_s = -x_s^2/2 - (3\delta/4) \, x_s$, a property that follows from the fact that $x_s$ is a critical point of $U(x)$. The dividing surface is non-invariant and has the local non-recrossing property. Moreover, it has the topology of a sphere $S^2$ with two hemispheres, the forward and backward dividing surfaces, given by:
\begin{equation}
\begin{split}
\mathcal{D}_{f}(H_0) &= \left\{(x,y,p_x,p_y) \in \mathbb{R}^4 \; \bigg| \; H_0 + \frac{x_s}{2} \left(x_s + \frac{3\delta}{2}\right) = \frac{1}{2}\left(p_x^2+p_y^2\right) + y^4 - y^2 \;,\; p_x > 0 \right\} \\[.2cm]
\mathcal{D}_{b}(H_0) &= \left\{(x,y,p_x,p_y) \in \mathbb{R}^4 \; \bigg| \; H_0 + \frac{x_s}{2} \left(x_s + \frac{3\delta}{2}\right) = \frac{1}{2}\left(p_x^2+p_y^2\right) + y^4 - y^2 \;,\; p_x < 0 \right\}
\end{split}
\;.
\end{equation}
The two hemispheres meet at the equator, which is a NHIM, or an UPO, described by:
\begin{equation}
\mathcal{N}(H_0) = \left\{(x,y,p_x,p_y) \in \mathbb{R}^4 \; \bigg| \; H_0 + \frac{x_s}{2} \left(x_s + \frac{3\delta}{2}\right) = \frac{1}{2} \, p_y^2 + y^4 - y^2 \;,\; p_x = 0 \right\} \;,
\end{equation}
with the topology of a circle $S^1$. The NHIM has stable and unstable manifolds:
\begin{equation}
\mathcal{W}^{u}(H_0) = \mathcal{W}^{s}(H_0) = \left\{\left(x,y,p_y,p_y\right) \in \mathbb{R}^4 \; \bigg| \; H_{x,s} = \frac{p_x^2}{2} + x^4 - x^2 - \delta x \;,\; H_{y,0} = \frac{p_y^2}{2} + y^4 - y^2 \right\} \;,
\end{equation}
and topologically they have the structure of $S^1 \times \mathbb{R}$, representing \textit{tube manifolds} or \textit{spherical cylinders}. Observe that for the asymmetric and uncoupled case that we are analyzing they have a homoclinic structure in phase space. The phase space structures defined above are responsible for the reaction mechanisms in phase space and characterize the bottleneck region that connects the upper-left and upper-right wells of the PES. Recall that these tube manifolds act as \textit{isomerizaton highways} or reactive conduits so that any initial condition chosen inside them will evolve back and forth between the upper-left and upper-right wells. Notice that in this context, the bottleneck connecting both wells opens when the energy level of the system is above that of the upper index-1 saddle, which yields the condition:
\begin{equation}
H_0 \geq H_{x,s} = V(x_s,\sqrt{2}/2) = - x_s^2/2 - (3\delta/4) \, x_s - 1/4 \;,
\end{equation}
Moreover, from Fig. \ref{bif_diag_delta} B) it is important to highlight that in the asymmetric Hamiltonian the phase space bottlenecks corresponding to the different index-1 saddles of the PES open in the following order as the energy level of the system is increased. First this occurs for the right index-1, then it takes place simultaneously for the saddles located at the top and bottom of the PES, and finally for that on the left region. Notice that in the symmetric system ($\delta = 0$), all bottlenecks open simultaneously for the same energy level, making all of the four stable isomer configurations represented by the potential wells of the PES accessible. On the other hand, when a symmetry-breaking perturbation in the $x$ DoF is introduced, its effect is to energetically favor the stable isomers corresponding to the lower and upper right potential wells. Since we are dealing with the case $\delta = 0.2$, the Hamiltonian system still has nine equilibrium points with the same stability character as those in the symmetric case. Therefore, the phase space dynamical behavior of the asymmetric system is qualitatively similar to the one we discussed earlier in the symmetric setup. The main influence of the symmetry-breaking perturbation is to deform and enlarge the phase space invariant manifolds present in the right part of the energy hypersurface.

\begin{figure}[!h]
	\begin{center}
		A)\includegraphics[scale=0.17]{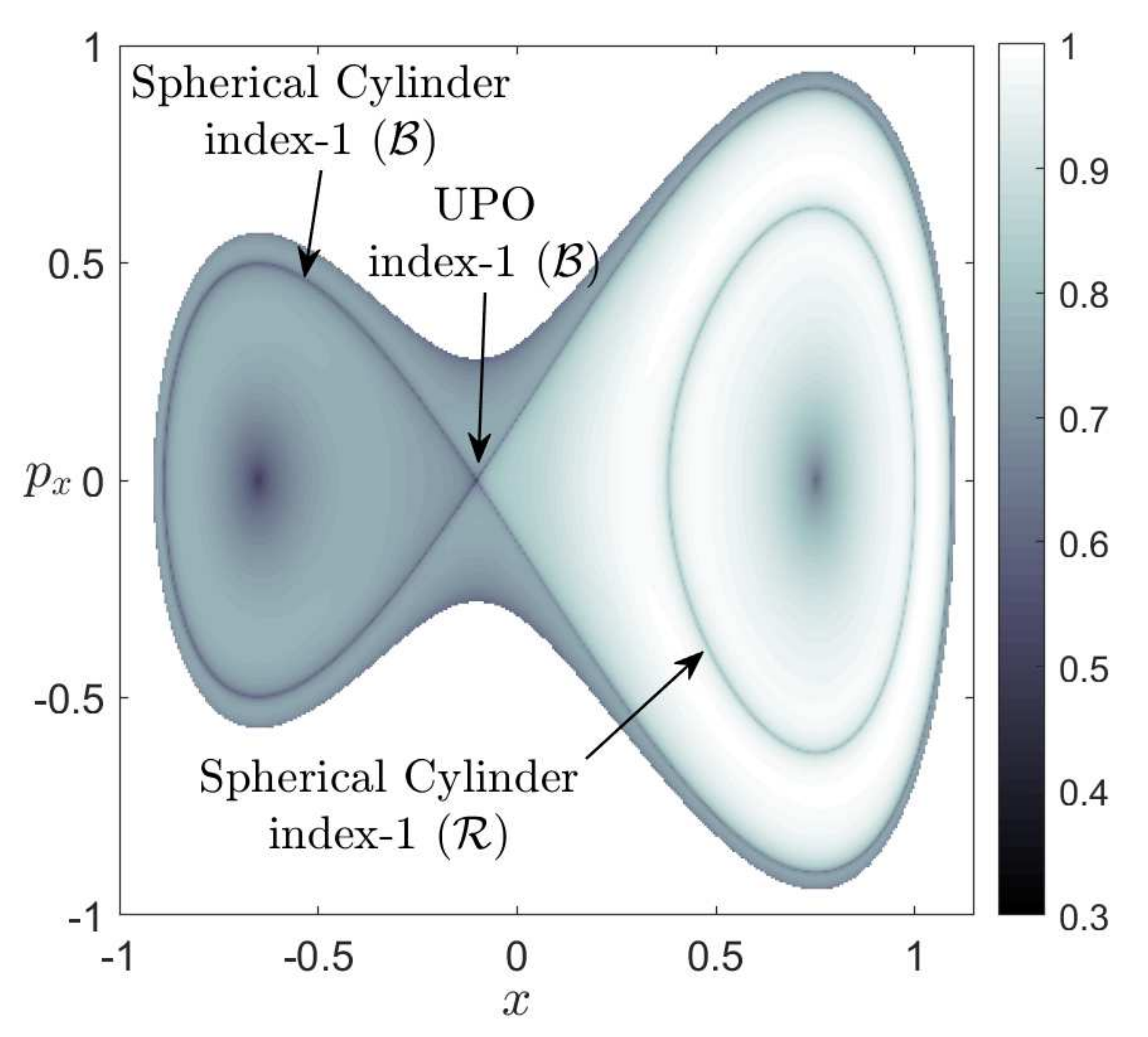}
		B)\includegraphics[scale=0.17]{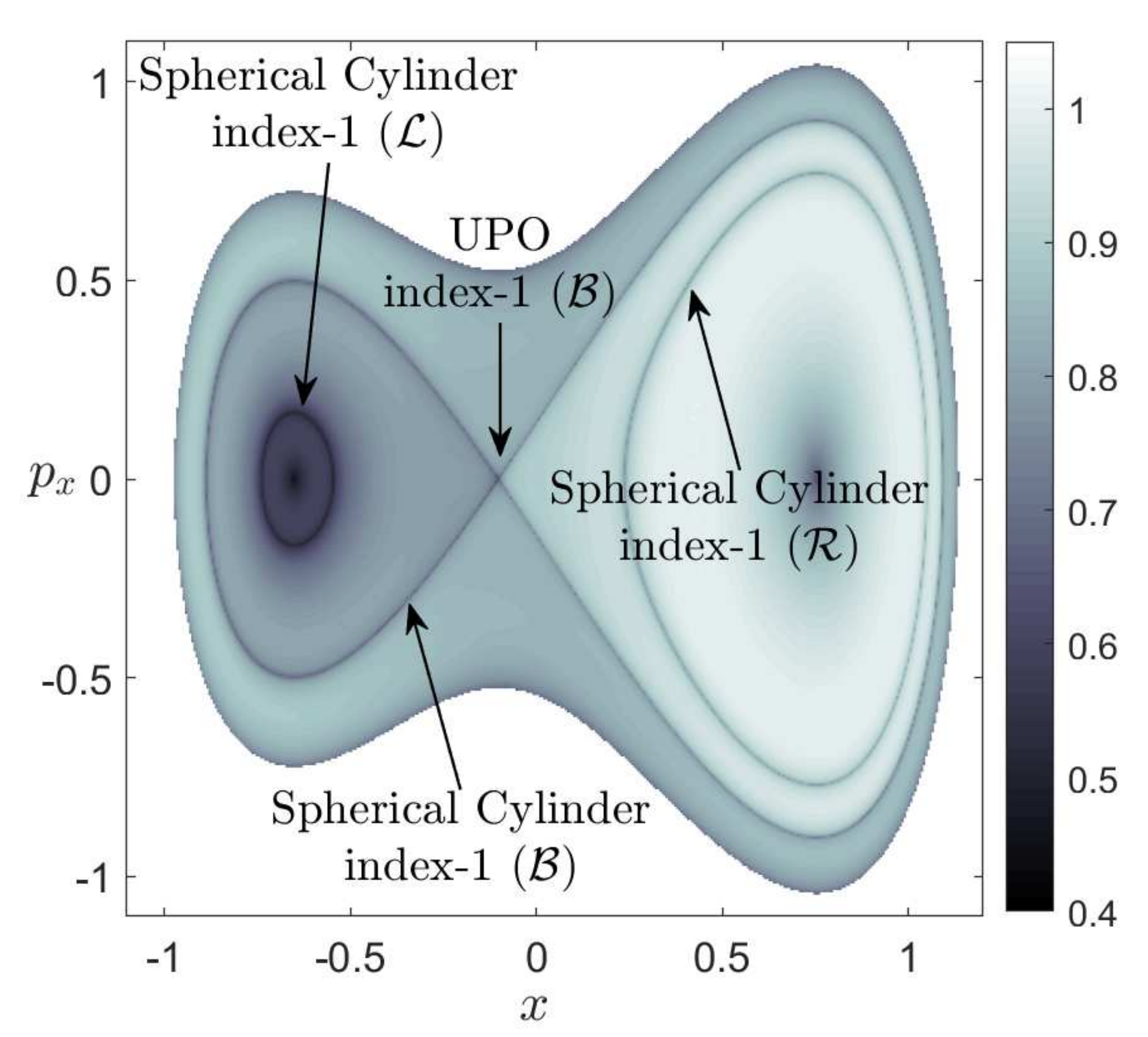}
		C)\includegraphics[scale=0.19]{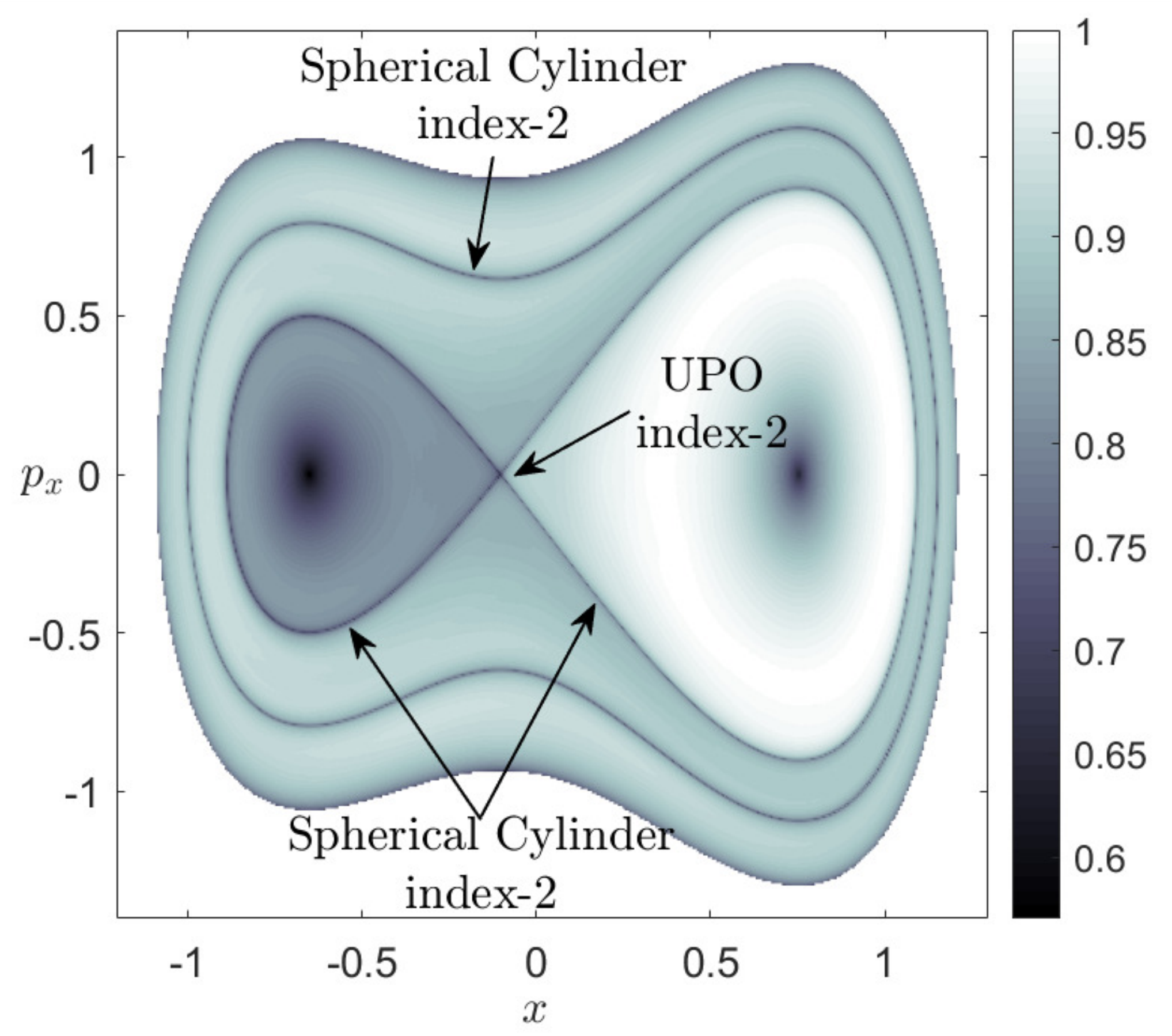}
	\end{center}
	\caption{Lagrangian descriptors calculated using $p = 1/2$ and $\tau = 5$ on the surface of section $y = -\sqrt{2}/2$ for the asymmetric uncoupled Hamiltonian system with $\delta = 0.2$. Energy levels: A) $H_0 = -0.2$, which is below the energy of the left index-1 saddle and above that of the bottom index-1 saddle; B) $H_0 = -0.1$, that is above the energy of the left index-1 saddle and below that of the index-2 saddle; C) $H_0 = 0.2$, above the energy of the index-2 saddle.}
	\label{LD_delta_02}
\end{figure}

In order to analyze the phase space structures that govern isomerization dynamics, we compute LDs on the PSOS given in Eq. \eqref{psos1}, for different energy levels. We begin by setting $H_0 = -0.2$, which is below the energy of the left index-1 saddle and above that of the right, top and bottom index-1 saddles. In this case, all the potential wells are energetically accessible, but the connection between the lower-left and upper-left wells through the phase space bottleneck associated to the left index-1 saddle is closed. This implies that there are no spherical cylinders present in the left region of the energy hypersurface where motion occurs. In Fig. \ref{LD_delta_02} A) we can clearly see how LDs successfully detect the tube manifolds associated to the bottom and right index-1 saddles. Observe that the intersection of the manifolds associated to the UPO of the bottom index-1 saddle with the PSOS appears as homoclinic trajectories, and at their crossing the method highlights the location of the UPO. Notice that initial conditions placed on the phase space region outside the homoclinic orbits will produce reactive trajectories that move between the lower-left and lower-right wells. In addition, the intersection of the spherical cylinders related to the UPO of the right index-1 saddle with this PSOS give rise to reactive islands with the form of closed curves (the cross-section of the tube manifolds). All initial conditions inside the reactive island will evolve back and forth between the lower-right and upper-right wells of the PES. It is evident from the picture in Fig. \ref{LD_delta_02} A) that the area enclosed by the reactive island of the spherical cylinders of the right index-1 saddle is larger than that outside the homoclinic trajectories, which represents the interior of the tube manifolds of the bottom index-1. Consequently, if we take an initial condition at random with energy $H_0 = -0.2$, there is a higher probability that the system will manifest sequential isomerization between the lower-right and upper-right wells.

We analyze next the case where the total energy of the system is $H_0 = -0.1$, that is  above the energy of all the index-1 saddles of the PES but below that of the index-2 saddle. In this situation all the wells of the PES are connected and the hilltop region is energetically forbidden, so that only sequential isomerization is possible in the system between neighboring wells. Observe that the reactive islands corresponding to the left index-1 saddle is now visible in the LD plot included in Fig. \ref{LD_delta_02} B). Moreover, if we choose initial conditions in the different phase space regions identified by LDs when calculated on the PSOS, see Fig. \ref{LD_delta_02} B), we obtain trajectories that evolve in the same way as those discussed in Fig. \ref{LD_sym_h_neg} for the symmetric system. Recall also that, since the system is uncoupled, the spherical cylinders (stable and unstable manifolds) of the UPOs associated to different index-1 saddles do not intersect. As a result, if we pick for example an initial condition on the lower-left well region of the PES and calculate its evolution, the trajectory cannot reach the upper-rgiht well. The two possible motions are: the trajectory is trapped forever in the lower-left well, or it moves back and forth between the lower-left and a neighboring well by crossing the bottleneck connecting both wells.  

The last energy level that we consider is $H_0 = 0.2$, which is taken above that of the index-2 saddle. In this case, the system can exhibit the two types of isomerizaton routes: sequential isomerization between neighboring wells where trajectories do not go over the hilltop of the index-1 saddle, or concerted isomerization for which trajectories can visit along their evolution all wells of the PES, since they are allowed to transit  the hilltop region. The third option available for the dynamical evolution of trajectories is that they remain trapped forever in a potential well. All the distinct phase space regions that give rise to these dynamical behaviors are nicely captured by LDs in Fig. \ref{LD_delta_02} C). Notice that the invariant manifolds obtained are analogous to those shown in Fig. \ref{LD_sym_h_pos} A) that we discussed in depth for the symmetric system. The main difference is the geometrical template of phase space structures is that in the asymmetric system the invariant manifolds are distorted due to the symmetry-breaking perturbation and they are larger and more predominant in the right part of the energy hypersurface.

% exhibit, display, etc...

We analyze next the situation where the asymmetry parameter takes the critical value $\delta_c(\alpha)  = (2\alpha/3)^{3/2}$. For this case, besides using LDs on the PSOS given in Eqs. \eqref{psos1} and \eqref{psos2} to explore the invariant manifolds that determine the system dynamics, we will also consider the phase space slice:
\begin{equation}
\mathcal{P}_3 = \left\{(x,y,p_x,p_y) \in \mathbb{R}^4 \;\big|\; p_y = 0 \; ,\; p_x > 0 \right\} \;.
\label{psos3}
\end{equation}  
The phase space of the Hamiltonian system undergoes at $\delta = \delta_c$ three  saddle-node bifurcations. This phenomenon occurs between three pairs of equilibrium points: the upper-left well and upper index-1 saddle, the left index-1 saddle and the index-2 saddle, and the lower-left well and bottom index-1 saddle. The critical points of the PES where the bifurcation manifests itself as a change in the topography are located at $(x_c,0)$ and $(x_c,\pm \sqrt{2}/2)$, where $x_c = -\sqrt{\alpha/6}$. Notice that these saddle-node bifurcations take place for two different values of the energy, given by the expressions:
\begin{equation}
V_1(\alpha) = V(-\sqrt{\alpha/6},0) = \dfrac{\alpha^2}{12} \quad.\quad V_2(\alpha) =  V(-\sqrt{\alpha/6},\pm \sqrt{2}/2) = \dfrac{\alpha^2}{12} - \dfrac{1}{4} = V_1(\alpha) - \dfrac{1}{4} \;.
\end{equation} 
In particular, for $\alpha = 1$ we have that $V_1 = 1/12$ and $V_2 = -1/6$, and therefore the saddle node bifurcations between the lower-left well and bottom index-1 saddle and that for the upper-left well and upper index-1 happen at the same energy level $V_2$, whereas the one occurring between the left index-1 and the index-2 takes place for a higher energy $V_1$ of the system. If we linearize Hamilton's equations in Eq. \eqref{ham_eqs} about the equilibrium points where the saddle-node bifurcations occur, it is straightforward to show that the resulting Jacobian matrix has a zero eigenvalue. This property implies that the stability of these points is said to be of parabolic type. Moreover, these parabolic points have invariant manifolds associated to them in phase space that play a relevant role in the dynamical evolution of the trajectories of the system \cite{casasayas1992,fontich1999,baldoma2004}. The manifolds attached to the parabolic points of the system, which are homoclinic trajectories that end at the equilibrium point in a cusp, are illustrated in the $x-p_x$ phase portrait in Fig. \ref{asym_pot} D).

\begin{figure}[!h]
	\begin{center}
		A)\includegraphics[scale=0.25]{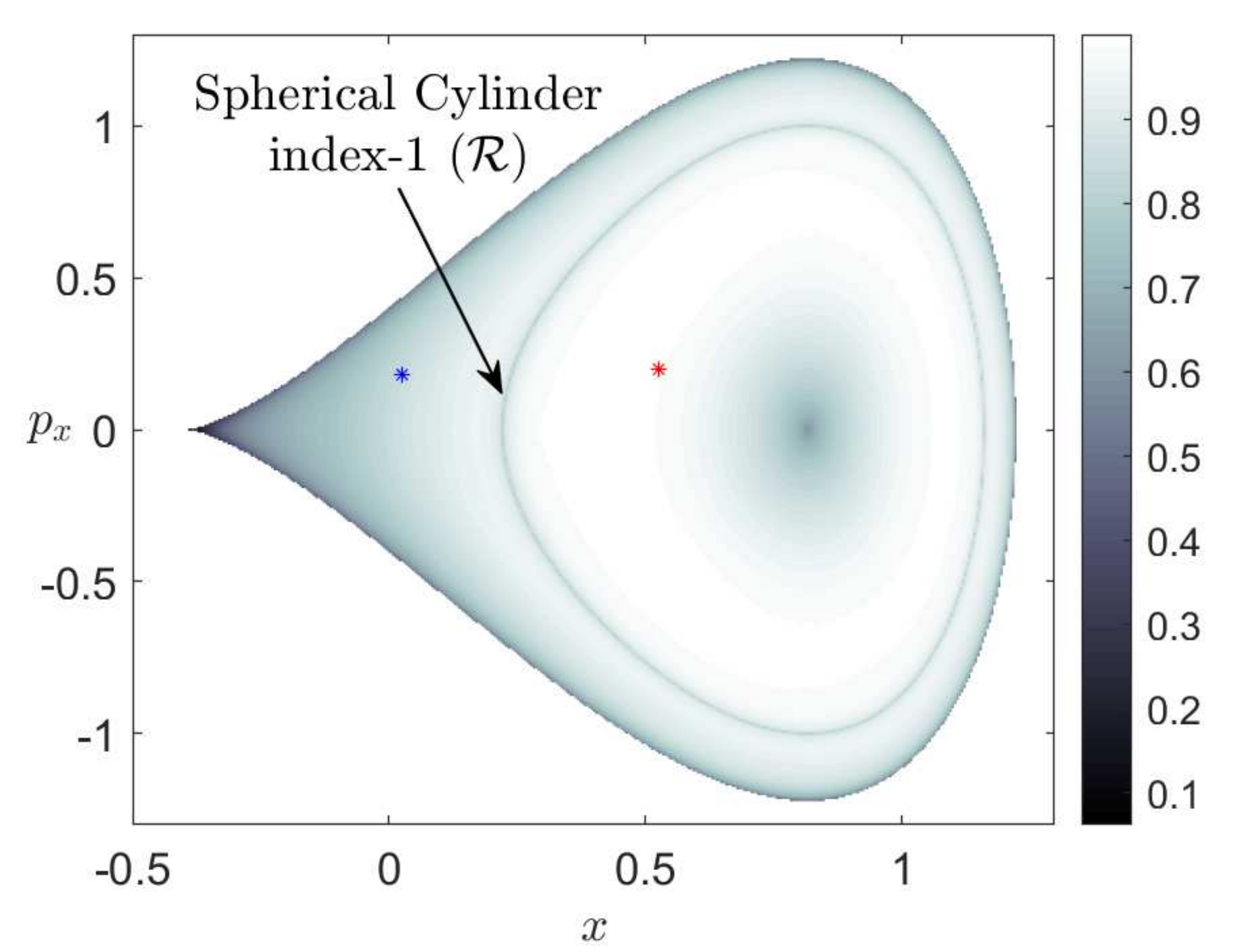}
		B)\includegraphics[scale=0.25]{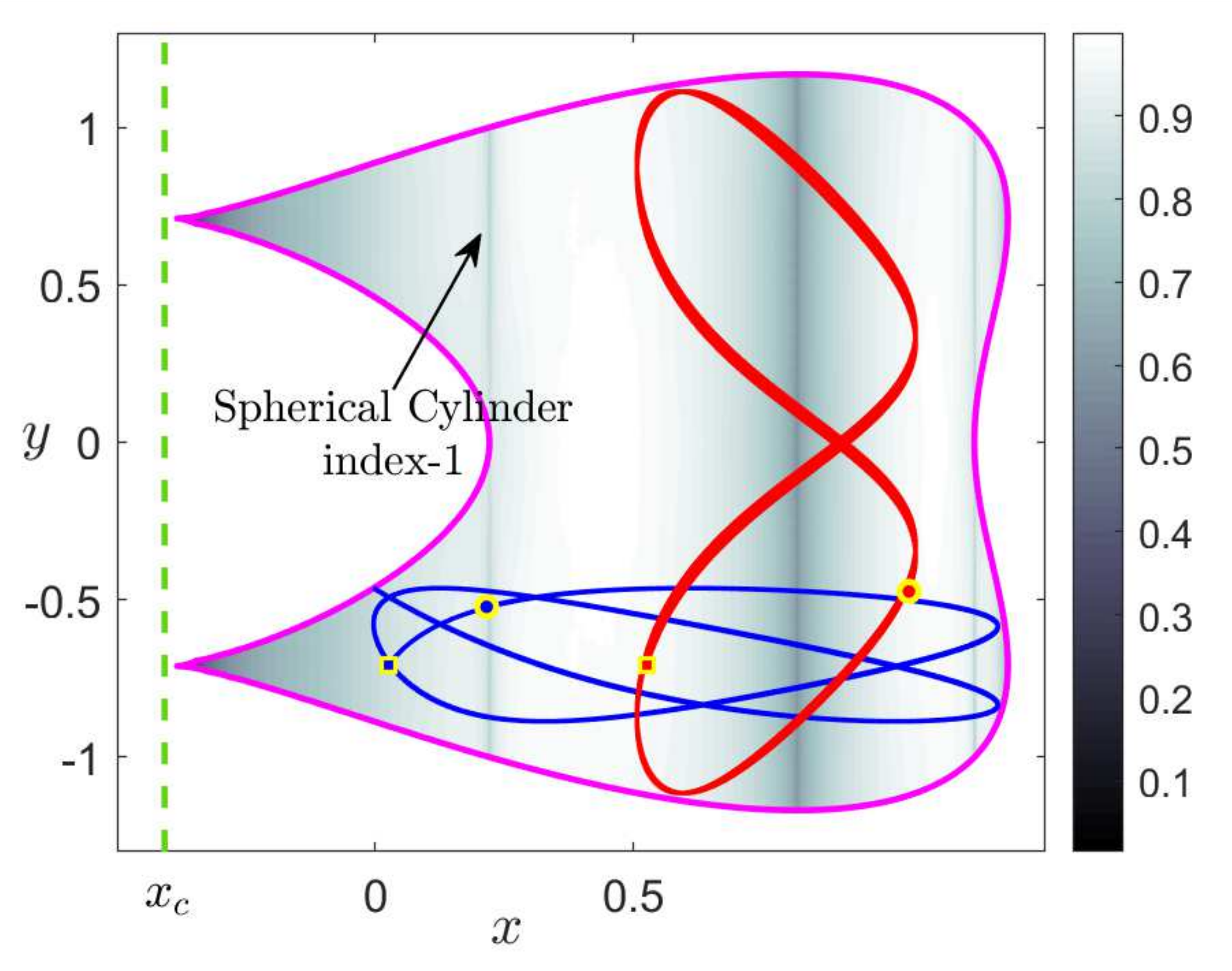}
		C)\includegraphics[scale=0.25]{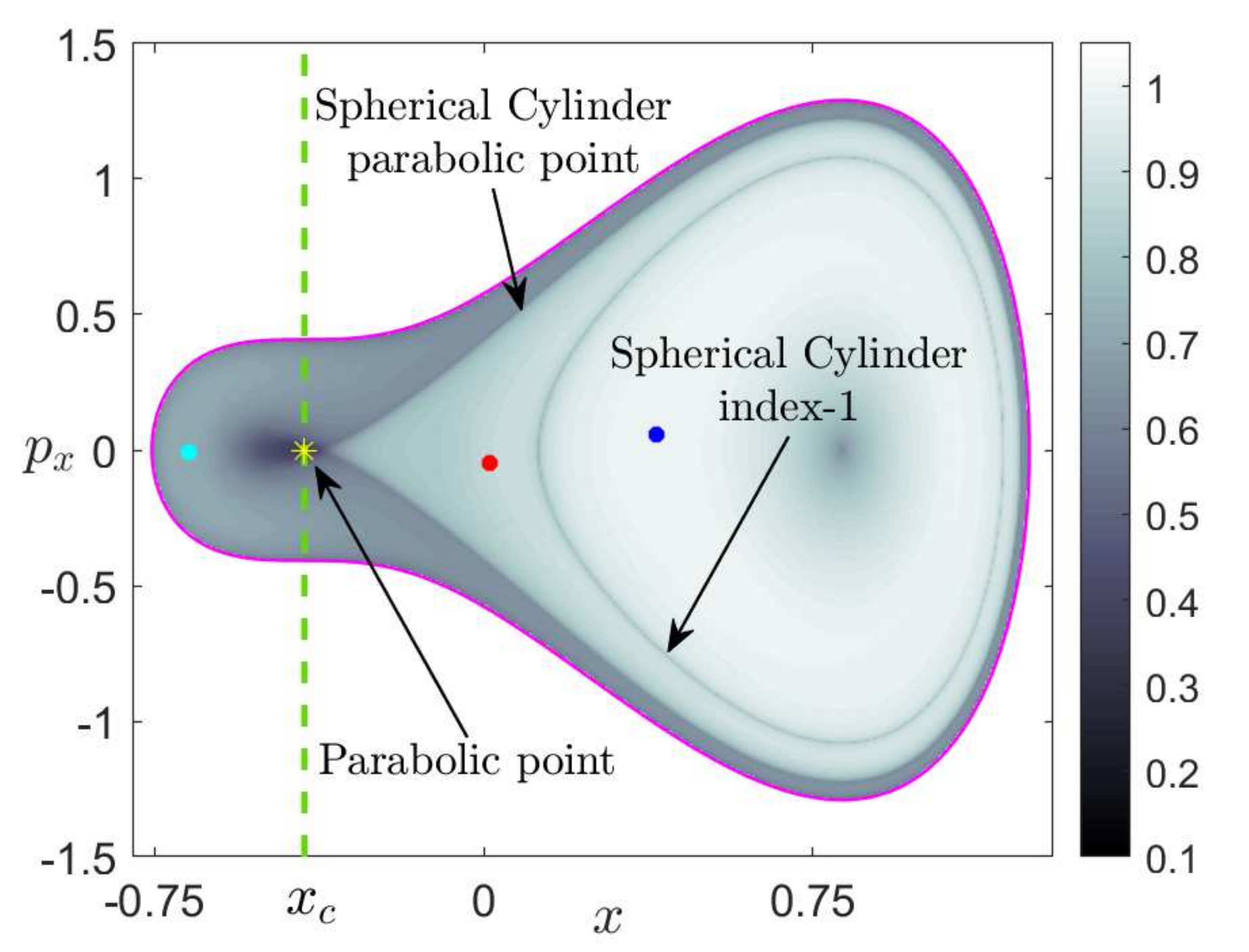}
		D)\includegraphics[scale=0.25]{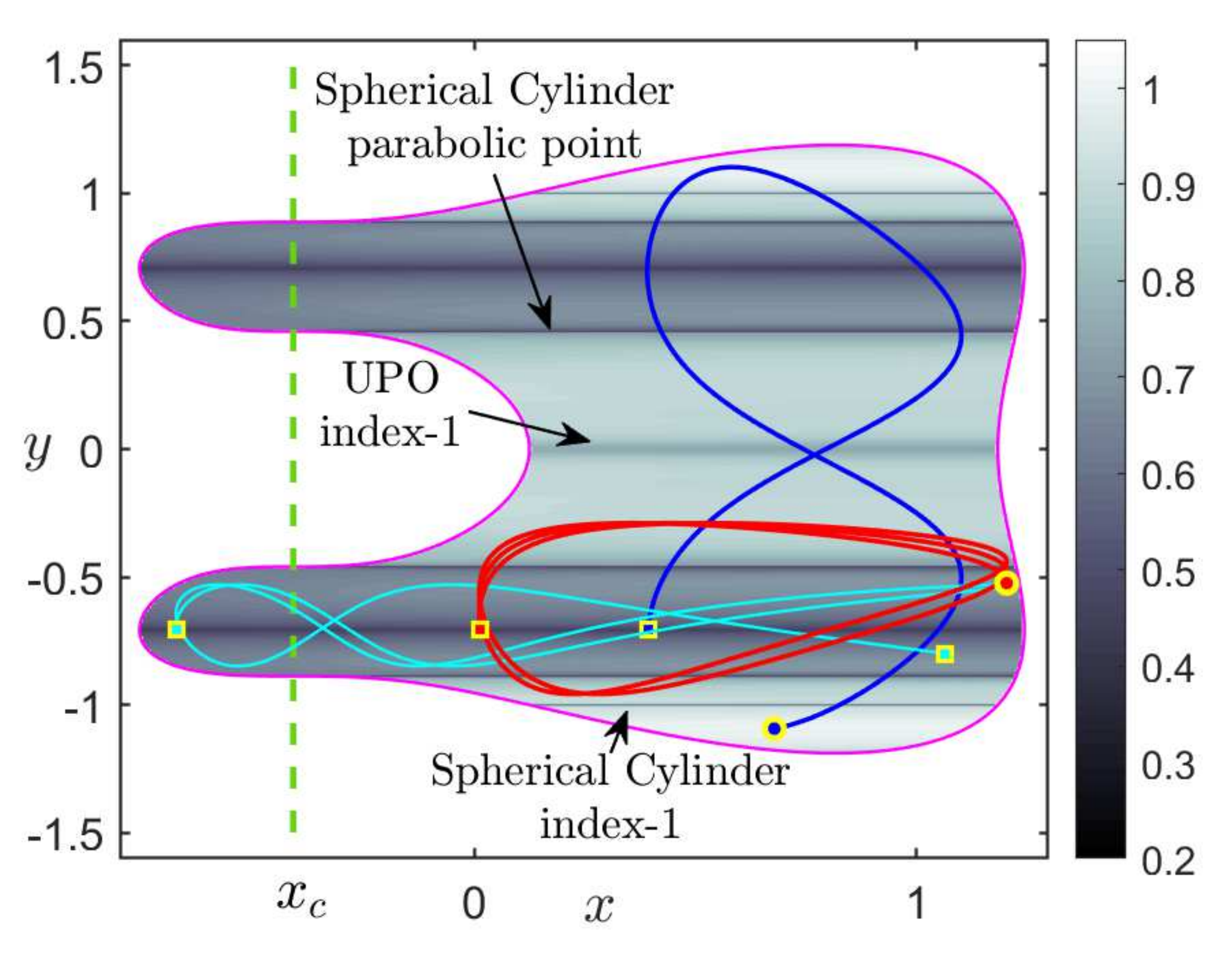}
		E)\includegraphics[scale=0.25]{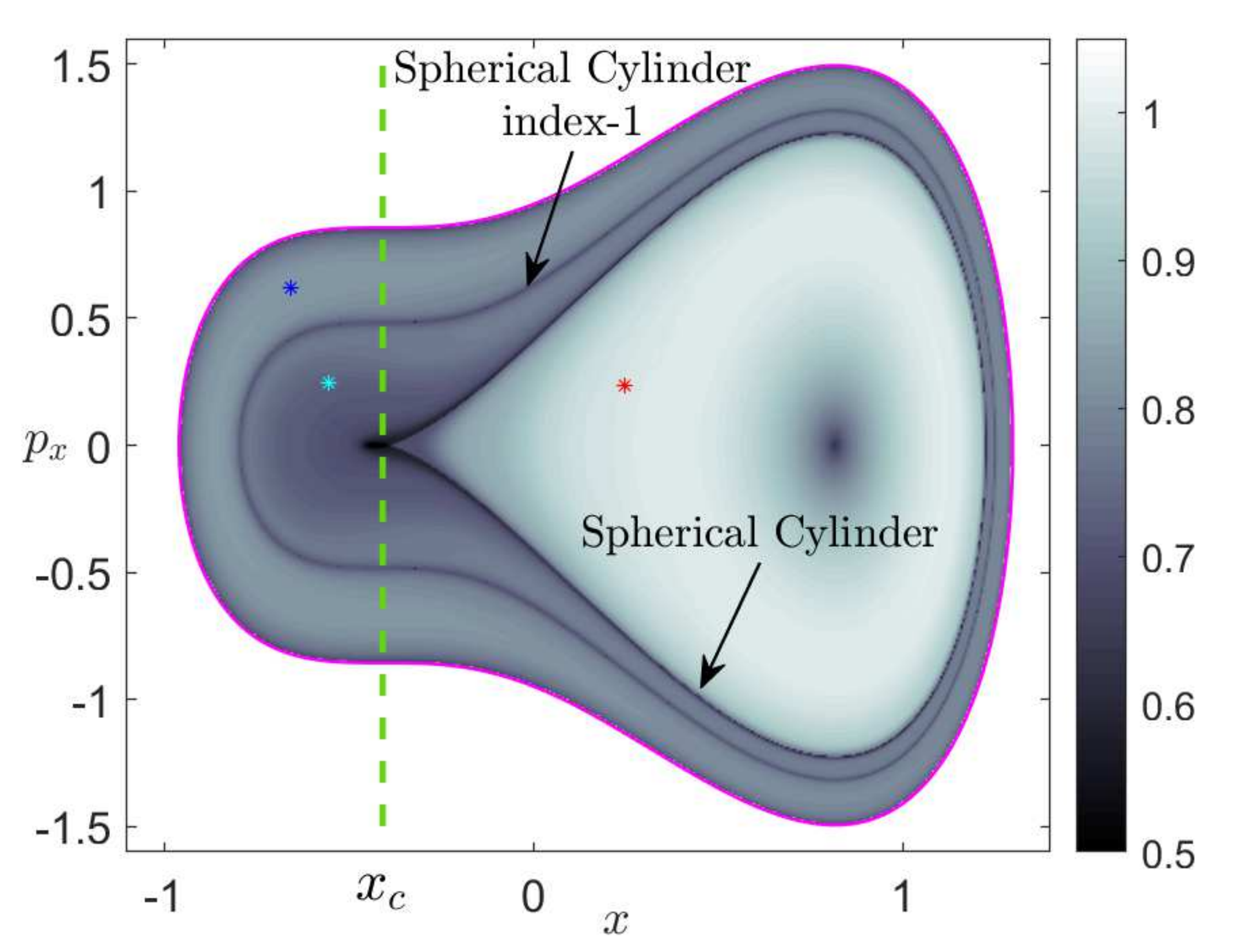}
		F)\includegraphics[scale=0.25]{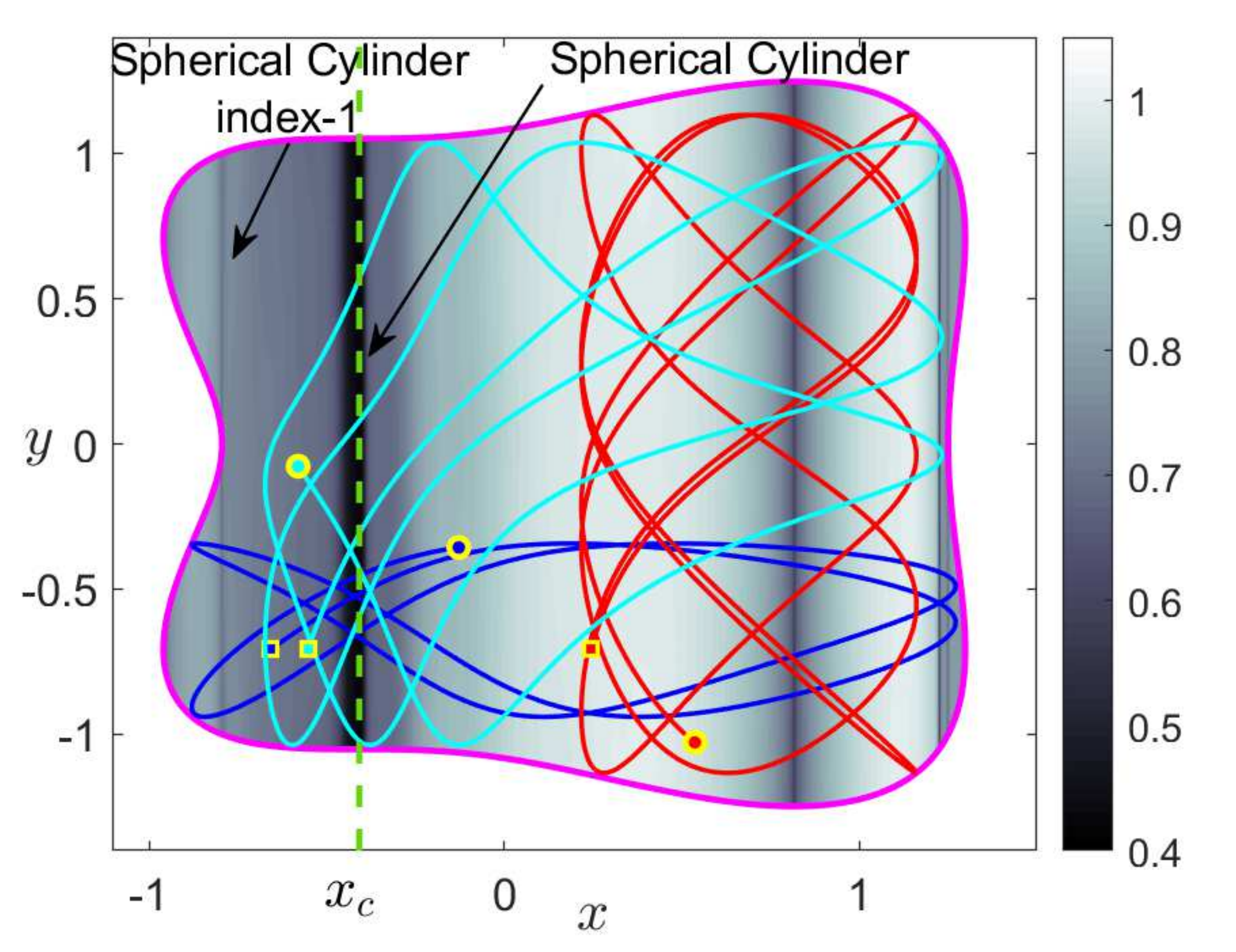}	
	\end{center}
	\caption{LDs calculated using $p = 1/2$ and $\tau = 10$ for the asymmetric and uncoupled Hamiltonian givn in Eq. \eqref{hamiltonian} using $\delta = \delta_c$. The left column corresponds to the phase space slice $y = -1/\sqrt{2}$, and the right column is for $p_x = 0$, except panel D) that is for $p_y = 0$. The energy of the system in each panel is: A) and B) use $H_0 = -1/6$; C) and D) are for $H_0 = -1/12$ ; E) and F) $H_0 = 0.2$.}
	\label{LD_delta_crit}
\end{figure}

The dynamical behavior of the Hamiltonian model subjected to the asymmetric perturbation at the critical value of the parameter $\delta = \delta_c$ is summarized in Fig. \ref{LD_delta_crit} for different energy levels. We select first a total energy $H_0 = -1/6$ that corresponds to that of the parabolic equilibrium points resulting from the saddle-node bifurcations at points with configuration coordinate $y = \pm \sqrt{2}/2$. The invariant manifolds of the parabolic points at this energy do not exist. Therefore, the spherical cylinders of the UPO associated to the index-1 left in the system are the only structures that influence the dynamics. In Fig. \ref{LD_delta_crit} A) and B) we show the computation of LDs using $\tau = 10$ on the PSOS given in Eqs. \eqref{psos1} and \eqref{psos2} respectively. Observe that the method highlights the reactive island that results from the intersection of the tube manifolds of the index-1 saddle with the SOS. In order to validate that initial conditions inside the reactive island will evolve back and firth between both potential wells, we select an initial condition marked with a red asterisk in panel A) and display in B) the projection of its trajectory onto configuration space. Moreover, we check that an initial condition outside the spherical cylinders is n-reactive, and thus remains trapped forever in one of the wells. To illustrate this behavior we pick an initial condition marked in blue. 

The next energy considered is $H_0 = -1/12$, which is still below the level energy of the parabolic point located on the line $y = 0$. However, in this situation the invariant manifolds of the other parabolic points are now present in the system. These manifolds also have the form of spherical cylinders, since they can be constructed as the cartesian product of the homoclinic orbit in the phase portrait $x-p_x$ and a closed trajectory of the $y-p_y$ plane. This representation is possible because the DoF of the system are decoupled. In Fig. \ref{LD_delta_crit} C) and D) we show the output of LDs overlaid with the evolution of three initial conditions selected in the different phase space regions revealed. In order to highlight the existence of he invariant manifolds corresponding to the parabolic points, we have decided to calculate LDs in panel D) on the phase space slice defined in Eq. \eqref{psos3}. The blue initial condition is chosen to lie inside the spherical cylinders corresponding to the UPO of the index-1 saddle and thus, it moves between both wells along its evolution. On the other hand, the red initial condition is outside the spherical cylinders associated to the bottom parabolic point and also to the index-1. Hence, it remains trapped in the lower potential well. Finally, the cyan trajectory is located inside the tube manifold of the lower parabolic point and thus it evolves in the lower region of the PES and, since it is outside the spherical cylinder of the index-1 saddle, it cannot cross the phase space bottleneck that connects both potential wells. This is reflected in the fact that the $y$ coordinate does not change sign along the evolution of the trajectory.

The last energy level that we take a look at is $H_0 = 0.2$, which is above the energy of the parabolic point located on the line $y = 0$. In this case, the dynamical behavior of the system is similar to the one we have previously discussed. However there is an important difference: the spherical cylinders associated to the parabolic point are contained inside the tube manifolds of the UPO of the index-1 saddle. This is clearly visible from the representations of the LD scalar field in Fig. \ref{LD_delta_crit} E) and F). Observe that the selection of the initial conditions to probe the dynamical behavior of the system, one in each of the three distinct regions that partition the energy hypersurface in phase space, is trivial since LDs give us all the information needed in relation to the boundaries between those regions. The blue initial condition, that is outside the invariant manifolds of the index-1 saddle, is trapped in the bottom well. On the other hand, the cyan and red initial conditions are both inside the spherical cylinders of the index-1 saddle, and therefore they give rise to reactive trajectories that evolve back and forth between both potential wells. Note that the cyan/red initial condition is located outside/inside the tube manifolds of the parabolic point. The dynamical consequence is that the red trajectory cannot cross the vertical line $x = x_c$ in configuration space, while the cyan trajectory in fact does cross it along its evolution.

We end this section by exploring the system dynamics post-bifurcation, and to illustrate this case we choose for the asymmetry parameter the value $\delta = 0.8 > \delta_c$. In this situation, three critical points of the PES remain: two wells separated by an index-1 saddle. As we will explain next, our numerical simulations for this case have evidenced that VRI points might be of dynamical significance for the analysis of trajectory evolution in phase space. Moreover, our analysis points to the possible existence of invariant manifolds in phase space associated to them. Despite we limit ourselves in this paper to briefly present the results obtained in this respect, this intriguing conjecture will be one of our goals for research in the near future. We recall that valley-ridge inflection points, depicted in Fig, \ref{asym_pot_dif_delta} E) and F), are given at locations where the conditions in Eq. \eqref{vri_conds} are satisfied. It is straightforward to verify that the configuration coordinates of the four VRI points are $(\pm \sqrt{\alpha/6},\pm \sqrt{6}/6)$. It is interesting to remark that they do not depend on the asymmetry. Of special interest are the two VRI points located on the left part of the PES, whose energy is: 

\begin{equation}
V(-\sqrt{\alpha/6},\pm \sqrt{6}/6) = \frac{6 \, \delta\sqrt{6\alpha} -5 (\alpha ^{2} +1)}{36} \;,
\end{equation}
and in particular, for the model parameters $\alpha = 1$ and $\delta = 0.8$ this yields the energy level $V_{ridge} \approx 0.04882$. 

\begin{figure}[!h]
	\begin{center}
		A)\includegraphics[scale=0.25]{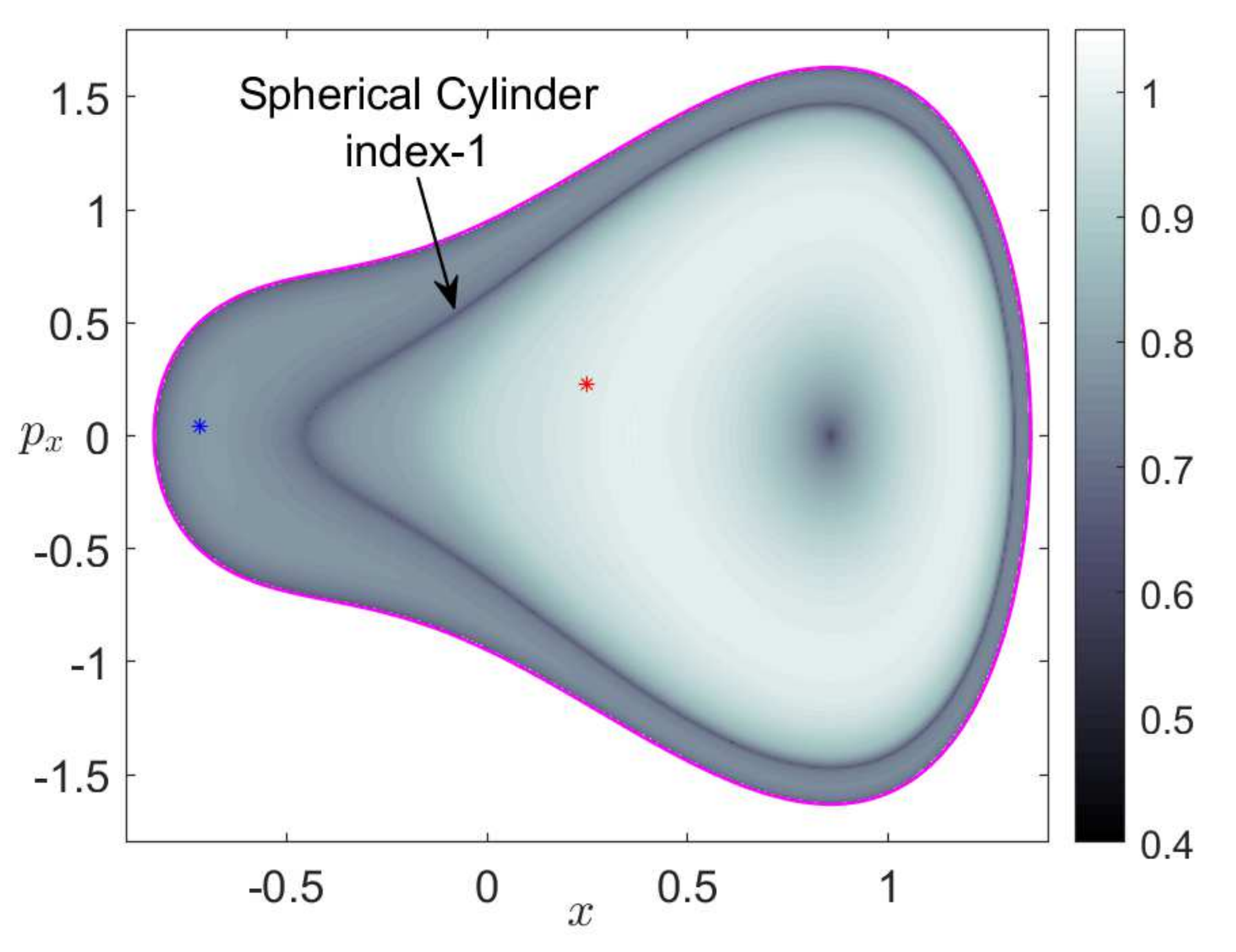}
		B)\includegraphics[scale=0.25]{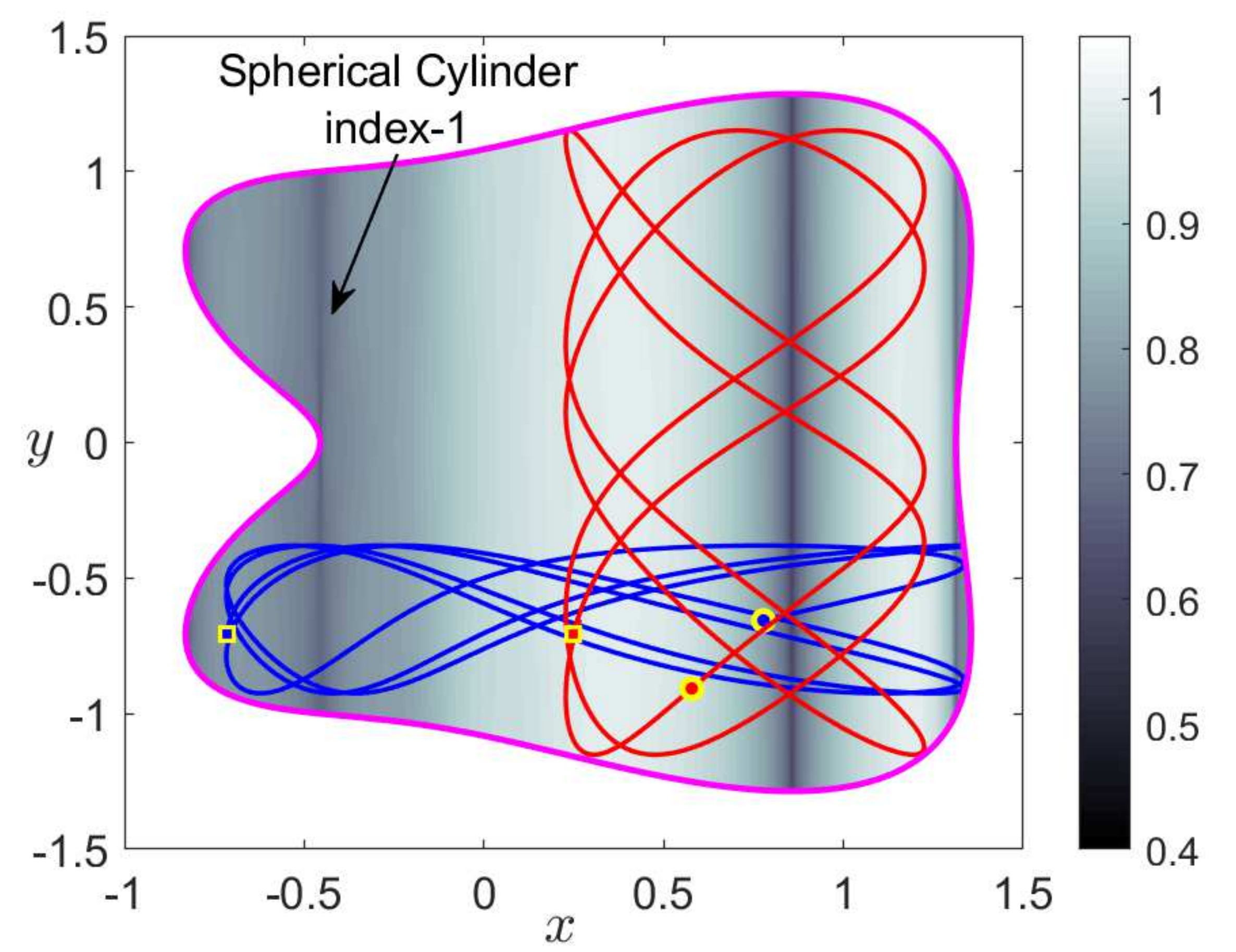}
	\end{center}
	\caption{LDs calculated using $p = 1/2$ and $\tau = 10$ for the asymmetric and uncoupled Hamiltonian with $\delta = 0.8 > \delta_c$ on the phase space slices $y = -1/\sqrt{2}$ and $p_x = 0$. The energy of the system is $H_0 = 0.2$. Red and blue asterisks in A) represent initial conditions, and the projection onto configurations space of their trajectories is shown in B).}
	\label{LD_delta_big}
\end{figure}

We compile in Fig. \ref{LD_delta_big} the phase space structures governing isomerization dynamics on the PES. We expect the classical picture of isomerization reactions that takes place between two potential wells separated by an index-1 saddle, where the tube manifolds of the UPO partition the phase space into reactive and non-reactive, i.e. trapped in a well, regions \cite{deleon1981,deleon1989,deleon1991a,deleon1991b,almeida1990}. To demonstrate that this is the case, we study the system for two energy levels, both of them selected above the energies corresponding to the VRI points and also to the index-1 saddle, so that the phase space bottleneck connecting both well regions is open. First, we consider $H_0 = 0.2$ and calculate LDs using $\tau = 10$ on the phase space slices $y = -1/\sqrt{2}$ and $p_x = 0$. The results are displayed in Fig. \ref{LD_delta_big} A) and B) where we can see that the method reveals the reactive island corresponding to the spherical cylinders of the UPO associated to the index-1 saddle. If we take two initial conditions inside/outside the tube manifolds, marked with a red/blue asterisk in Fig. \ref{LD_delta_big} A), the red trajectory gives rise to sequential isomerization while the blue one is trapped in the lower well. 

\begin{figure}[!h]
	\begin{center}
		A)\includegraphics[scale=0.175]{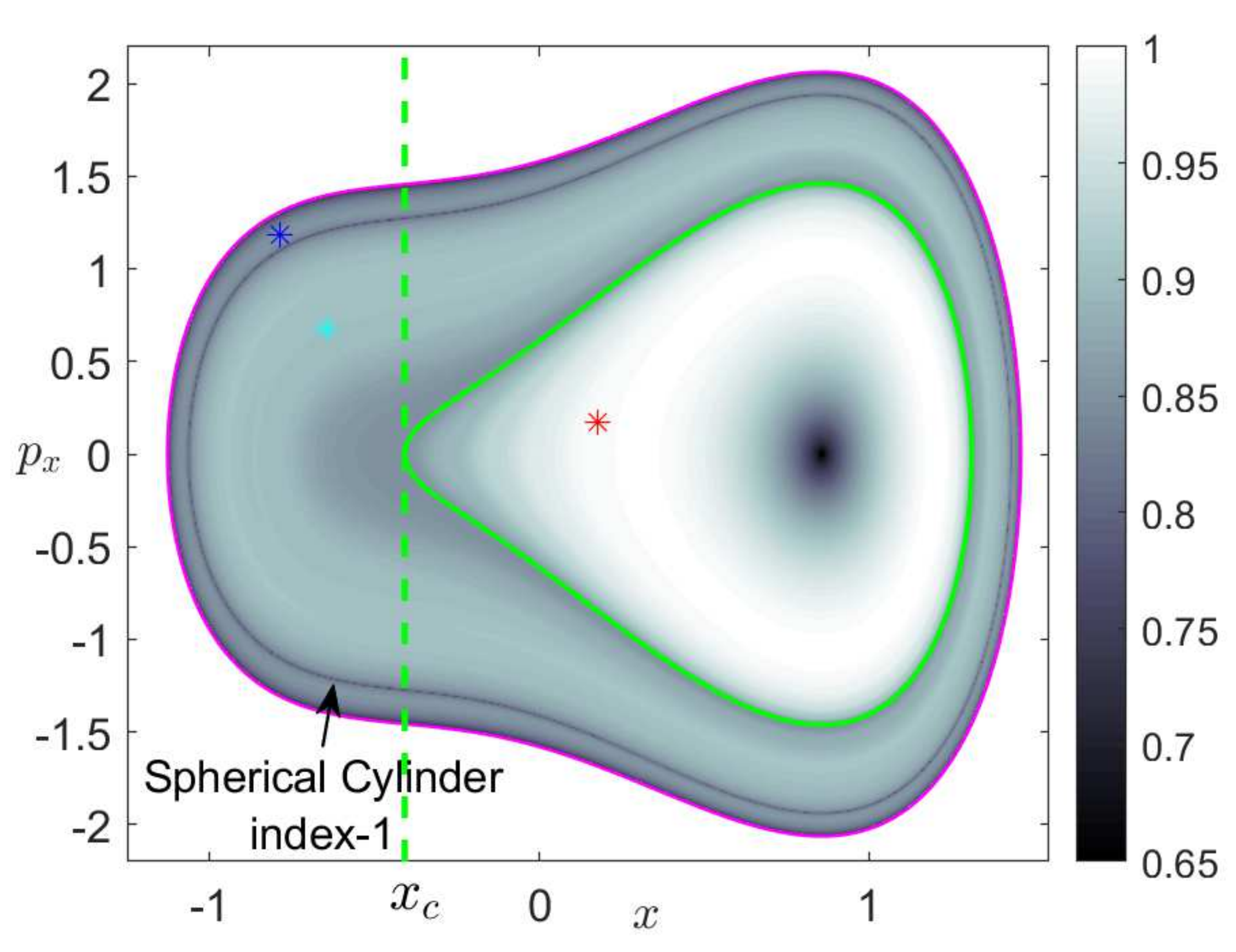}
		B)\includegraphics[scale=0.21]{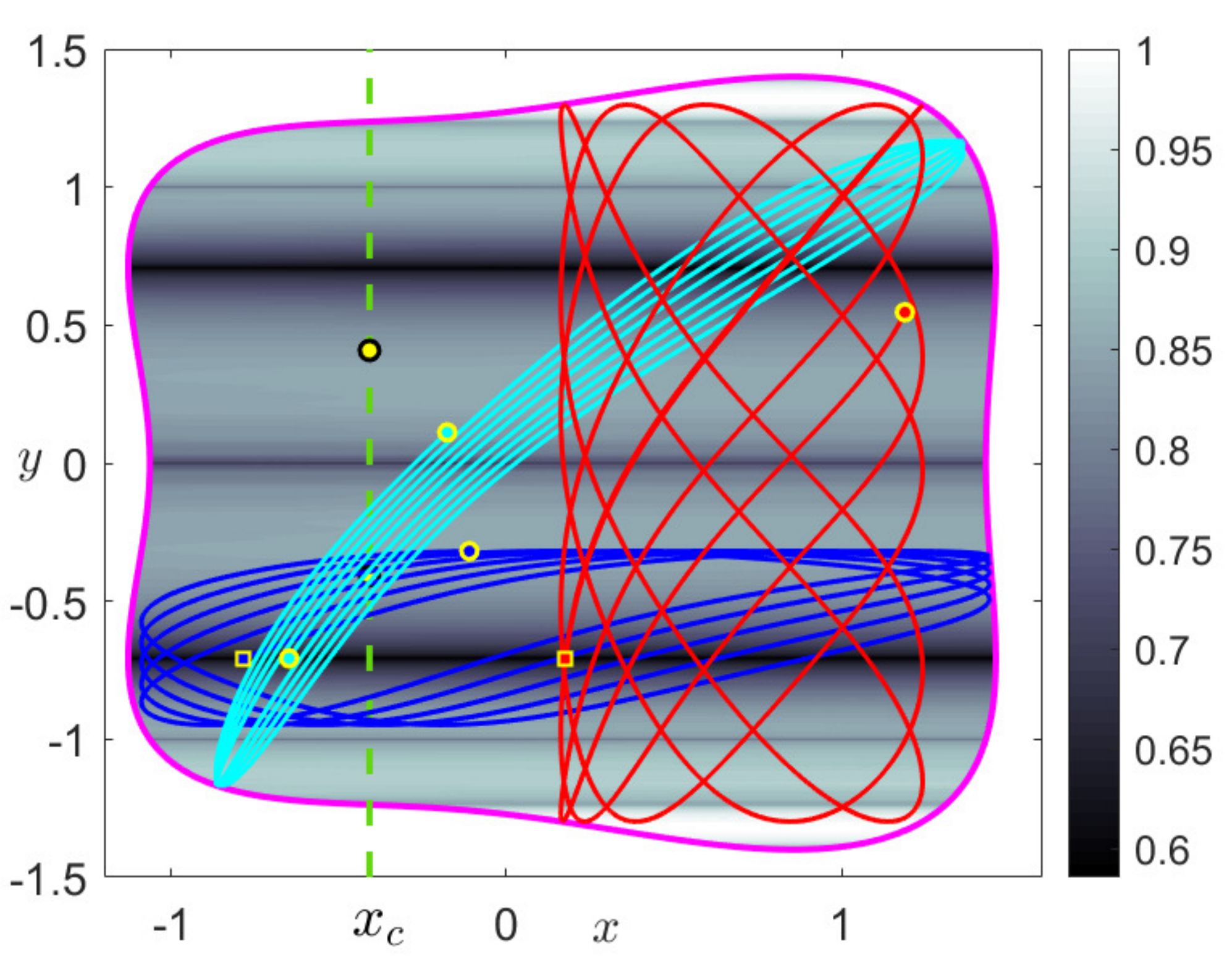}
		C)\includegraphics[scale=0.25]{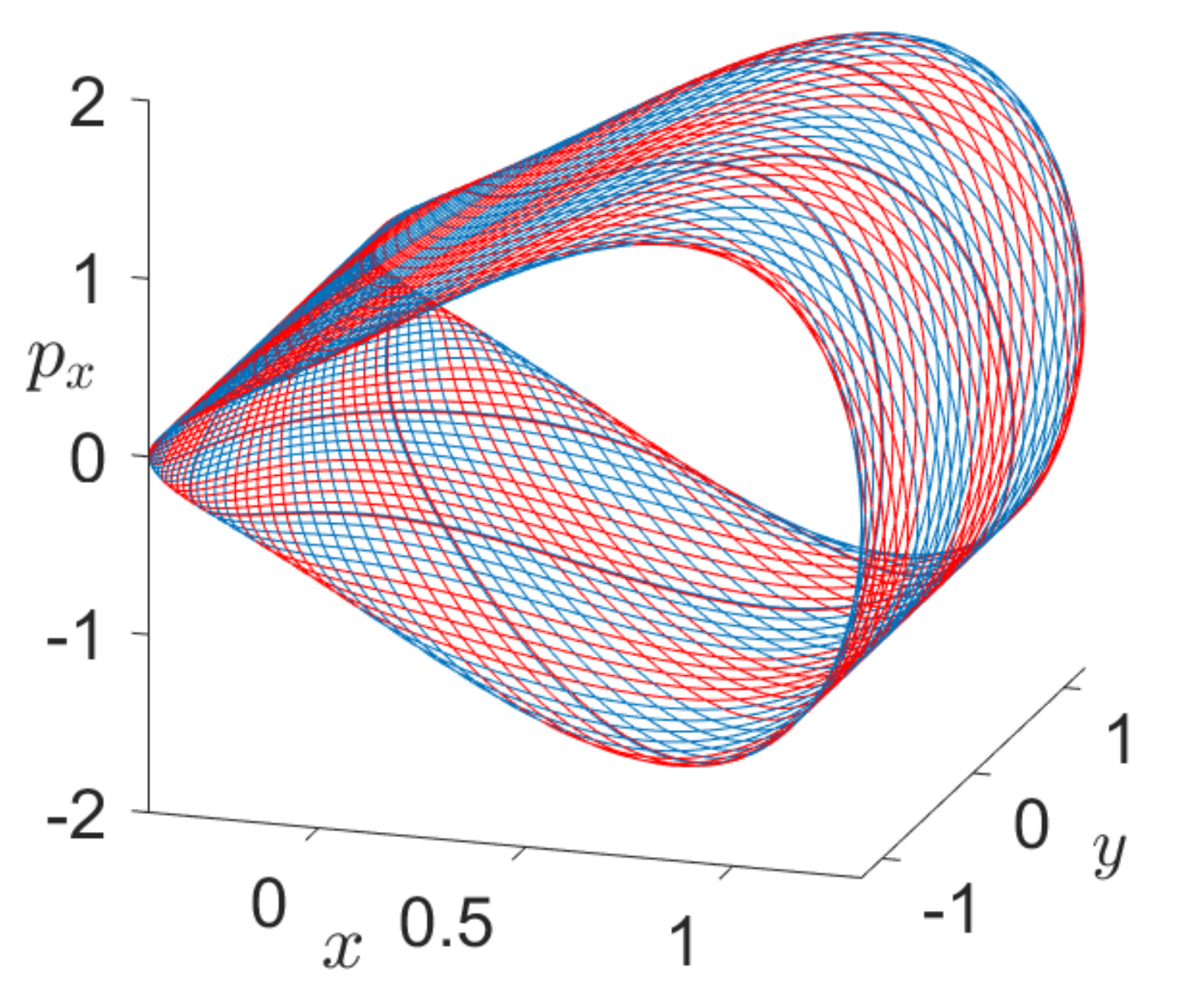}
	\end{center}
	\caption{LDs calculated using $\tau = 25$ for the asymmetric and uncoupled Hamiltonian with $\delta = 0.8 > \delta_c$ and energy $H_0 = 1$. A) corresponds to the phase space slice $y = -1/\sqrt{2}$; B) is for $p_y = 0$. Red, blue and cyan asterisks represent initial conditions. The solid green curve in A) marks the location of the phase space structure related the VRI points, highlighted as yellow dots in B). The invariant manifold related to the VRI points is illustrated in panel C).}
	\label{LD_delta_big2}
\end{figure}

\noindent We increase now the energy of the system further to $H_0 = 1$. This change would imply that the bottleneck grows, and therefore more trajectories become reactive. This effect is visible in Fig. \ref{LD_delta_big2} A), where LDs have been calculated on the PSOS $y = -1/\sqrt{2}$ using $\tau = 25$. Observe that the phase space region outside the spherical cylinder of the index-1 saddle shrinks, so that fewer trajectories are trapped in one of the wells. Interestingly, besides the reactive island associated to the tube manifolds of the index-1 saddle, we can observe in Fig. \ref{LD_delta_big2} that LDs also detect another type of phase space structure that we have marked with a green curve. Recall that LDs reveal invariant manifolds in phase space at points where the scalar field that the method produces is non-differentiable, see Appendix \ref{sec:appA} for more details. The existence of this structure is also apparent from the computation of LDs on the surface of section displayed in Fig. \ref{LD_delta_big2} B). in order to probe the dynamical behavior of the system we choose three initial conditions marked with blue, cyan and red asterisks in Fig. \ref{LD_delta_big2} A). While the blue trajectory is non-reactive and hence is trapped in the lower potential well, the other two conditions are reactive. On the other hand, the cyan trajectory crosses the line $x = x_c$ along its evolution, whereas the red trajectory moves back and forth between both wells but is constrained to the right of the line $x = x_c$. This clearly suggests that the structure that we are observing for the VRI points is acting as a transport barrier in phase space. To finish, and as a validation that the phase space structure that we have revealed for the VRI points is in fact invariant, we have selected an initial condition on it and evolved it forward (blue) and backward (red) in time. The three-dimensional visualization of the resulting trajectories is depicted in Fig. \ref{LD_delta_big2} C), supporting our claim. 

\section{Analysis of the Coupled System}
\label{sec:sec3}

We finish our analysis of isomerization dynamics in the four-well Hamiltonian model by discussing the effect that coupling between the DoF of the system brings into the geometrical template of phase space structures. We will focus on the particular case where the PES is symmetric, which is obtained by setting $\delta = 0$. For the uncoupled system that we studied at the beginning of this work, the system is integrable as energy is conserved in each DoF separately. Hence, besides the invariant stable and unstable manifolds attached to the UPOs of the index-1 saddles and also those for the index-2 saddle, the phase space is foliated by tori. Moreover, motion is regular and no chaotic behavior is allowed. When the energy of the system is above that of the index-1 saddles, but below the energy level of the index-2, the system can only display sequential isomerization between neighboring wells. Reactive trajectories that start for instance on the lower-left well of the PES can transit to the upper-left or lower-right wells through the corresponding phase space bottlenecks, following the isomerization highways defined by the tube manifolds of the UPOs of the index-1 saddles. However, the spherical cylinders of different UPOs do not intersect in phase space, so that initial conditions can only evolve back and forth between two wells, and they are forbid to visit the other wells. The only way that trajectories can visit all the wells is when the energy of the system is above that of the index-2 saddle, since in this situation both sequential and concerted isomerization routes are possible. For trajectories that undergo concerted isomerization, i.e. those that cross the PES hilltop, will visit all the wells along their evolution.

This picture changes dramatically when coupling among the DoF is considered. From the KAM theorem we know that tori in phase space will be destroyed as the perturbation strength is increased. This generates the classical image of islands of regularity corresponding to the tori that survive the perturbation, interspersed in a sea of chaos, which is nicely captured when applying Poincar\'e maps to reveal the spreading of chaos in the system. One of the fundamental reasons that gives rise to this chaotic behavior in the coupled system is that the invariant stable and unstable manifolds of the UPOs associated to the index-1 saddles of the PES begin to intersect, forming intricate homoclinic and heteroclinic tangles that help to spread chaos throughout the energy hypersurface where motion takes place. Recall that a homoclinic connection results from the intersections of the invariant manifolds corresponding to a given UPO, while heteroclinic connections are established between the spherical cylinders of \textit{different} UPOs. The heteroclinic intersections established in the coupled system create connections between the isomerization channels and consequently, some of the initial conditions that undergo sequential isomerization can visit \textit{all} the potential wells along their evolution, a behavior that is impossible to achieve in the uncoupled system. Our goal in this section is to demonstrate how LDs provides us with an extremely powerful tool to reveal the phase space regions that characterize the distinct dynamical evolution of the coupled and symmetric system.

\begin{figure}[!h]
	\begin{center}
		A)\includegraphics[scale=0.31]{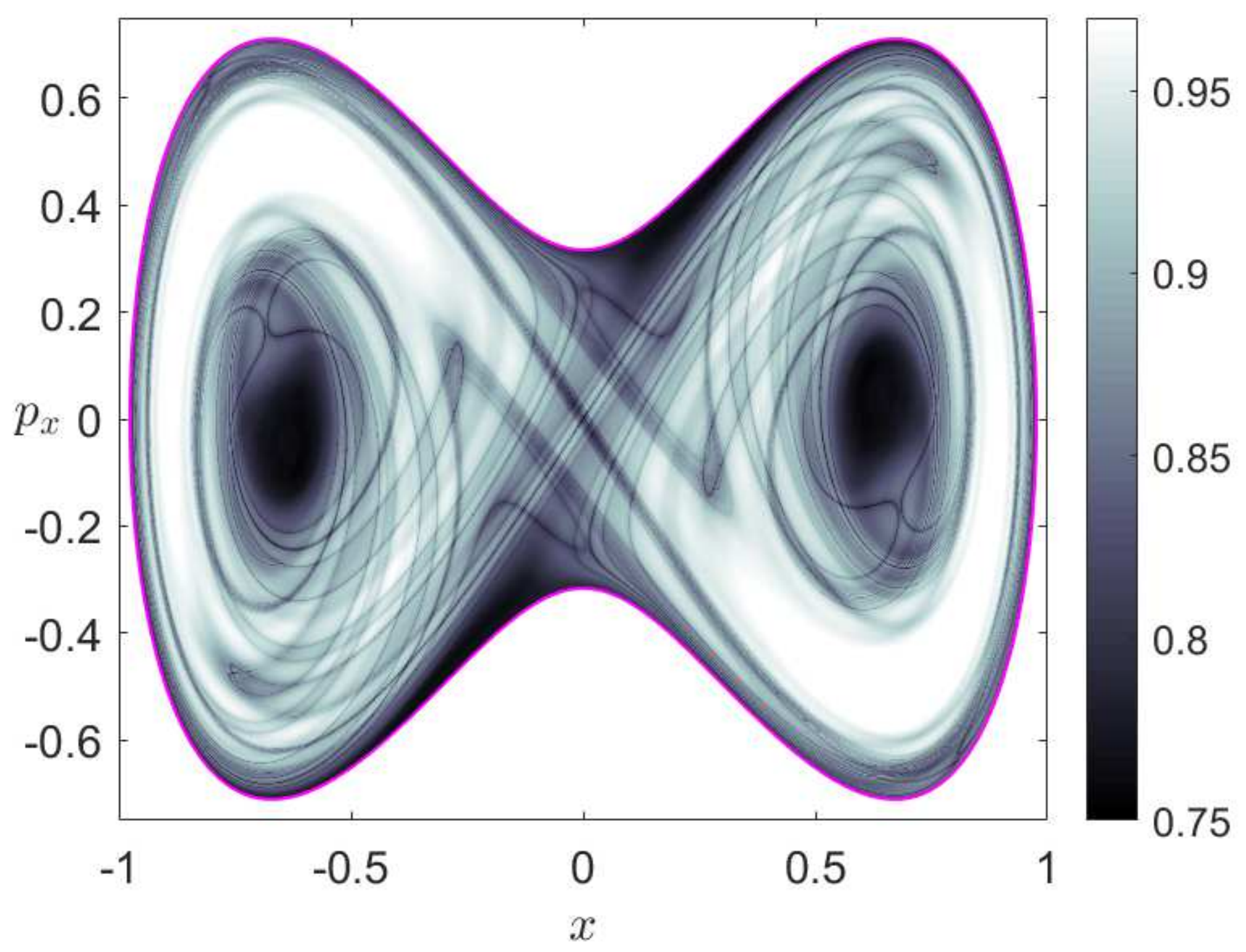}
		B)\includegraphics[scale=0.31]{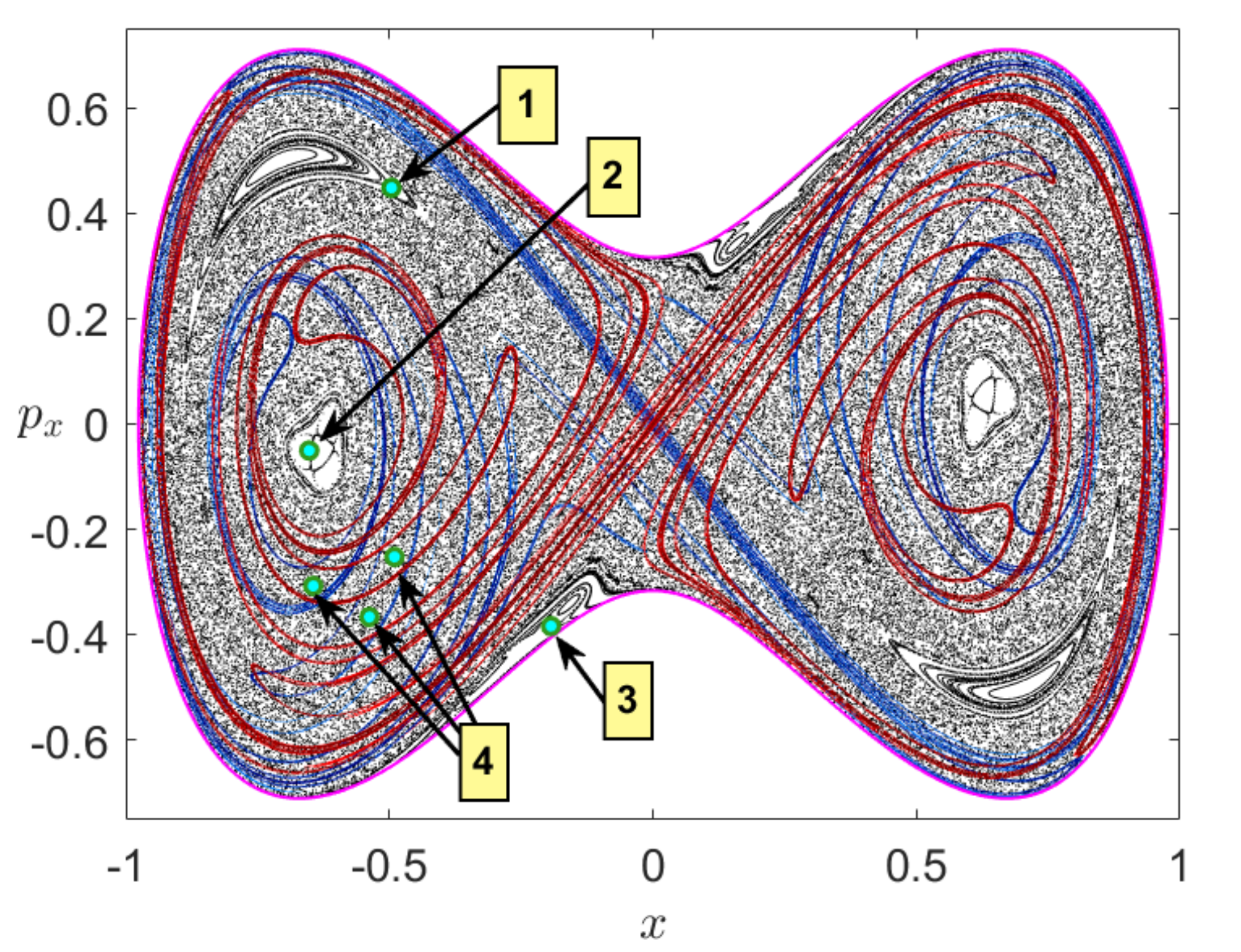}
		C)\includegraphics[scale=0.47]{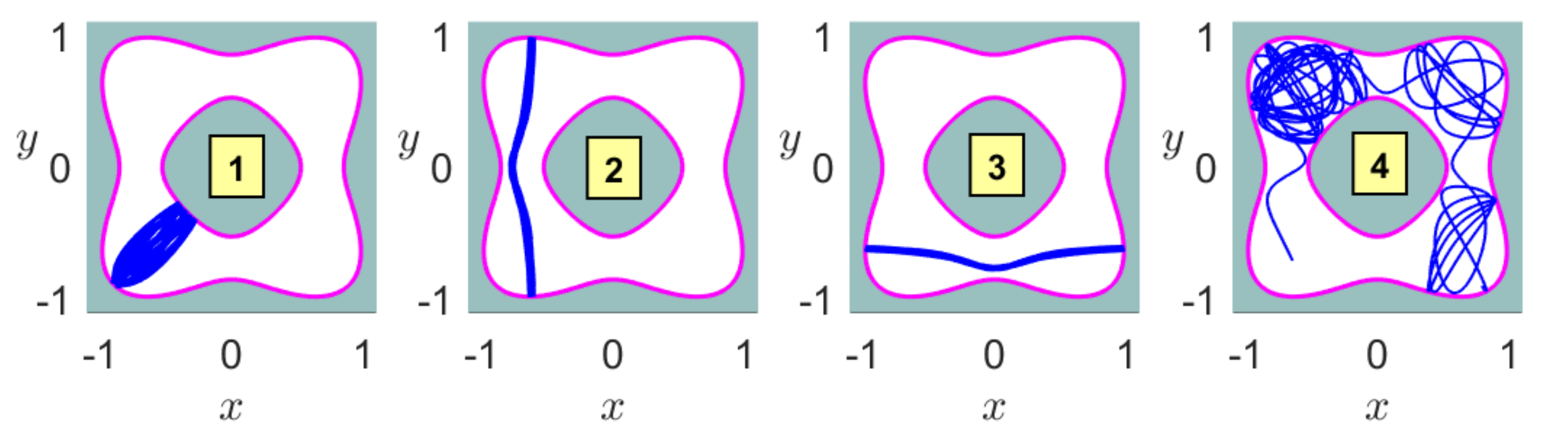}
	\end{center}
	\caption{Phase space structures on the surface of section given in Eq. \eqref{psos1}  at an energy $H_0 = -0.2$ and dynamical evolution of initial conditions for the symmetric and coupled Hamiltonian with $\alpha = 1$, $\delta = 0$ and $\beta = 0.2$. A) LDs calculated using $p = 1/2$ and $\tau = 12$. B) Poincar\'e map superimposed with the stable (blue) and unstable (red) manifolds extracted from the gradient of the LD function. We have also marked in the picture four different types of initial conditions. C) Dynamical evolution of the initial conditions selected in panel B). In all diagrams we have indicated the energy boundary with a magenta curve.} 
	\label{coupDyn_belowIdx2}
\end{figure}

To begin the analysis, we take the coupling strength as $\beta = 0.2$ and consider first the situation where the symmetric ($\delta = 0$) system is at an energy level $H_0 = -0.2$ above that of the index-1 saddles but below the index-2. All the phase space bottlenecks that interconnect the four potential wells are open in this case. Furthermore, the the Hill's region excludes a neighborhood of the index-2 at the origin, since the hilltop is energetically forbidden. In order to unveil the geometrical skeleton of invariant manifolds in phase space that determines isomerization we apply LDs using $\tau = 12$ on the slice given in Eq. \eqref{psos1} that passes through the index-1 saddle at the bottom of the PES. We complement this analysis by computing also the Poincar\'e map to help us distinguish between regular and chaotic regions. We summarize in Fig. \ref{coupDyn_belowIdx2} the dynamical behavior exhibited by the system, depending on the phase space region where the initial conditions are chosen. The method of Lagrangian descriptors allows us to successfully address the challenging task of identifying these regions, since we can recover the location of the stable and unstable manifolds from the gradient of the scalar field produced. We can do so with arbitrarily high-resolution without the a priori computation of the UPOs in the system, and subsequently globalizing the manifolds from the knowledge of UPOs, which is a non-trivial procedure to implement with a high computational cost. A detailed explanation on LDs and how to apply this technique to reveal the phase space of Hamiltonian systems is included in Appendix \ref{sec:appA}.
 
In Fig. \ref{coupDyn_belowIdx2} A) we depict the output of LDs and Fig. \ref{coupDyn_belowIdx2} B) displays the Poincar\'e map superimposed with the stable (blue) and unstable (red) manifolds extracted from the gradient of the scalar field generated by LDs. Notice that there is a correlation between the dynamics that both methods reveal. However, there is a fundamental difference: chaotic regions of the phase space, visible as a sea of random points in the Poincar\'e map, hide the structures of the stable and unstable manifolds, while LDs help us reveal the manifolds responsible for the chaotic behavior. Once the geometrical template of manifolds and their lobe intersections is recovered, the problem of analyzing the system dynamics reduces to running trajectores from initial conditions selected on the different regions identified. Observe that even though the coupling strength is small, i.e. $\beta = 0.2$, the system dynamics has become chaotic in a large volume of the energy hypersurface, and only small regions of regular behavior survive. This is clear from the Poincar\'e map in Fig. \ref{coupDyn_belowIdx2} B). It is important to highlight that those regions displaying regular behavior are those that are unaffected by the homoclinic and heteroclinic connections between the spherical cylinders of the different UPOs associated to the index-1 saddles. We probe the system dynamics by choosing four initial conditions and we label the different cases in Fig. \ref{coupDyn_belowIdx2} B). The corresponding trajectories, projected onto configuration space, are shown in Fig. \ref{coupDyn_belowIdx2} C). Initial conditions $1,2,3$ are selected in regular regions of the phase space that are outside the lobes formed by the heteroclinic and homoclininc connections, while the initial condition $4$ is inside a lobe corresponding to a heteroclinic intersection. The initial condition $1$ is located outside the spherical cylinders of both UPOs associated to the left and bottom index-1 saddles. Therefore, it will give rise to trapped motion in the lower-left well of the PES. Initial condition $2$ is inside the spherical cylinders of the left index-1 and thus it results in a reactive trajectory that moves back and forth between the lower-left and upper-left wells. Similarly, initial condition $3$ is inside the tube manifolds of the bottom index-1 and hence the system undergoes sequential isomerization that involves the lower-left an lower-right wells. Finally, initial condition $4$ also generates a reactive trajectory, but the difference with those obtained from $2$ and $3$ is that it visits all the wells of the PES along its evolution. This is possible because it is inside a heteroclinic intersection, so that it can transit through the bottlenecks corresponding to the different index-1 saddles.

\begin{figure}[htbp]
	\begin{center}
		A)\includegraphics[scale=0.25]{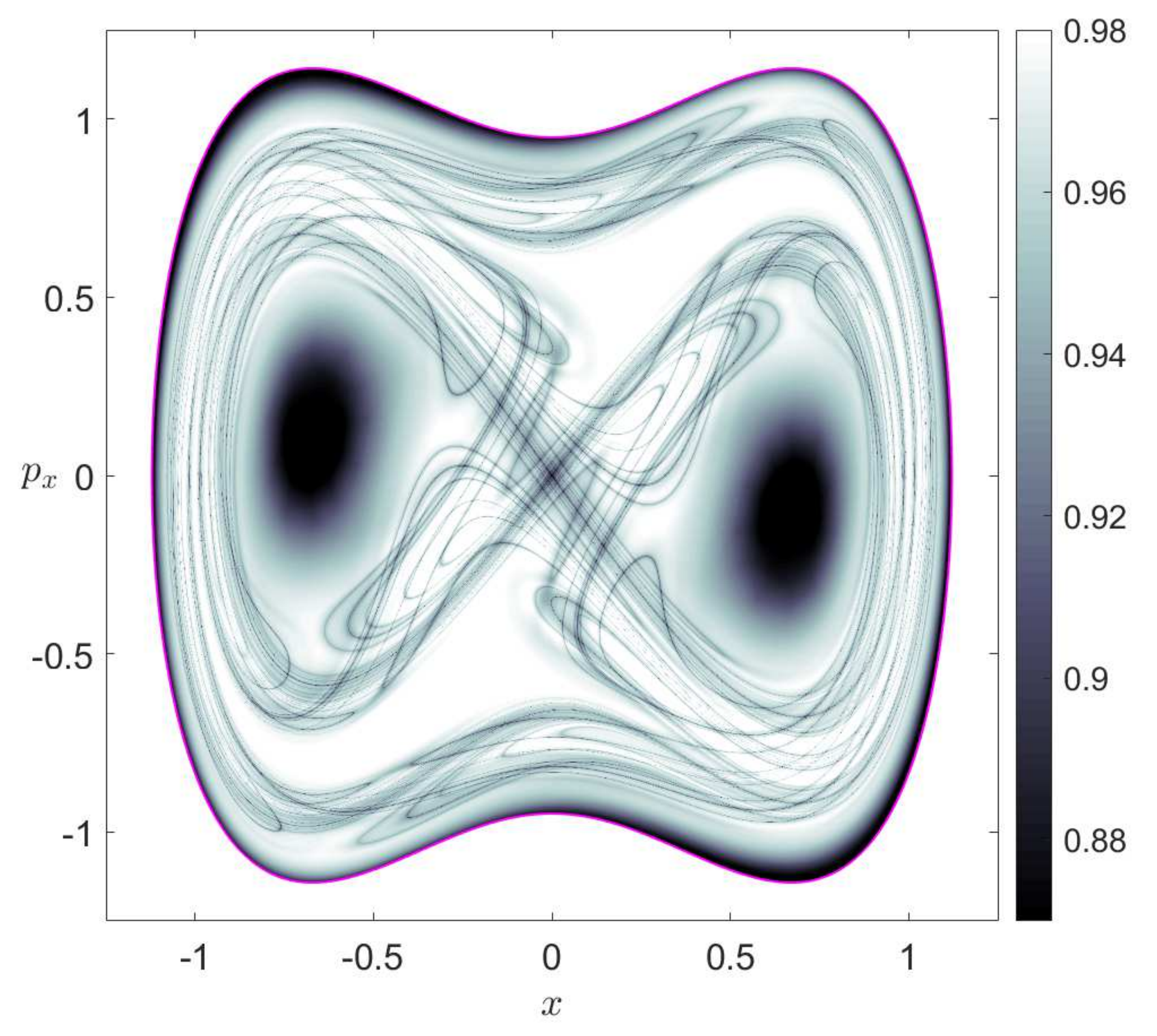}
		B)\includegraphics[scale=0.26]{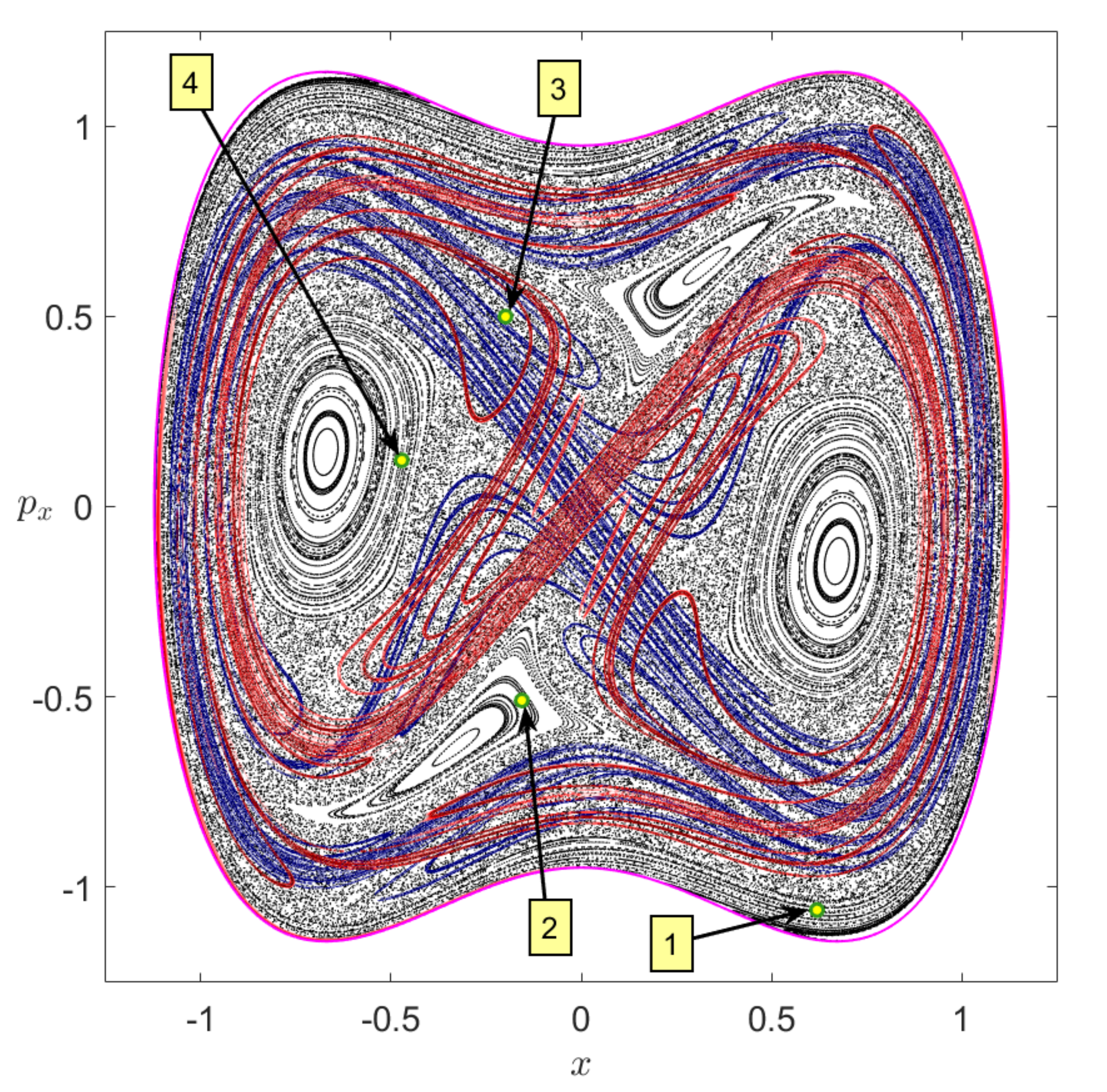}
		C)\includegraphics[scale=0.43]{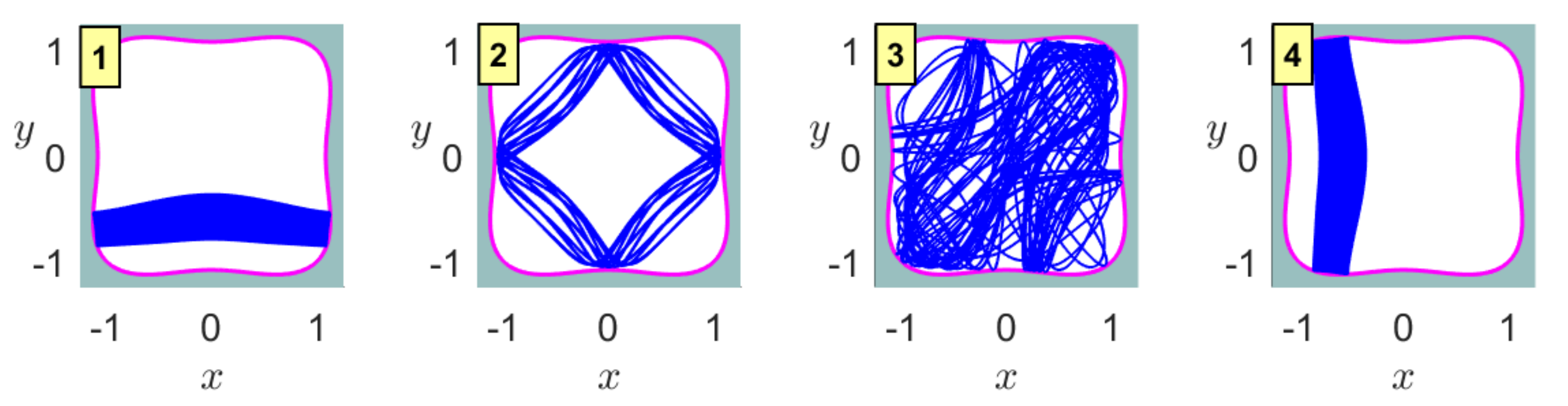}
	\end{center}
	\caption{Phase space structures on the surface of section given in Eq. \eqref{psos1}  at an energy $H_0 = 0.2$ and dynamical evolution of initial conditions for the symmetric and coupled Hamiltonian with $\alpha = 1$, $\delta = 0$ and $\beta = 0.2$. A) LDs calculated using $p = 1/2$ and $\tau = 12$. B) Poincar\'e map superimposed with the stable (blue) and unstable (red) manifolds extracted from the gradient of the LD function. We have also marked in the picture four different types of initial conditions. C) Dynamical evolution of the initial conditions selected in panel B). In all diagrams we have indicated the energy boundary with a magenta curve.} 
	\label{coupDyn_aboveIdx2}
\end{figure}

We finish our analysis of the coupled case by looking at an energy level $H_0 = 0.2$, which is above that of the index-2 saddle. Now, both sequential and concerted isomerization processes are possible. In order to detect the phase space regions that characterize the distinct dynamical behavior of the system, we follow the same approach of computing both LDs and the Poincar\'e map on the surface of section $y = -\sqrt{2}/2$. We present the results obtained with these methods in Fig. \ref{coupDyn_aboveIdx2}. Interestingly, an increase in the energy of the system results in a reduction of the chaotic region in comparison to what we observed in the previous case, see Fig. \ref{coupDyn_aboveIdx2} B). We select four initial conditions, three of them in the regularity regions, and another one in the homoclininc intersection between the stable and unstable manifolds of the index-2 saddle. The projections onto configuration space of the trajectories generated by these initial conditions are depicted in Fig. \ref{coupDyn_aboveIdx2} C). Initial condition $1$ is located inside the spherical cylinders that mediate transport between the wells in the lower region of the PES, and thus the system undergoes sequential isomerization. A similar dynamical behavior is observed for initial condition $4$, but the isomerization reaction occurs between the lower-left and upper-left wells. We take now initial condition $2$ on a tori corresponding to an island of regularity. In the uncoupled system this initial condition would give rise to concerted isomerization, see Fig. \ref{LD_sym_h_pos}, but when coupling is introduced into the system, tori break up and the invariant manifolds of the index-2 saddle act as transport barriers blocking the entrance of trajectories in the neighborhood of the hilltop. Hence, this reactive trajectory will follow sequential isomerization instead, visiting all the wells of the PES. Finally, if we choose an initial condition, labelled by $2$, inside one of the lobes formed by the homoclinic intersection of the invariant manifolds of the index-2 saddle, the trajectory will be chaotic and the system displays concerted isomerization. This is a consequence of the fact that the chaotic region is interconnected throughout the energy hypersurface and covers completely a neighborhood of the origin where the hilltop of the PES is located. For a better understanding on the effect that the coupling has on the invariant manifolds of the index-2 saddle, compare the phase space structures displayed in Fig. \ref{coupDyn_aboveIdx2} with those discussed for the uncoupled system in Fig. \ref{LD_sym_h_pos}. 

\section{Conclusions}
\label{sec:conc}

In this work we have applied the method of Lagrangian descriptors in order to reveal the geometrical template of phase space structures that characterizes the existing isomerization routes on a potential energy surface with an index-2 saddle and four wells separated by index-1 saddles. For the model Hamiltonian system discussed, we have studied the effects of simple  symmetry-breaking and nonlinear coupling perturbations on the phase space invariant manifolds and how these structures influence the distinct dynamical behavior of trajectories. Our analysis, which demonstrates the advantages that this technique brings for the exploration of dynamics in high-dimensional Hamiltonian systems, allows us to identify regions of phase space that correspond to initial conditions that undergo sequential or concerted isomerization. Furthermore, the approach followed has proven to successfully recover the phase space structures associated to index-2 saddle points, unveiling the dynamical transport mechanisms that they give rise to in phase space. We are confident that the investigations carried out here would help to shed some light on the dynamical significance of index-2, and higher order, saddle points for isomerization chemical reactions.

\section*{Acknowledgments}

The authors would like to acknowledge the financial support provided by the EPSRC Grant No. EP/P021123/1 and the Office of Naval Research Grant No. N00014-01-1-0769.

\bibliographystyle{natbib}
\bibliography{SNreac}

\appendix

%%%%%%%%%%%%%%%%%%%%%%%%%%%%%%%%%%%%%%%%%%%%%%%%%%%%%%%
%%%%%%%%%%%%%%%%%%%  APPENDICES %%%%%%%%%%%%%%%%%%%%%%%
%%%%%%%%%%%%%%%%%%%%%%%%%%%%%%%%%%%%%%%%%%%%%%%%%%%%%%%

\section{Lagrangian Descriptors}
\label{sec:appA}

The method of Lagrangian descriptors (LDs) is a trajectory-based scalar diagnostic from Dynamical Systems Theory that has the capability of revealing the geometrical template of phase space structures that characterizes qualitatively distinct dynamical behavior, a long-sought goal envisioned by Henri Poincar\'e in his pioneering work on the three body problem of Celestial Mechanics \cite{hp1890}. This technique was introduced a decade ago in \cite{madrid2009} to analyze Lagrangian transport and mixing in Geophysical flows \cite{mendoza2010}, and was originally defined by means of computing the arclength of the trajectories of initial conditions as they evolve forward and backward in time \cite{mancho2013lagrangian}. The approach provided by Lagrangian descriptors to unveil the skeleton of dynamical processes has found a myriad of applications in different scientific areas. For instance, in the context of Oceanography it has been used to plan transoceanic autonomous underwater vehicle missions by taking advantage of the underlying dynamical structure of ocean currents \cite{ramos2018}. Recently, this tool has also received recognition in the field of Chemistry, in particular in Transition State Theory \cite{craven2015lagrangian,craven2016deconstructing,craven2017lagrangian,revuelta2019unveiling}, where the computation of chemical reaction rates relies on the knowledge of the phase space structures that separate reactants from products. These high-dimensional structures characterizing reaction dynamics are typically related to normally hyperbolic invariant manifolds (NHIMs) and their stable and unstable manifolds that occur in Hamiltonian systems.
 
One of the biggest challenges when exploring the high-dimensional phase space of a dynamical system is to describe the behavior of ensembles of initial conditions, and to recover from their trajectory evolution the underlying geometrical phase space structures that govern the dynamical mechanisms of the flow. The problem that naturally arises in a   high-dimensional phase space is that the trajectories of ensembles of initial conditions that start nearby might get ``lost'' with respect to each other very quickly, making the use of classical nonlinear dynamics techniques that rely on tracking the location of neighboring trajectories computationally expensive and very difficult to interpret. On the other hand, the method of Lagrangian descriptors provides a revolutionary and radically different approach that resolves this issue, as it focuses on integrating a positive scalar function along the trajectory of any initial condition of the system instead of tracking their phase space location. This is probably one of the key ideas behind the success of this technique, since the phase space geometry is encoded in the initial conditions themselves.

Consider a dynamical system with general time-dependence in the form:
\begin{equation}
\dfrac{d\mathbf{x}}{dt} = \mathbf{v}(\mathbf{x},t) \;,\quad \mathbf{x} \in \mathbb{R}^{n} \;,\; t \in \mathbb{R} \;,
\label{gtp_dynSys}
\end{equation}
where the vector field $\mathbf{v}(\mathbf{x},t) \in C^{r}  (r \geq 1)$ in $\mathbf{x}$ and continuous in time. In order to explore the phase space structures of the dynamical system given by Hamilton's equations in Eq. \eqref{ham_eqs}, we have used in this work the $p$-norm definition of Lagrangian descriptors introduced in \cite{lopesino2017} with $p = 1/2$. Given an initial condition $\mathbf{x}_0$ at time $t_0$, take a fixed integration time $\tau > 0$ and $p \in (0,1]$. The method is defined as follows:
\begin{equation}
M_p(\mathbf{x}_{0},t_0,\tau) = \int^{t_0+\tau}_{t_0-\tau} \, \sum_{i=1}^{n} |v_{i}(\mathbf{x}(t;\mathbf{x}_0),t)|^p \; dt = M_p^{(b)}(\mathbf{x}_{0},t_0,\tau) + M_p^{(f)}(\mathbf{x}_{0},t_0,\tau) \;,
\label{Mp_function}
\end{equation}
where $M_p^{(b)}$ and $M_p^{(f)}$ represent, respectively, backward and forward integration of the initial condition, that is:
\begin{equation}
M_p^{(b)}(\mathbf{x}_{0},t_0,\tau) = \int^{t_0}_{t_0-\tau} \sum_{i=1}^{n} |v_{i}(\mathbf{x}(t;\mathbf{x}_0),t)|^p \; dt \quad,\quad M_p^{(f)}(\mathbf{x}_{0},t_0,\tau) = \int^{t_0+\tau}_{t_0} \sum_{i=1}^{n} |v_{i}(\mathbf{x}(t;\mathbf{x}_0),t)|^p \; dt
\end{equation}
It is important to highlight that with this definition of LDs one can mathematically prove in Hamiltonian systems that NHIMs and their stable and unstable manifolds are detected as singularities of the $M_p$ scalar field, that is, points at which the function is non-differentiable and thus its gradient takes very large values \cite{lopesino2017,demian2017,naik2019a}. Moreover, in this context it has also been shown that:
\begin{equation}
\mathcal{W}^u(\mathbf{x}_{0},t_0) = \textrm{argmin } M_p^{(b)}(\mathbf{x}_{0},t_0,\tau) \quad,\quad \mathcal{W}^s(\mathbf{x}_{0},t_0) = \textrm{argmin } M_p^{(f)}(\mathbf{x}_{0},t_0,\tau)
\label{min_LD_manifolds}
\end{equation}
where $\mathcal{W}^u$ and $\mathcal{W}^s$ are, respectively, the unstable and stable manifolds calculated at time $t_0$ and $\textrm{argmin}(\cdot)$ denotes the phase space coordinates $\mathbf{x}_0$ that minimize the function $M_p$. In addition, NHIMs at time $t_0$ can be calculated as the intersection of the stable and unstable manifolds:
\begin{equation}
\mathcal{N}(\mathbf{x}_{0},t_0) = \mathcal{W}^u(\mathbf{x}_{0},t_0) \cap \mathcal{W}^s(\mathbf{x}_{0},t_0) = \textrm{argmin} M_p(\mathbf{x}_{0},t_0,\tau)
\label{min_NHIM_LD}
\end{equation}

As we have mentioned before, NHIMs and their associated stable and unstable manifolds play a crucial role for the analysis of transition dynamics across index-1 saddles that separate two neighboring wells of a PES. This situation is representative for example in chemical isomerization problems, where two wells corresponding to different stable configurations of a given molecule are separated by an energy barrier that the system has to cross in order to undergo an isomerization reaction. In this context, several numerical studies have been carried out recently by means of Lagrangian descriptors to analyze escaping dynamics on open PESs \cite{demian2017,naik2019b,GG2019}.

At this point, it becomes clear from our discussion that the development of nonlinear techniques that have the capability of unveiling the high-dimensional phase space structures that characterize reaction mechanisms is of paramount importance. The methodology offered by LDs in this respect has been shown to have many advantages. For instance, it is straightforward to implement and computationally inexpensive when applied to systems with two or three DoF. But probably the most important feature of this tool is that it allows to produce a complete and detailed geometrical \textit{phase space tomography} in high dimensions by means of using low-dimensional phase space probes to extract the intersections of the phase space invariant manifolds with these slices \cite{demian2017,naik2019a,naik2019b,GG2019}. 

To finish this section we will illustrate how LDs can be used to detect the geometrical phase space structures, that is, the NHIMs and their stable and unstable invariant manifolds that characterize reaction dynamics between different wells of a PES separated by index-1 saddles. In particular, we will focus on the extraction of the phase space structures for the Hamiltonian system given in Eq. \eqref{ham_eqs} using the model parameters $\alpha = 1$ and considering the symmetric PES, that is $\delta = 0$. We will analyze the phase space structures for the system with energy $H = -0.2$ in the following Poincar\'e surface of section (SOS):
\begin{equation}
\mathcal{P}^{+}_{x,p_x} = \left\{ (x,y,p_x,p_y) \in \mathbb{R}^4 \;\big|\; y = -1/\sqrt{2} \; ,\; p_y > 0 \right\} \;.
\label{psos_ld}
\end{equation}
In this case, the energy of the system is below that of the index-2 saddle at the origin but above the energies of the four index-1 saddles surrounding it and connecting the four wells of the PES. Therefore, transport can take place between wells through the phase space bottlenecks that are open in the neighborhood region of the index-1 saddles. The transport mechanism in phase space between different wells is mediated by the stable and unstable manifolds (spherical cylinders) of the unstable periodic orbits (UPOs) corresponding to each index-1 saddle. These geometrical structures act as a 'reactive highway' allowing the system to transit from well to well which would correspond to a given molecule undergoing an isomerization reaction. Suppose that a given molecule starts as an isomer that can be identified with the lower-left well of the PES. The question is: Is it possible to determine the initial conditions of the state of the molecule for which it will transform  into an isomer associated to an adjacent well (sequential isomerization)? The answer to this question is 'yes' and the method of Lagrangian descriptors provides us with the key to unveil the phase space transport routes of this problem. 
Notice that since the chosen energy is below that of the index-2 saddle, the region in the neighborhood of the index-2 saddle is forbidden and thus the system cannot exhibit concerted isomerization, that is, transit from the lower-left well to the upper-right well across the index-2 saddle of the PES. When the system is uncoupled or the coupling is very small, the stable and unstable manifolds of the UPOs of the index-1 saddles of the PES do not give rise to heteroclinic intersections, so that the dynamics of the system remains trapped in the lower-left well or it can exhibit sequential isomerization to the lower-right or upper-left wells. On the other hand, if the DoF of the system are strongly coupled, heteroclinic connections arise between the manifolds of the UPOs of the different index-1 saddles of the PES and the system also admits sequential isomerization from the lower-left to the upper-right well. 

\begin{figure}[htbp]
	\centering
	A)\includegraphics[scale=0.2]{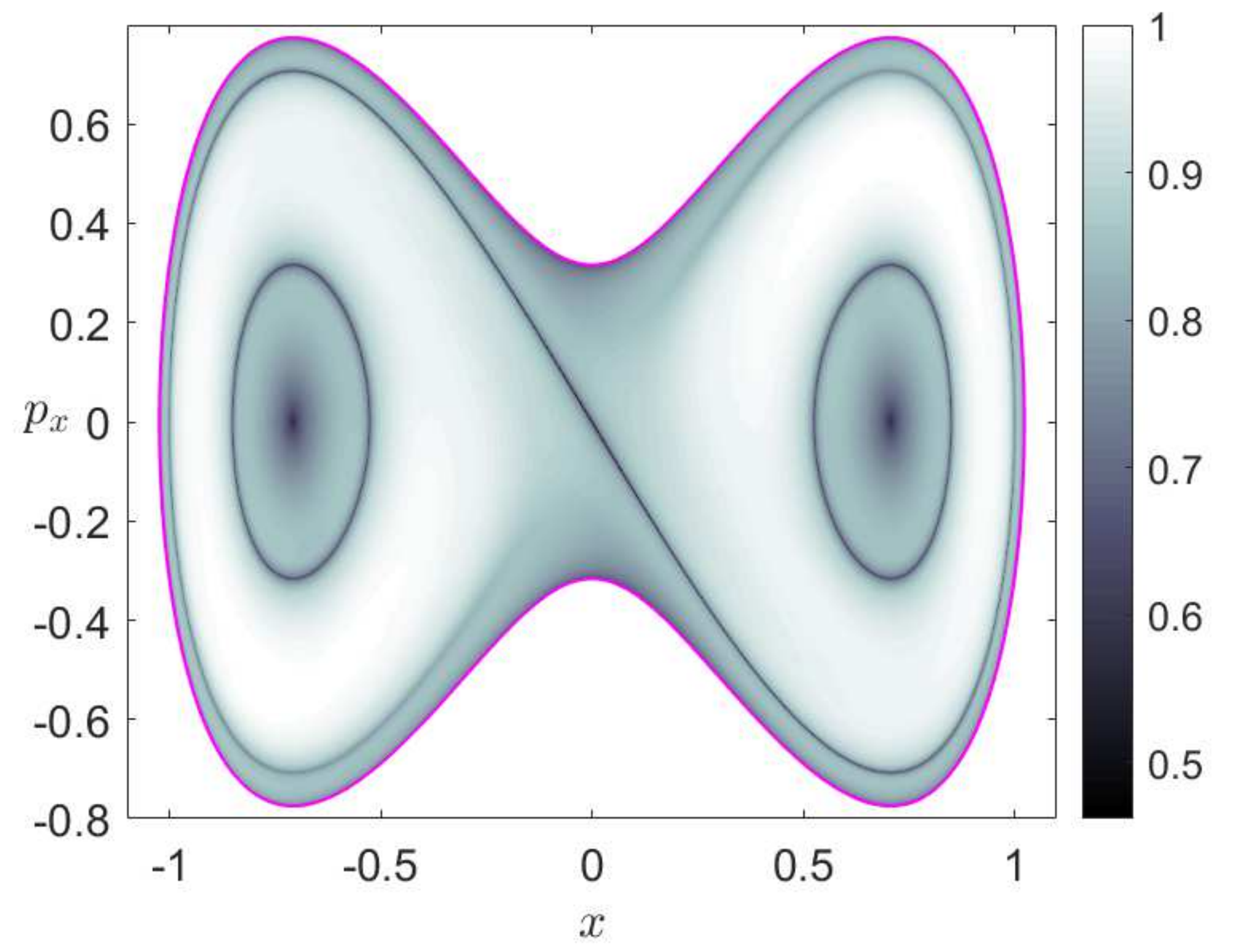}
	B)\includegraphics[scale=0.2]{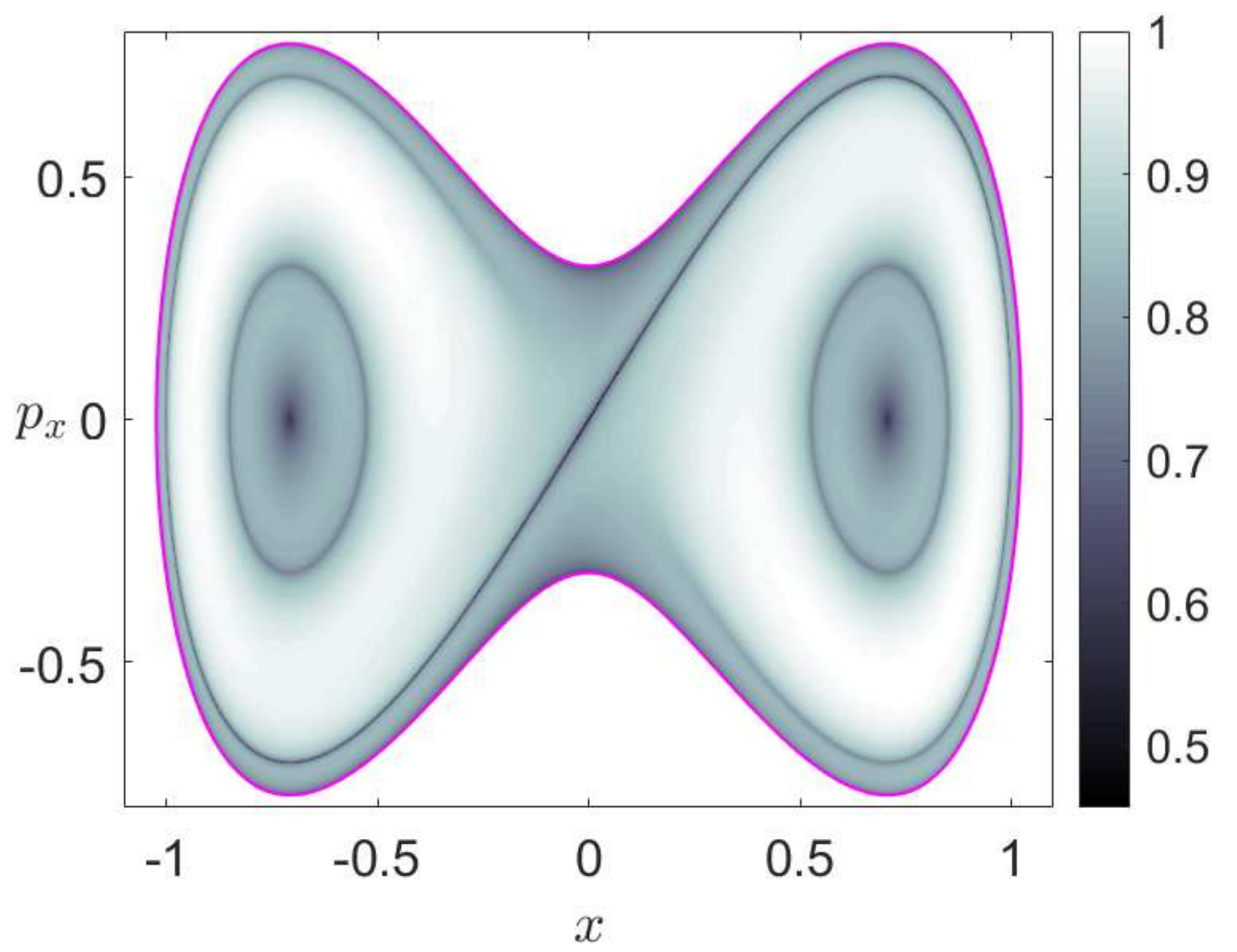}
	C)\includegraphics[scale=0.2]{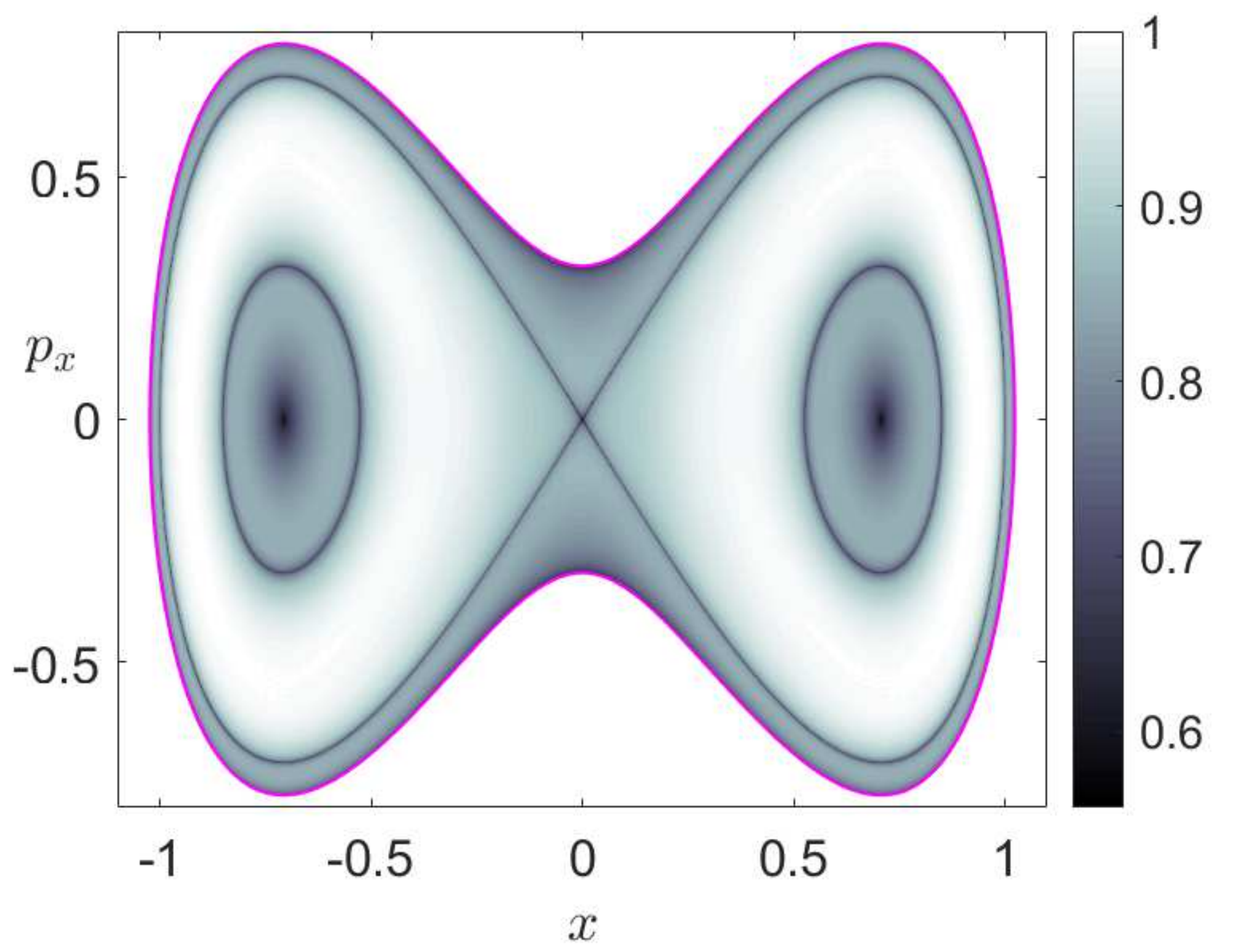}
	\caption{Computation of LDs in the PSOS in Eq. \eqref{psos_ld} using $\tau = 5$ and $p = 1/2$. A) Forward integration LDs; B) Backward integration LDs; C) The sum of forward and backward integration LDs. The energy boundary is represented in magenta.}
	\label{fig:LD_tau5}
\end{figure}

\begin{figure}[htbp]
\centering
A)\includegraphics[scale=0.28]{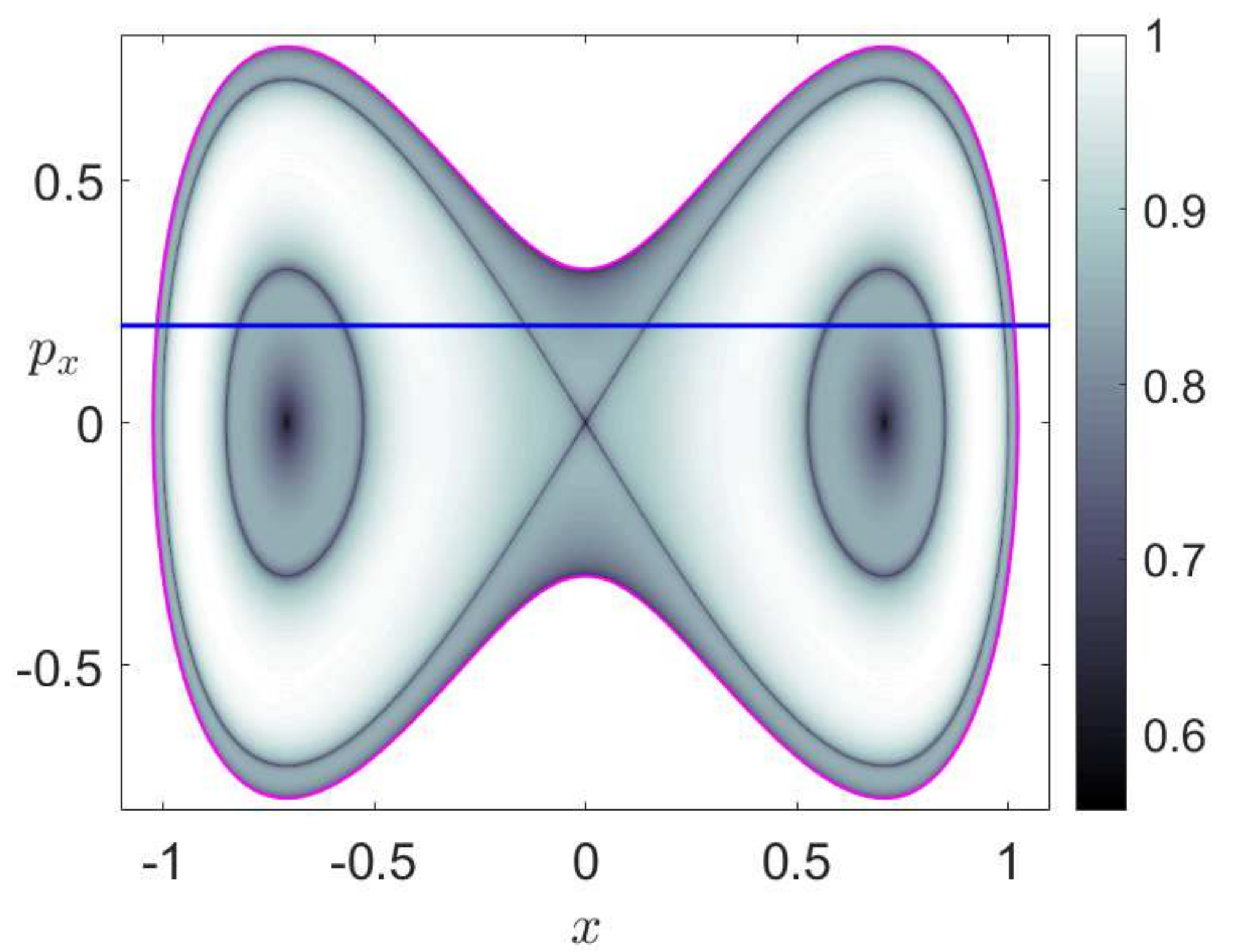}
B)\includegraphics[scale=0.275]{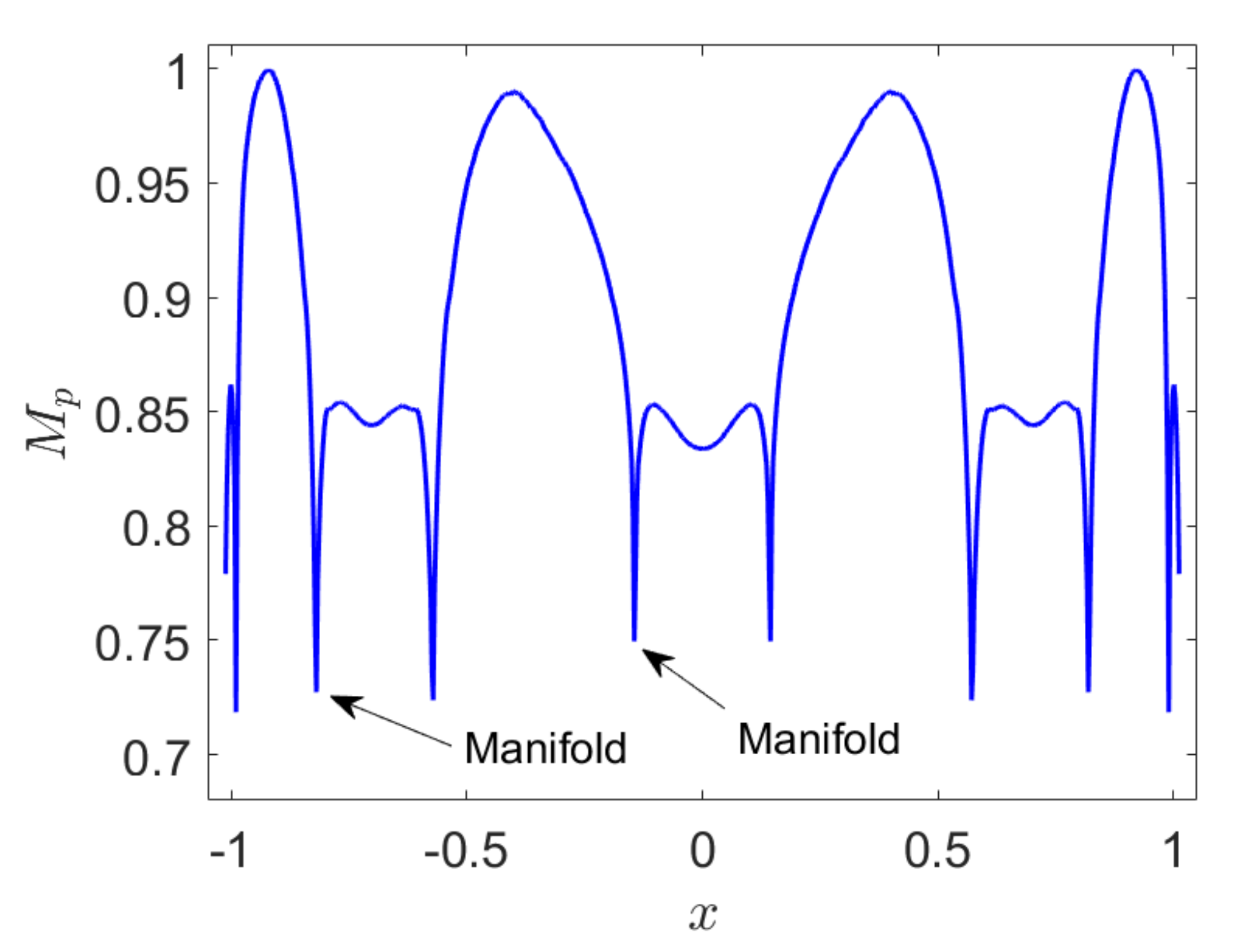}
\caption{Detection of the stable and unstable manifolds at phase space points where the LD scalar function is non-differentiable. A) LDs calculated on the Poincar\'e SOS defined in Eq. \eqref{psos_ld} using $\tau = 5$ and $p = 1/2$; B) Value of LDs along the line $p_x = 0.2$ depicted in blue in panel A).}
	\label{fig:LD_tau5_maniDetect}
\end{figure}

\begin{figure}[!h]
\begin{center}
A)\includegraphics[scale=0.3]{LD_tau_5_H_-02_y_-1sqrt2_alpha_1_delta_0_beta_0}
B)\includegraphics[scale=0.31]{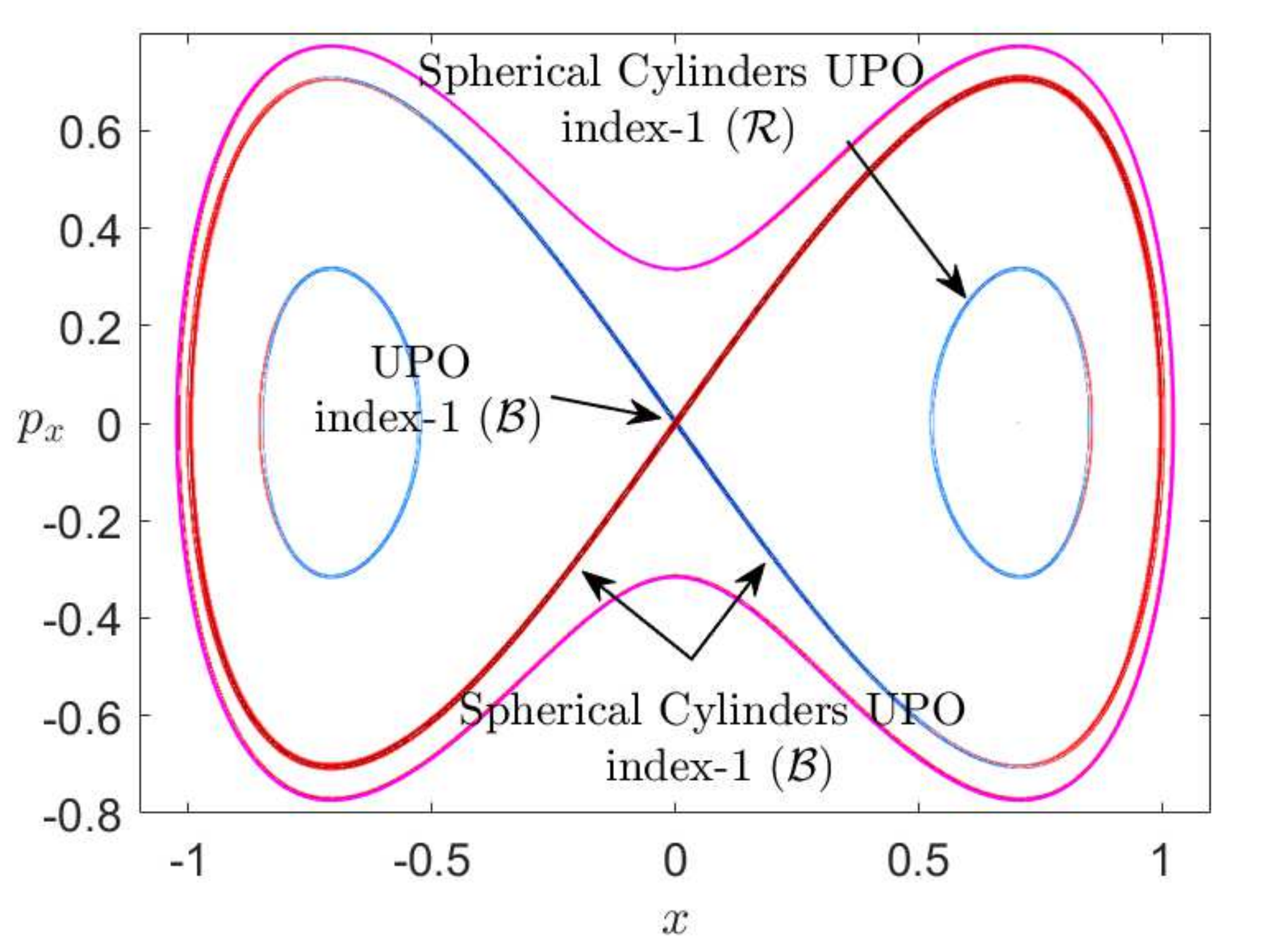}
C)\includegraphics[scale=0.3]{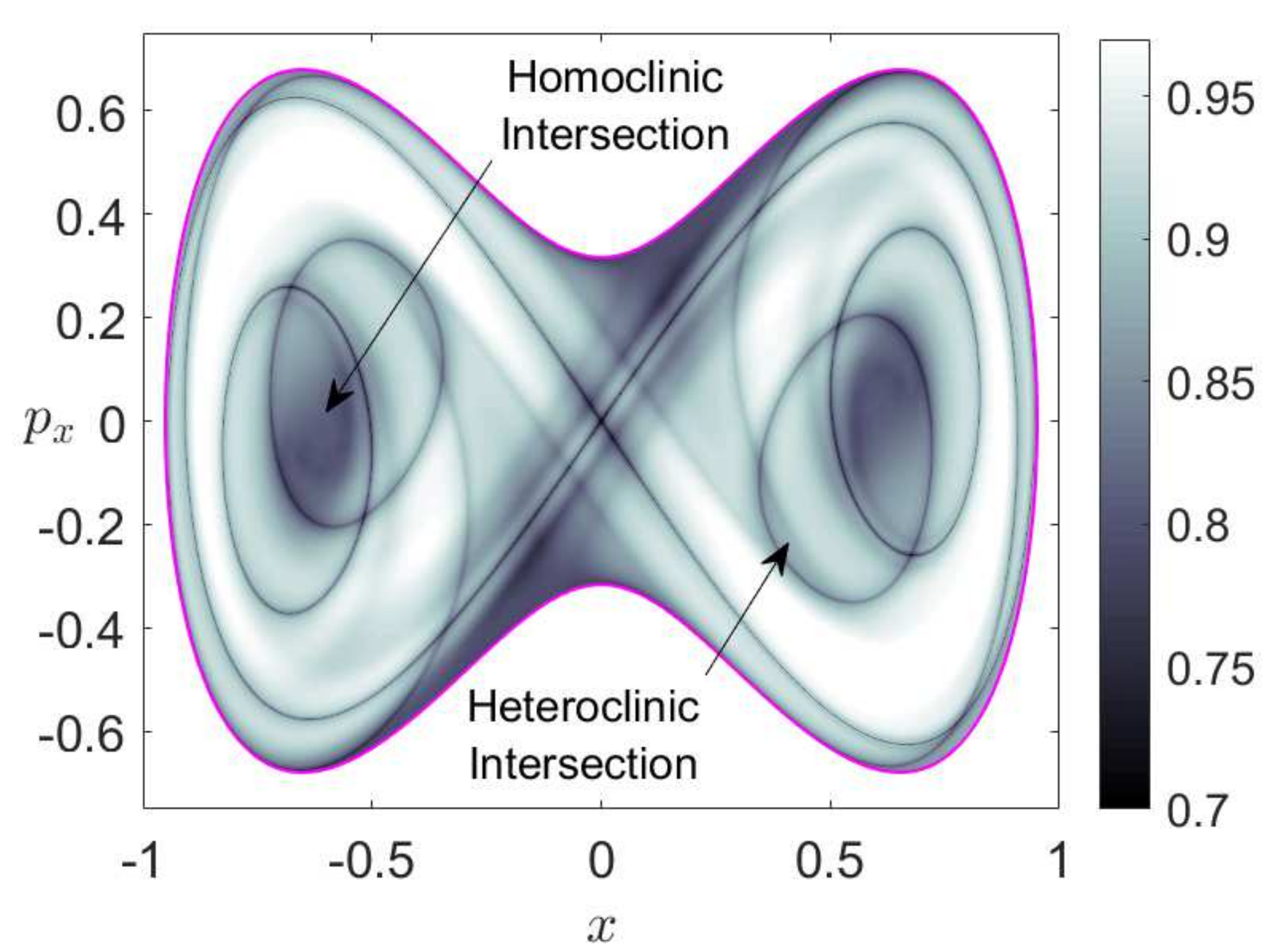}
D)\includegraphics[scale=0.3]{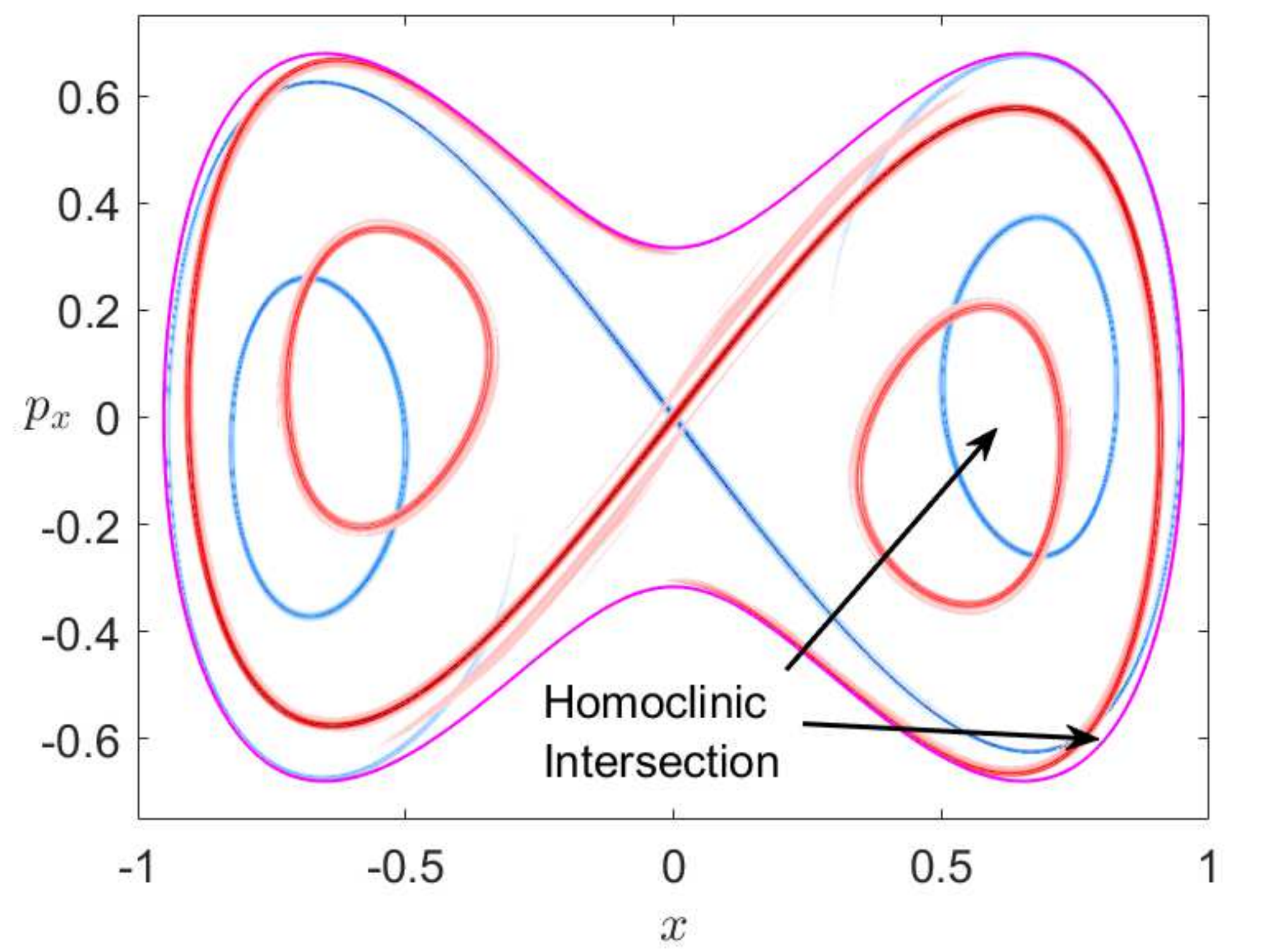}
E)\includegraphics[scale=0.3]{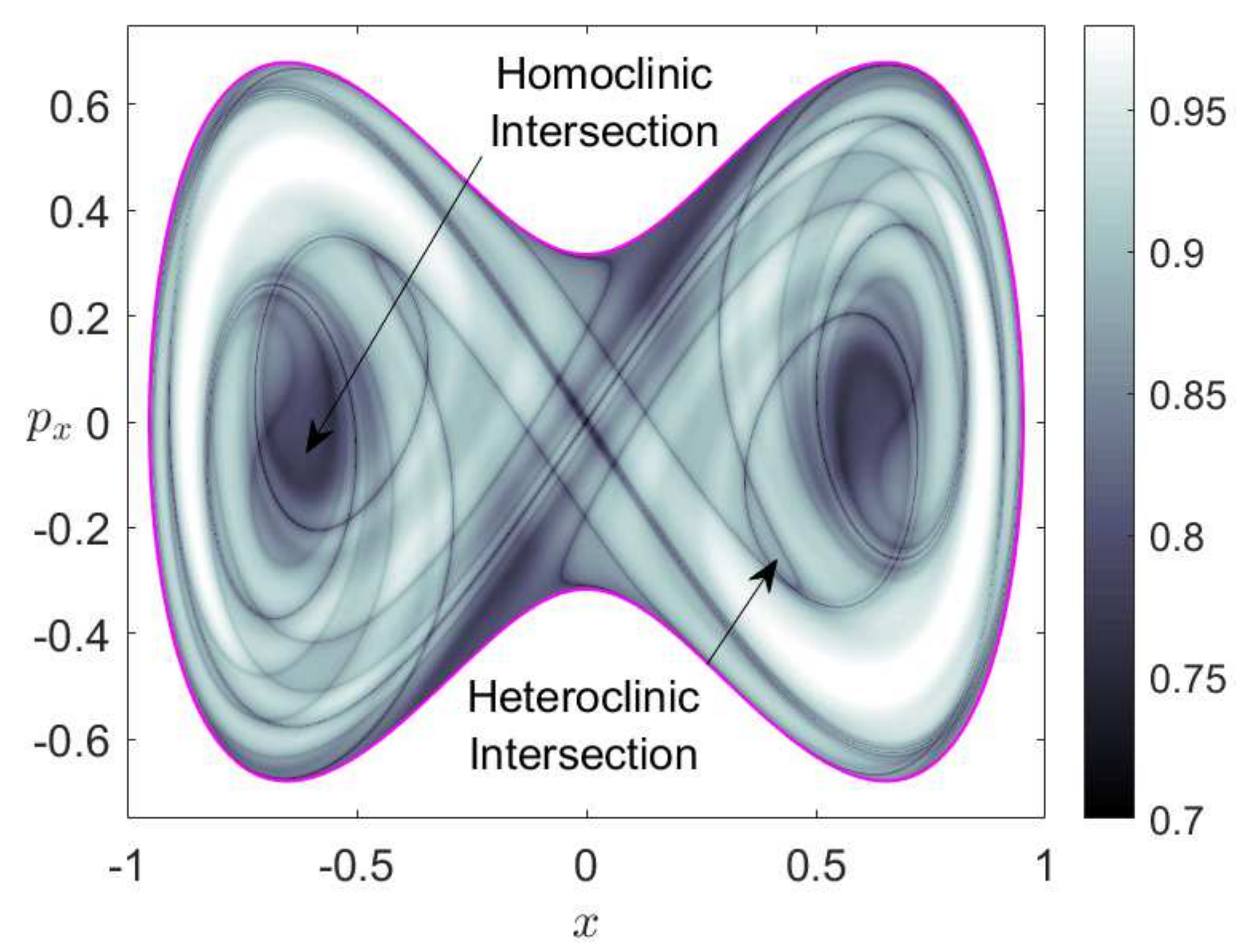}
F)\includegraphics[scale=0.31]{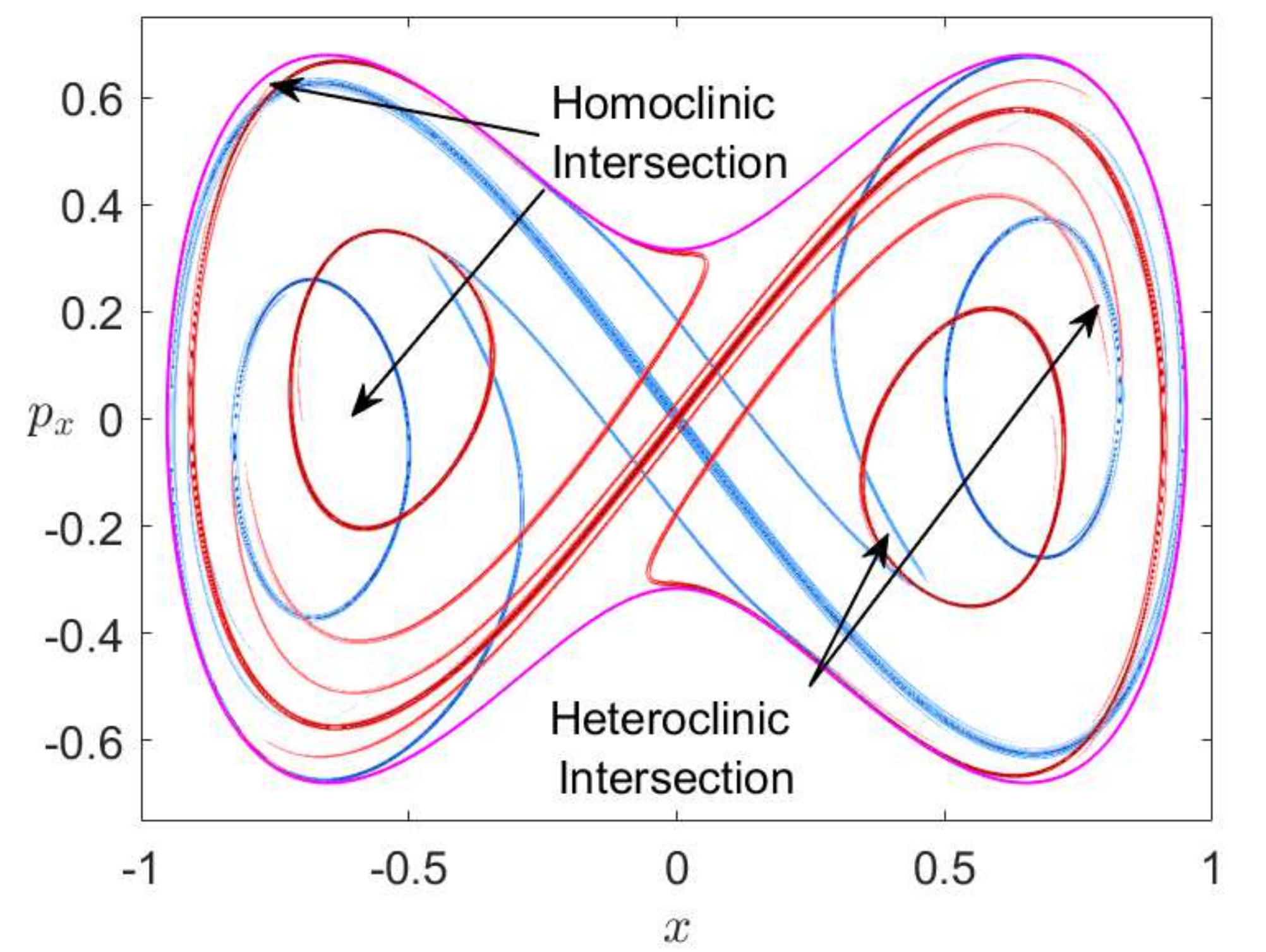}
\end{center}
\caption{(Left column) LDs calculated on the SOS given in Eq. \eqref{psos_ld} using: A) $\tau = 5$ and $\beta = 0$ (uncoupled system); B) $\tau = 6$ and $\beta = 0.3$ (coupled system); C) $\tau = 8$ and $\beta = 0.3$ (coupled system). (Right column) invariant stable (blue) and unstable (red) manifolds extracted from the gradient of the scalar function $M_p$. The magenta curve represents the energy boundary.}
\label{fig:LD_mani_extract}
\end{figure}

We start our explanation on how to use the method of Lagrangian descriptors by computing it on the Poincar\'e SOS in Eq. \eqref{psos_ld} for the uncoupled system, i.e. $\beta = 0$, using a small integration time $\tau = 5$. Once we have fixed the phase space slice where we want to calculate LDs, we select a grid of initial conditions and, after discarding those that are energetically unfeasible, we integrate the remaining conditions both forward and backward in time, and compute LDs using the definition in Eq. \eqref{Mp_function} with $p = 1/2$. The result is that if we plot the LDs values obtained from the forward/backward integration, the scalar field will reveal the stable/unstable manifolds in the SOS under consideration. Moreover, if we plot the combined sum of forward and backward integration, the method highlights both stable and unstable manifolds simultaneously. This is shown in Fig. \ref{fig:LD_tau5}, where the values of LDs for forward/backward integration is displayed in panel A)/B) and the combination of both is depicted in C). We can clearly see that the manifolds are detected at points where the LD scalar function is non-differentiable. To demonstrate this mathematical property, we represent in Fig. \ref{fig:LD_tau5_maniDetect} the values taken by the LD function along the line $p_x = 0.2$. Notice the jumps in the values of the function, which indicate non-differentiability by means of very large gradient values. Therefore, we can directly extract the invariant stable and unstable manifolds in the Poincar\'e SOS from the gradient, that is, using $||\nabla \mathcal{M}_p||$. This is illustrated in Fig. \ref{fig:LD_mani_extract}  where different values for the integration time have been used to compute LDs, in particular $\tau = 5$ for the uncoupled system, that is $\beta = 0$, and $\tau = 6$, $\tau = 8$ for the coupled system with $\beta = 0.3$. Observe in Fig. \ref{fig:LD_mani_extract}A)-B) that for the uncoupled system, the spherical cylinders of the UPOs of the left and right index-1 saddles do not intersect with the manifolds of the UPO of the bottom index-1 saddle. Moreover, since the system is uncoupled the stable and unstable manifolds for a given UPO coincide. All this shows that if an initial condition starts at the lower-left well of the PES, it can only transit to the lower-right  (upper-left) well if located inside the spherical cylinders of the UPO of the bottom (left) index-1. Once the DoF of the system are coupled, the stable and unstable manifolds of any UPO break apart and, if the coupling is strong enough, they can give rise to homoclinic and heteroclinic connections. The lobes formed by the intricate tangle of the manifolds, in particular, those formed by the manifolds of the UPO of the bottom index-1 with those of the UPOs of the left and right index-1 saddles will have the dynamical effect of allowing an initial condition that starts on the lower-left well of the PES reach, by means of sequential isomerization, the upper-right well. We illustrate in Fig. \ref{fig:LD_mani_extract}C)-F) how the method of Lagrangian descriptors has the capability of unveiling the lobe structure which is crucial for the understanding of the complex dynamical evolution of the system.

It is important to note here the crucial role that the integration time $\tau$ plays when it comes to revealing the invariant manifolds in phase space. As shown in Fig. \ref{fig:LD_mani_extract}, when we increase the value for the integration time, richer and more complex details of the underlying geometrical template of phase space structures is unveiled. This behavior is expected, since an increase of the integration time would imply incorporating more information about the past and future dynamical history of particle trajectories in the computation of LDs. This means that $\tau$ is intimately related to the time scales of the dynamical phenomena that take place in the model under consideration and thus, it is a parameter that is problem-dependent. Consequently, there is no general ``golden'' rule for selecting its value for exploring phase space, and thus it is usually selected from the information obtained by performing several numerical experiments. One needs to always bare in mind that there is a compromise between the complexity of the structures that one would like to reveal to explain a certain dynamical mechanism, and the interpretation of the intricate manifolds displayed in the LD scalar output. 

As a final remark, there is a key point that needs to be highlighted with regard to the application of LDs and that demonstrates the real potential of this tool with respect to other classical nonlinear dynamics techniques. We have seen that one can extract the stable and unstable manifolds from the gradient of the $M_p$ function and therefore obtain \textit{all} the manifolds coming from \textit{any} NHIM in phase space \textit{at the same time}. This is of course a tremendous advantage in comparison to the classical approach of computing stable and unstable manifolds that relies on locating first the NHIMs in phase space individually, and for every NHIM globalize the manifolds separately, for which a knowledge of the eigendirections is crucial. Consequently, the application of LDs offers the capability of recovering \textit{all} the relevant phase space structures in one \textit{shot} without having to study the local dynamics about equilibrium points of the dynamical system.

\section{Equilibrium Points of the Hamiltonian Model}
\label{sec:appB}

In this appendix we will describe the equilibrium points of the Hamiltonian model:
\begin{equation}
H(x,y,p_x,p_y) = \frac{1}{2}\left(p_x^2 + p_y^2\right) + V(x,y) = \frac{1}{2}\left(p_x^2 + p_y^2\right) + x^4 - \alpha x^2 - \delta x + y^4 - y^2 + \beta x^2y^2
\label{ham_model}
\end{equation}
where we consider that the masses in each DoF are $m_x = m_y = 1$. This Hamiltonian consists of two coupled DoF, $x$ and $y$, where the potential energy of the $y$ DoF is a symmetric double well, $W(y) = y^4-y^2$, while the $x$ DoF is defined by an asymmetric double well, $U(x) = x^4 - \alpha x^2 - \delta x$, where $\delta \geq 0$ is the asymmetry parameter and $\alpha > 0$ measures the barrier height, i.e. the vertical distance from one of the wells to the local maximum at the origin $x = 0$, in the symmetric case ($\delta = 0$). The term that defines the interaction between both DoF has the form $\beta x^2 y^2$, where $\beta \geq 0$ represents the coupling strength. In order to obtain the equilibrium points of the Hamiltonian in Eq. \eqref{ham_model} as a function of the model parameters, we need to analyze Hamilton's equations:
\begin{equation}
\begin{cases}
\dot{x} = \dfrac{\partial H}{\partial p_x} = p_x \\[.4cm]
\dot{y} = \dfrac{\partial H}{\partial p_y} = p_y \\[.4cm]
\dot{p}_x = -\dfrac{\partial H}{\partial x} = -4 x^3 + 2 \alpha x + \delta - 2 \beta \, x y^2 \\[.4cm]
\dot{p}_y = -\dfrac{\partial H}{\partial y} = -4 y^3 + 2 y - 2 \beta \, x^2 y
\end{cases}
\label{ham_eqs_model}
\end{equation}
It is straightforward to show that any equilibrium point $\mathbf{x}_e = (x_e,y_e,p_x^e,p_y^e)$ of above equations is contained in configuration space, i.e. $p_x^e = p_y^e = 0$, and its configuration coordinates are critical points of the PES, hence $\nabla V(x_e,y_e) = 0$. The critical points of the PES in the symmetric case $(\delta = 0)$ have been studied in \cite{collins2011}, where it was shown that there are nine critical points: four minima (potential wells), four index-1 saddles, and an index-2 saddle (hilltop) at the origin. The locations and potential energies of these points is summarized below:
\begin{align}
(x_e,y_e) =& \left(\pm \sqrt{\dfrac{2\alpha - \beta}{4 - \beta^2}},\pm \sqrt{\dfrac{2 - \alpha\beta}{4 - \beta^2}}\right) \hspace{-2.2cm} &,& \quad V(x_e,y_e) = -\dfrac{\alpha^2 + 1 - \alpha\beta}{4 - \beta^2} \hspace{-3cm} & \text{(minima)} \\[.3cm]
(x_e,y_e) =& \left(\pm \dfrac{\sqrt{2\alpha}}{2},0\right) \hspace{-2.2cm} &,& \quad V(x_e,y_e) = -\dfrac{\alpha^2}{4} \hspace{-3cm} & \text{(index-1 saddles)} \\[.3cm]
(x_e,y_e) =& \left(0,\pm \dfrac{\sqrt{2}}{2}\right) \hspace{-2.2cm} &,& \quad V(x_e,y_e) = -\dfrac{1}{4} \hspace{-3cm} & \text{(index-1 saddles)} \\[.3cm]
(x_e,y_e) =& \left(0,0\right) \hspace{-2.2cm} &,& \quad V(x_e,y_e) = 0 \hspace{-3cm} & \text{(index-2 saddle)}
\end{align}
In this work we focus our attention on the asymmetric case $(\delta > 0)$. Depending on whether the DoF of the system are coupled or not, we have the following analysis:

\begin{itemize}
	\item \underline{Uncoupled system ($\beta = 0$):} The equilibrium points $\mathbf{x}_e = (x_e,y_e,p_x^e,p_y^e)$ satisfy the equations:
	\begin{equation}
	p_x^e = p_y^e = 0 \quad , \quad  -4 y_e^3 + 2 y_e = 0 \quad , \quad 4 x_e^3 - 2 \alpha x_e - \delta = 0 \;.
	\end{equation}
	Solving for the $y$ DoF we get $y_e \in \lbrace -\sqrt{2}/2,0,\sqrt{2}/2 \rbrace$, and the $x$ DoF gives a cubic polynomial $f(x) = 4x^3 - 2 \alpha x - \delta$ whose roots can be obtained analytically using the formulas in \cite{brizzard2015}. If we take $\gamma$ such that $f^{\prime\prime}(\gamma) = 0$ and define:
	\begin{equation}
	\mu = \sqrt{-\frac{1}{3}f^{\prime}(\gamma)} \quad,\quad \phi = \arccos\left(-\dfrac{f(\gamma)}{\mu^3}\right)
	\label{cubic_rules}
	\end{equation}
	then for this cubic polynomial we have that $\gamma = 0$ and therefore:
	\begin{equation}
	\mu = \sqrt{\dfrac{2\alpha}{3}} \quad,\quad \phi = \arccos \left( \left(\dfrac{3}{2\alpha} \right)^{3/2} \delta \right)
	\end{equation}
	The roots of the cubic polynomial have the form:
	\begin{equation}
	\begin{cases}
	x_1 = \gamma + \mu \cos\left(\dfrac{\phi}{3}\right) = \mu \cos\left(\dfrac{\phi}{3}\right) \\[.4cm]
	x_{2,3} = \gamma - \mu \cos\left(\dfrac{\pi \pm \phi}{3}\right) = - \mu \cos\left(\dfrac{\pi \pm \phi}{3}\right) = -\dfrac{x_1}{2} \pm \dfrac{\mu\sqrt{3}}{2} \sin \left(\dfrac{\phi}{3}\right)
	\end{cases}
	\;.
	\label{roots_cubic_asymm}
	\end{equation}
	Observe that the two roots $x_{2,3}$ merge into one when $\phi = 0$, i.e. $x_{2,3} = -x_1/2$, and this occurs for:	
	\begin{equation}
	\left(\dfrac{3}{2\alpha} \right)^{3/2} \delta = 1 \quad \Leftrightarrow \quad \delta_c(\alpha) = \left(\dfrac{2\alpha}{3} \right)^{3/2} \;.
	\label{crit_asym_uncoup}
	\end{equation}
	Consequently, there is a saddle-node bifurcation in the geometry of the potential energy of the $x$ DoF at the value of the asymmetry parameter $\delta = \delta_c(\alpha)$. For $0 < \delta < \delta_c$ we have three real roots, and at the critical value $\delta = \delta_c$ the roots $x_2$ and $x_3$ merge and the cubic polynomial has only two real roots left: 
	\begin{equation}
	x_1 = \mu \quad,\quad x_2 = x_3 = -\mu / 2 = -\sqrt{\alpha/6} \;.
	\end{equation}
	Moreover, for values of the asymmetry $\delta > \delta_c$ only the root $x_1$ remains. In summary, when $0 \leq \delta < \delta_c$  the PES has 9 critical points. For $\delta = \delta_c$ only 6 critical points remain, since in each line $y = -\sqrt{2}/2, \, 0, \, -\sqrt{2}/2$ two of the roots merge as a result of a saddle-node bifurcation. Therefore, at this critical value of the asymmetry, three simultaneous saddle-node bifurcations occur in the system. Finally, for $\delta > \delta_c$ we are left with 3 critical points of the PES. We give below the energies of the critical points of the PES:
	\begin{eqnarray}
	(x_k,0) \quad ,& \quad  &V(x_k,0) = -\dfrac{x_k}{2} \left(\alpha x_k + \dfrac{3\delta}{2}\right) \;,\quad k \in \lbrace 1,2,3 \rbrace \\[.2cm]
	\left(x_k,\pm \sqrt{2}/2\right) \quad ,& \quad &V(x_k,\pm \sqrt{2}/2) = -\dfrac{x_k}{2} \left(\alpha x_k + \dfrac{3}{2}\delta \right) - \frac{1}{4} \;,\quad k \in \lbrace 1,2,3 \rbrace
	\end{eqnarray}
	In particular, the energies of the critical points where the saddle-node bifurcations take place are:
	\begin{equation}\label{energyeq}
	V(-\sqrt{\alpha/6},0) = \frac{\alpha^2}{12} \quad.\quad V(-\sqrt{\alpha/6},\pm \sqrt{2}/2) = \frac{\alpha^2-3}{12} \;.
	\end{equation}
	
	\item \underline{Coupled system ($\beta > 0$):} The equilibrium points $\mathbf{x}_e = (x_e,y_e,p_x^e,p_y^e)$ satisfy the equations:
	\begin{equation}
	p_x^e = p_y^e = 0 \quad , \quad  y_e \left(4 y_e^2 - 2 + 2\beta \, x_e^2\right) = 0 \quad , \quad 4 x_e^3 - 2 \alpha x_e - \delta + 2\beta \, x_e y_e^2 = 0 \;.
	\label{eqcond_coupl}
	\end{equation}
	First, we consider the case where $y_e = 0$. This yields that $4 x_e^3 - 2 \alpha x_e - \delta = 0$ and therefore the values of $x_e$ that solve this cubic polynomial have  already been discussed in the uncoupled case, see Eq. \eqref{roots_cubic_asymm}. Second, consider the condition $2\beta \, x_e^2 + 4 y_e^2 = 2$. This implies that the critical points lie in an ellipse of semiaxes $1/\sqrt{\beta}$ and $1/\sqrt{2}$. Solving for $y_e$ and substituting in the third equation of Eq. \eqref{eqcond_coupl} yields:
	\begin{equation}
	y_e^2 = \dfrac{1 - \beta \, x_e^2}{2} \quad , \quad \left(4 - \beta^2\right) x_e^3 - \left(2\alpha - \beta\right)x_e - \delta = 0 \;.
	\label{eqcond_coupl2}
	\end{equation}
	Now, if the coupling strength parameter is taken as $\beta = 2$ then the critical points are:
	\begin{equation}
	x_e = \dfrac{\delta}{2\left(1-\alpha\right)} \quad,\quad y_e = \pm \dfrac{\sqrt{2\left(1 - \alpha \right)^2 - \delta^2}}{2\,|1-\alpha|} \quad,\quad \alpha \neq 1 \;.
	\end{equation}
	We discuss next the case where $\beta < 2$. We can rewrite Eq. \eqref{eqcond_coupl2} as:
	\begin{equation}
	f(x) = 4 x^3 - \dfrac{4\left(2\alpha - \beta\right)}{4 - \beta^2} x - \dfrac{4\delta}{4-\beta^2} = 0 \;,
	\end{equation}	
	so that we can apply again the formulas in \cite{brizzard2015} to obtain the roots of the cubic polynomial, provided that $\alpha \geq \beta/2$. Using the rules in Eq. \eqref{cubic_rules} with $\gamma = 0$ gives:	
	\begin{equation}
	\mu = \frac{2\sqrt{3}}{3} \sqrt{\dfrac{2\alpha - \beta}{4 - \beta^2}} \quad,\quad \phi = \arccos\left( \dfrac{3\delta}{2\left(2\alpha - \beta\right)} \sqrt{\dfrac{3\left(4-\beta^2\right)}{2\alpha-\beta}}\right) \;,
	\end{equation}
	so that the roots of the cubic can be calculated using the expressions in Eq. \eqref{roots_cubic_asymm}. Notice that in this case, the merging of roots that correspond to bifurcations (saddle-node) in the geometry of the PES occur for the critical value of the asymmetry parameter:
	\begin{equation}
	\widetilde{\delta}_{c}(\alpha,\beta) = \frac{2}{3}\left(2\alpha - \beta\right) \sqrt{\dfrac{2\alpha - \beta}{3\left(4 - \beta^2\right)}} \;.
	\label{crit_asym_coup}
	\end{equation}	
	At this point it is important to highlight that, when the system is coupled, the merging of critical points of the PES occurs for different values of the asymmetry parameter. In fact, for $y_e = 0$ this takes place for $\delta_c$ given in Eq. \eqref{crit_asym_uncoup}, while the bifurcation of critical points at the locations $y_e = \pm \sqrt{(1-\beta x_e^2)/2}$ occurs for $\widetilde{\delta}_{c}$ in Eq. \eqref{crit_asym_coup}. In particular, for the range of parameter values that we are going to explore in this work, that is $\alpha = 1$, then $\widetilde{\delta}_{c} \leq \delta_c$ holds for any value of the coupling parameter $\beta \in (0,2)$, and therefore the bifurcations at the points $y_e = \pm \sqrt{(1-\beta x_e^2)/2}$ occur first. We finish this analysis by calculating the value of the energy at any of the critical points $(x_e,y_e)$ of the PES. To do so, we use the relationships between the configuration coordinates of the critical points established in Eqs. \eqref{eqcond_coupl} and \eqref{eqcond_coupl2}. This yields:
	\begin{eqnarray}
	(x_e,0) \quad ,& \quad  &V(x_e,0) = -\dfrac{x_e}{2} \left(\alpha x_e + \dfrac{3\delta}{2}\right) \\[.2cm]
	\left(x_e,y_e\right) \quad ,& \quad &V(x_e,y_e) = -\dfrac{x_e}{2} \left(\left(\alpha - \frac{\beta}{2}\right) x_e + \dfrac{3\delta}{2}\right) -\frac{1}{4}
	\end{eqnarray}
\end{itemize}

\end{document}